\documentclass[useAMS,usenatbib]{mn2e}
\usepackage{multirow}
\usepackage{graphicx}
\usepackage{amssymb} 
\def\aap{A\&A}

\def\apj{ApJ}
\def\apjs{ApJS}

\def\mnras{MNRAS}
\def\aj{AJ}
\def\nat{Nature}

\def\prd{Phys. Rev. D}

\def\araa{ARA\&A}       % Annual Review of Astron and Astrophys
\def\jcap{J. Cosmol. Astropart. Phys.}

\def \mpc {\,h^{-1}{\rm Mpc}}

\begin{document}

\title[Cosmological implications of the BOSS-CMASS $\xi_{\perp}(s)$ and $\xi_{\parallel}(s)$]
{
The clustering of galaxies in the SDSS-III Baryon Oscillation
 Spectroscopic Survey: cosmological constraints from the full shape of the clustering wedges 
}
\author[A.G. S\'anchez et al.]
{\parbox[t]{\textwidth}{
Ariel~G. S\'anchez$^{1}$\thanks{E-mail: arielsan@mpe.mpg.de},
Eyal~A. Kazin$^{2,3}$,
Florian Beutler$^{4}$,
Chia-Hsun Chuang$^{5}$,
Antonio~J. Cuesta$^{6}$,
Daniel~J. Eisenstein$^{7}$,
Marc Manera$^{8}$, 
Francesco Montesano$^{1}$,
Bob Nichol$^{8}$,
Nikhil Padmanabhan$^{6}$,
Will Percival$^{8}$,
Francisco Prada$^{5,9,10}$,
Ashley~J. Ross$^{8}$,
David~J. Schlegel$^{4}$,
Jeremy Tinker$^{11}$,
Rita Tojeiro$^{8}$,
David~H. Weinberg$^{12}$,
Xiaoying Xu$^{13}$,
J. Brinkmann$^{14}$,
Joel~R. Brownstein$^{15}$,
Donald~P. Schneider$^{16,17}$
and Daniel Thomas$^{8}$
}
\vspace*{6pt} \\ 
$^{1}$ Max-Planck-Institut f\"ur extraterrestrische Physik, Postfach 1312, Giessenbachstr., 85741 Garching, Germany\\ 
$^{2}$ Centre for Astrophysics and Supercomputing, Swinburne University of Technology, P.O. Box 218, Hawthorn, Victoria 3122, Australia\\
$^{3}$ ARC Centre of Excellence for All-sky Astrophysics (CAASTRO)\\
$^{4}$ Lawrence Berkeley National Laboratory, 1 Cyclotron Rd, Berkeley, CA 94720, USA\\
$^{5}$ Instituto de F\'isica Te\'orica, (UAM/CSIC), Universidad Aut\'onoma de Madrid, Cantoblanco, E-28049 Madrid, Spain\\
$^{6}$ Department of Physics, Yale University, 260 Whitney Ave, New Haven, CT 06520, USA\\
$^{7}$ Harvard-Smithsonian Center for Astrophysics, 60 Garden St., Cambridge, MA 02138, USA \\
$^{8}$ Institute of Cosmology \& Gravitation, University of Portsmouth, Dennis Sciama Building, Portsmouth PO1 3FX, UK\\
$^{9}$ Campus of International Excellence UAM+CSIC, Cantoblanco, E-28049 Madrid, Spain\\
$^{10}$ Instituto de Astrof\'isica de Andaluc\'ia (CSIC), Glorieta de la Astronom\'ia, E-18080 Granada, Spain\\
$^{11}$ Center for Cosmology and Particle Physics, New York University, NY 10003, USA\\
$^{12}$ Dept. of Astronomy and CCAPP, Ohio State University, Columbus, OH, USA\\
$^{13}$ Department of Physics, Carnegie Mellon University, 5000 Forbes Ave., Pittsburgh, PA 15213, USA\\
$^{14}$ Apache Point Observatory, P.O. Box 59, Sunspot, NM 88349-0059, USA\\
$^{15}$ Department of Physics and Astronomy, The University of Utah, 115 S 1400 E, Salt Lake City, UT 84112, USA\\
$^{16}$ Department of Astronomy and Astrophysics, The Pennsylvania State University, University Park, PA 16802, USA\\
$^{17}$ Institute for Gravitation and the Cosmos, The Pennsylvania State University, University Park, PA 16802, USA\\
}
\date{Submitted to MNRAS}
%\date{draft version}

\maketitle
%\clearpage
\begin{abstract}
We explore the cosmological implications of the clustering wedges,
$\xi_{\perp}(s)$ and $\xi_{\parallel}(s)$, of the CMASS Data Release 9 (DR9) sample of the SDSS-III
Baryon Oscillation Spectroscopic Survey (BOSS). These clustering wedges are defined by averaging the full
two-dimensional correlation function, $\xi(\mu,s)$, over the ranges $0<\mu<0.5$ and $0.5<\mu<1$,
respectively. These measurements allow us to constrain the parameter combinations
$D_{\rm A}(z)/r_{\rm s}(z_{\rm d})=9.03\pm0.21$ and
$cz/(r_{\rm s}(z_{\rm d})H(z))=12.14\pm0.43$ at the mean redshift of the sample, $z=0.57$. 
We combine the information from the clustering wedges with recent measurements
of CMB, BAO and type Ia supernovae to obtain constraints on the cosmological parameters of the
standard $\Lambda$CDM model and a number of potential extensions.
The information encoded in the clustering wedges is most useful when the dark energy equation of state is 
allowed to deviate from its standard $\Lambda$CDM value.
The combination of all datasets shows no evidence of a deviation from a constant dark energy equation of state,
in which case we find $w_{\rm DE}=-1.013\pm0.064$, in complete agreement with a cosmological constant.
We explore potential deviations from general relativity by constraining the growth rate $f(z)={\rm d}\ln D(a)/{\rm d}\ln a$,
in which case the combination of the CMASS clustering wedges with CMB data implies $f(z=0.57)=0.719_{-0.096}^{+0.092}$,
in accordance with the predictions of GR. 
Our results clearly illustrate the additional constraining power of anisotropic clustering measurements
with respect to that of angle-averaged quantities.
\end{abstract}
\begin{keywords}
cosmological parameters, large scale structure of the universe
\end{keywords}

%\pagebreak
\section{Introduction}
\label{sec:intro}

Observations of the large-scale structure (LSS) of the Universe have shaped our current understanding
of cosmic history, playing a central role at establishing the $\Lambda$CDM model as the current cosmological paradigm 
\citep{Davis1983,Maddox1990,Percival2001,Tegmark2004,Cole2005,Eisenstein2005,Anderson2012}.
The information encoded in the large-scale galaxy distribution, usually characterized in terms of two-point statistics like the power spectrum, $P(k)$, or the 
correlation function, $\xi(s)$, is highly complementary to that of cosmic microwave background (CMB) measurements,
as it helps to break the degeneracies between various cosmological parameters which are inherent to this dataset \citep{Efstathiou1999}.
The combination of CMB and LSS datasets has been used to place tight constraints on the basic set of cosmological parameters,
restricting the range of possible deviations from the of the $\Lambda$CDM model
\citep[e.g.][]{Percival2002,Percival2010,Tegmark2004,Sanchez2006,Sanchez2009,Sanchez2012,Spergel2007,Komatsu2009,Komatsu2011,
Reid2010,Blake2011,Montesano2012,Anderson2012,Parkinson2012,Samushia2013}.

A particularly important source of cosmological information contained in the large-scale galaxy clustering pattern
is the signature of the baryon acoustic oscillations (BAO).
These are the remnants of the acoustic waves that propagated through the photon-baryon fluid prior to
recombination. The signature of the BAO appears as a broad peak in the correlation function, located at a scale
closely related to the size of the sound horizon at the drag redshift, $r_{\rm s}(z_{\rm d})\simeq150\,{\rm Mpc}$ \citep{Matsubara2004}.
In the power spectrum, the Fourier transform of $\xi(s)$, the BAO signal appears as an oscillatory amplitude 
modulation, whose wavelength is related to $\lambda_{\rm s}\simeq2\pi/r_{\rm s}(z_{\rm d})$ \citep{Eisenstein1998,Meiksin1999}.
As CMB observations provide accurate measurements of $r_{\rm s}(z_{\rm d})$ \citep[e.g.][]{Komatsu2011,Hinshaw2012}, 
the acoustic scale inferred from the galaxy clustering in the direction parallel and 
perpendicular to the line-of-sight can be used as a standard ruler to 
measure the redshift evolution of the Hubble parameter, $H(z)$, and the angular diameter distance, $D_{\rm A}(z)$,
through the Alcock--Paczynski test \citep{Alcock1979,Blake2003,Linder2003}.
The BAO feature was first detected in the correlation function of the luminous red galaxy (LRG)
sample of the Sloan Digital Sky Survey \citep[SDSS,][]{York2000} by \citet{Eisenstein2005} and the 
power spectrum of the Two-degree Field Galaxy Redshift survey \citep[2dFGRS,][]{Colless2001,Colless2003} by \citet{Cole2005}.
This detection has been confirmed with increasing precision using a variety of datasets and techniques
\citep{Padmanabhan2007,Percival2007,Hutsi2010,Percival2010,Cabre2009,Gaztanaga2009a,Gaztanaga2009b,
Kazin2010,Beutler2011,Blake2011,Seo2012,Anderson2012}.

By offering a powerful method to probe the expansion history of the Universe, LSS observations are 
among the most promising tools to obtain new clues on one of the greatest unanswered questions 
in physics today: what is the origin of cosmic acceleration?
This phenomenon might be driven by the repulsive effect of an unknown energy component, called dark energy,
with an equation of state parameter, defined as the ratio of its pressure to density, satisfying $w_{\rm DE}< -1/3$. 
The most simple explanation of this component is that it is due to vacuum energy or a cosmological constant, 
characterized by $w_{\rm DE}=-1$.
As this hypothesis is consistent with all current cosmological observations, it has become the standard 
model for dark energy. However, a variety of alternative models have been proposed
\citep[for a review see e.g., ][]{Peebles2003,Frieman2008}.
Alternatively, cosmic acceleration could be the signature of the
breakdown of general relativity (GR) on cosmological scales. This possibility can be distinguished from the dark
energy scenario by simultaneous measurements of the expansion history of the Universe and the growth of density
fluctuations. 
A detection of a deviation from $w_{\rm DE}=-1$ or from the predictions of general relativity, at any
time in cosmic history, would have strong implications on our understanding of cosmic acceleration.

To date, most BAO analyses have focussed on angle-averaged measurements. 
However, these measurements are only sensitive to the combination $D_{\rm A}(z)^2/H(z)$ \citep{Eisenstein2005},
providing degenerate constraints on $H(z)$ and $D_{\rm A}(z)$.
The full constraining power of the BAO test can be exploited by means of anisotropic clustering measurements
\citep{Hu2003,Wagner2008,Shoji2009}, such as the two-dimensional correlation function $\xi(\mu,s)$, where $\mu$ is the cosine of
the angle between the separation vector $\mathbf{s}$ and the line-of-sight direction. 
Although some studies have attempted to extract cosmological information from this measurement
\citep{Okumura2008,Blake2011,Chuang2012}, even for large-volume surveys the expected signal-to-noise ratio in the large-scale
two-dimensional correlation function is low. In addition to this limitation, the use of the full $\xi(\mu,s)$ poses problems related to the size
of its covariance matrix, whose robust estimation and inversion becomes problematic.

 Fortunately, the information content in $\xi(\mu,s)$ can be condensed
into a reduced number of one-dimensional projections that can be measured with higher signal-to-noise ratio, and
whose covariance matrices can be managed more easily. \citet{Padmanabhan2008} proposed to use the
first multipoles from the expansion of $\xi(\mu,s)$ in terms of Legendre polynomials. The joint
analysis of the angle-averaged correlation function (monopole) and the next non-zero multipole (quadrupole) 
provides measurements of the combinations $D_{\rm A}(z)^2/H(z)$ and 
$D_{\rm A}(z)\,H(z)$, from which the values of $H(z)$ and $D_{\rm A}(z)$ can be derived.
Alternatively, \citet*{Kazin2012} proposed to use the clustering wedges statistic, $\xi_{\Delta\mu}(s)$,
defined as the average of $\xi(\mu,s)$ over a given interval $\Delta\mu$. As shown by \citet{Kazin2012}, the 
use of two wide clustering wedges, $\xi_{\perp}(s)$ and $\xi_{\parallel}(s)$, defined for 
$0<\mu<0.5$ and $0.5<\mu<1$, respectively, can break the degeneracies obtained from the angle-averaged quantities,
providing separate constraints on $H(z)$ and $D_{\rm A}(z)$.

The high constraining power of LSS observations has led to the construction of a new generation of galaxy surveys.
By probing much larger volumes than their predecessors, these surveys can provide more accurate views of the
large-scale galaxy clustering pattern than ever before.
Examples of these new surveys include the completed WiggleZ \citep{Drinkwater2010}, %the future HETDEX \citep{cita},
and the ongoing Baryon Oscillation Spectroscopic Survey \citep[BOSS,][]{Dawson2013}.
BOSS is one of the four components of SDSS--III \citep{Eisenstein2011}. BOSS is
designed to provide high-precision BAO measurements at intermediate redshifts ($z\simeq0.5$)
from the large-scale galaxy clustering, and at high redshift ($z\simeq2.5$) from the Ly$\alpha$ forest
signal inferred from a quasar sample.

The first results from BOSS, based on the galaxy and quasar samples from SDSS Data Release 9
\citep[DR9,][]{Ahn2012}, have shown a clear detection of the BAO feature \citep{Anderson2012,Busca2012}. 
This information has been used to place constraints on cosmological parameters 
\citep{Anderson2012,Sanchez2012,Reid2012,Samushia2013,Ross2013,Zhao2012}.
In particular, \citet{Sanchez2012} explored the cosmological implications of the full shape of the angle-averaged 
correlation function of a high-redshift galaxy sample from BOSS DR9.
In this paper we extend this analysis by exploring the cosmological implications of the
full shape of the clustering wedges of the same sample.
We combine this information with recent measurements
of CMB, BAO and type Ia supernovae data. We derive constraints on the parameters of the standard
$\Lambda$CDM model, and on a number of potential extensions. We place particular emphasis on the effect
of the additional information contained in the clustering wedges with respect to that 
of the angle-averaged correlation function. 
 \citet{Reid2012} and \citet{Samushia2013} used the full shape of the monopole and quadrupole correlation
functions of the same galaxy sample to constrain the angular diameter distance, the Hubble expansion rate,
and the growth rate of structure, and explored the cosmological implications of these measurements.
We compare our results with these studies to assess the consistency between
our analysis techniques.

Our analysis is part of a series of papers examining the information in the anisotropic clustering
pattern of the CMASS sample of BOSS DR9. \citet{Chuang2013} present an analysis of the cosmological implications
of the full shape of the monopole-quadrupole pair of this galaxy sample. \citet{Kazin2013} perform a detailed analysis 
of the geometrical information that can be derived from the BAO signal in these measurements and the clustering wedges
in a model-independent fashion.
Finally, \citet{Abalone2013}
%\footnote{The authors of this paper are listed alphabetically. The first author might 
%change by the time of submission.} 
use the BAO-only results obtained from clustering wedges and multipoles to 
derive constraints on cosmological parameters.

The outline of this paper is as follows. In Section~\ref{sec:data} we describe our galaxy sample,
the procedure we follow to measure the clustering wedges, and the
additional datasets that we include in our analysis.
Our model of the full shape of the clustering wedges is described in Section~\ref{sec:model}.
Section~\ref{sec:method} describes our methodology to obtain cosmological constraints and the tests we have 
performed by applying it to a set of mock catalogues. In Section~\ref{sec:results} we present
the constraints on cosmological parameters obtained from different combinations of datasets and parameter
spaces. Finally, Section~\ref{sec:conclusions} contains our main conclusions.

\section{The data}
\label{sec:data}

\subsection{The clustering wedges of the BOSS-CMASS galaxies}
\label{sec:clustering}

BOSS targets two separate luminous galaxy samples, LOWZ and CMASS, designed to have 
a roughly constant number density $n\simeq3 \times 10^{-4} h^3{\rm Mpc}^{-3}$
over the redshift range $0.2<z<0.7$ \citep[][Padmanabhan et al. in preparation]{Eisenstein2011,Dawson2013}.
The selection criteria of these samples are based on the multicolour
SDSS imaging done with the dedicated 2.5-m Sloan Telescope \citep{Gunn2006} located at Apache Point
Observatory, using a drift-scanning mosaic CCD camera \citep{Gunn1998}.
The redshift of the galaxies in the LOWZ and CMASS samples are measured by applying the minimum-$\chi^2$
template-fitting procedure described in \citet{Aihara2011}, with templates and methods updated for
BOSS data as described in \citet{Bolton2012} to the spectra obtained with the double-armed BOSS
spectrographs \citep{Smee2012}. 

Our analysis is based on the CMASS sample corresponding to the SDSS Data Release 9 (DR9) \citep{Ahn2012}.
The CMASS sample can be described as approximately complete down to a limiting stellar mass \citep{Maraston2012}, 
and is dominated by early type galaxies, although it contains a significant
fraction of massive spirals \citep[$\sim$26 per cent,][]{Masters2011}.
Most of the galaxies in this sample are central galaxies, with a $\sim$10 per cent satellite
fraction \citep{White2011,Nuza2012}.

\citet{Anderson2012} presents a detailed description of the construction of a catalogue
for LSS studies based on the CMASS sample. Different aspects of the clustering properties of this sample 
have been analysed by \citet{Anderson2012}, \citet{Sanchez2012}, \citet{Reid2012}, \citet{Tojeiro2012},
and \citet{Nuza2012}.
In particular, \citet{Sanchez2012} analysed the large-scale angle-averaged correlation function 
to infer constraints on cosmological parameters. Here we extend this analysis by 
analysing the clustering properties of the same sample by means of the clustering wedges
statistic, as defined in \citet{Kazin2012}.

A general clustering wedge $\xi_{\Delta\mu}(s)$ can be obtained by averaging the full
two-dimensional correlation function $\xi(\mu,s)$ over a given interval
$\Delta\mu=\mu_{\rm max}-\mu_{\rm min}$, that is
\begin{equation}
\xi_{\Delta\mu}(s)\equiv \frac{1}{\Delta \mu}\int^{\mu_{\rm max}}_{\mu_{\rm min}}{\xi(\mu,s)}\,{{\rm d}\mu}.
\label{eq:wedges}
\end{equation}
We use two wide clustering wedges, $\xi_{\perp}(s)$ and $\xi_{\parallel}(s)$, defined 
for the intervals $0 \le \mu \le 0.5$ and $0.5 \le \mu \le 1$ respectively. 
The basic procedure implemented to obtain these measurements from the CMASS sample is analogous
to that of \citet{Anderson2012} and \citet{Sanchez2012}. Here we summarize the most important
points and refer the reader to these studies for more details.

We convert the observed redshifts into distances assuming a flat $\Lambda$CDM fiducial cosmology
with matter density, in units of the critical density, of $\Omega_{\rm m}=0.274$.
This is the same fiducial 
cosmology assumed by the recent clustering analyses of the CMASS DR9 sample
\citep{Anderson2012,Sanchez2012,Manera2012,Ross2012,Reid2012,Tojeiro2012}.
The effect of the fiducial cosmology on the measurements of the clustering wedges will be discussed in
Section~\ref{sec:da_h}.
  
We compute the full correlation function $\xi(\mu,s)$ using the estimator of \citet{Landy1993},
with a random sample containing 50 times more objects than the original CMASS catalogue, constructed to follow
the same selection function. We infer the clustering wedges $\xi_{\perp}(s)$ and $\xi_{\parallel}(s)$
by integrating the full $\xi(\mu,s)$ according to equation~(\ref{eq:wedges}).

\begin{figure}
\centering
\centerline{\includegraphics[width=0.45\textwidth]{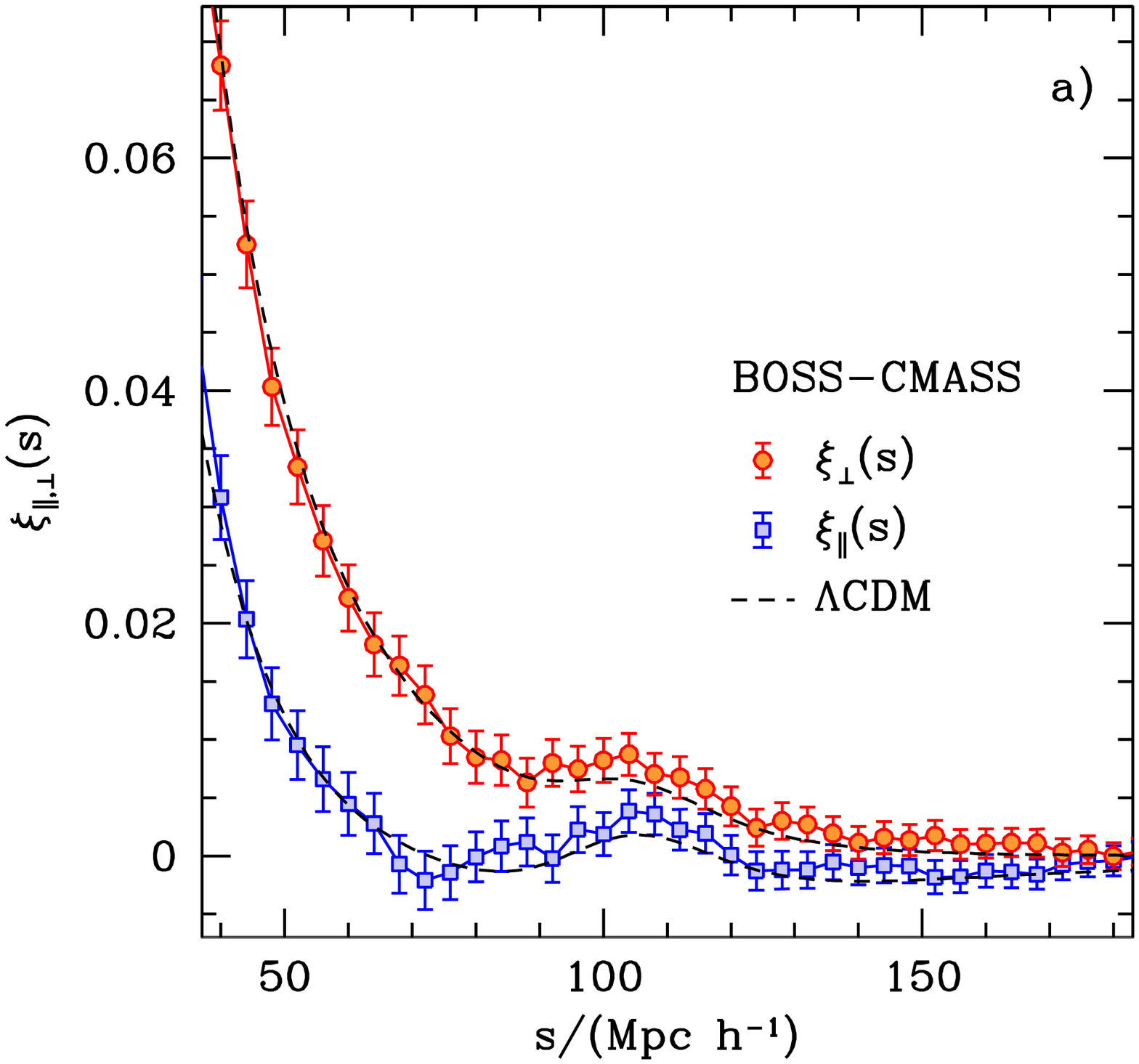}}
\centerline{\includegraphics[width=0.45\textwidth]{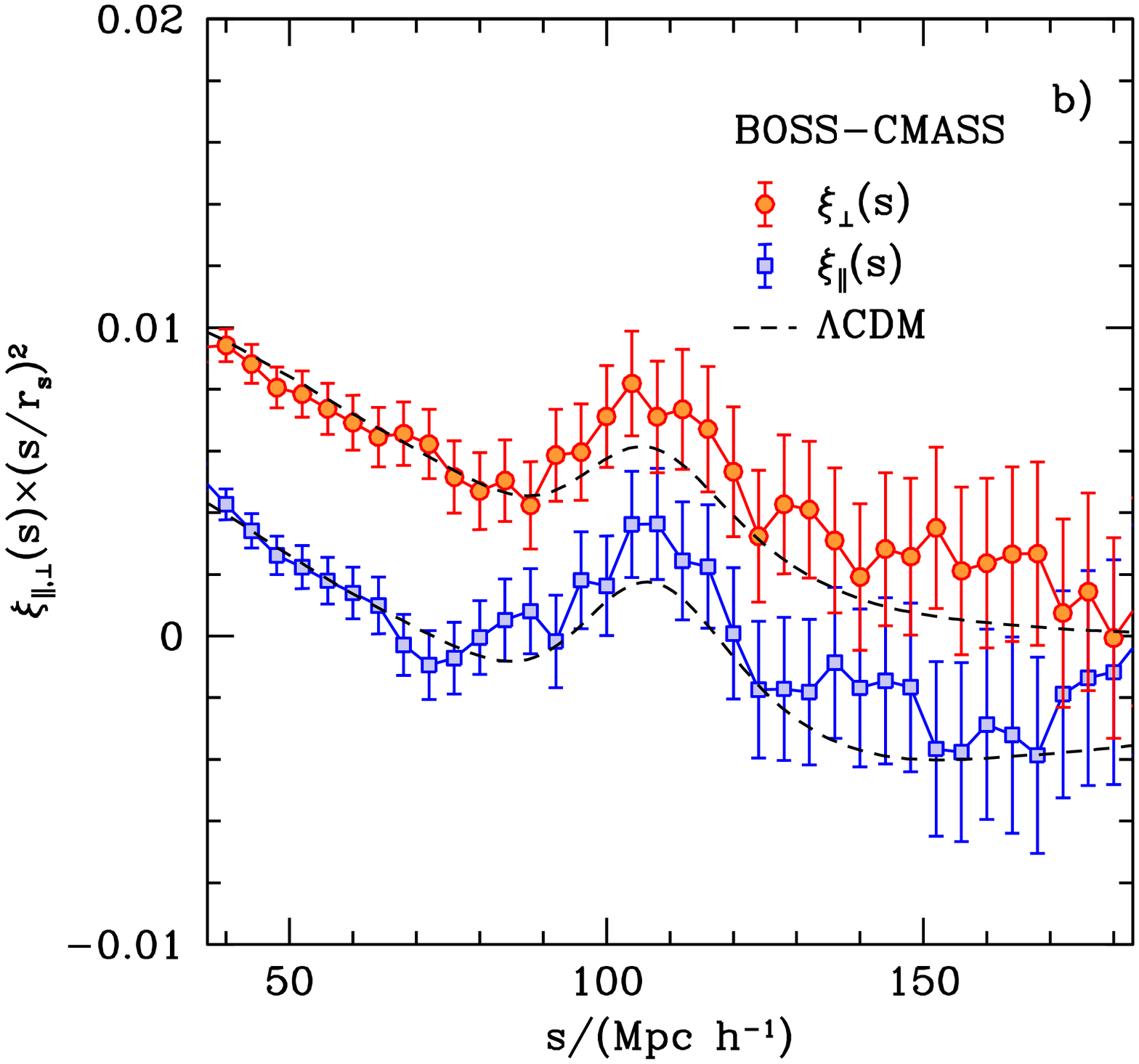}}
\caption{
Panel (a): clustering wedges $\xi_{\perp}(s)$ (circles) and $\xi_{\parallel}(s)$ (squares) of the BOSS-DR9 
CMASS sample. The errorbars were obtained from a set of 600 mock catalogues with the same selection function of the
survey \citep{Manera2012}. The dashed line corresponds to the best-fitting $\Lambda$CDM model obtained
from the combination of the shape of these measurements and our CMB dataset (see Section \ref{sec:lcdm}).
Panel (b): same format as panel (a), but rescaled by $(s/r_{\rm s})^2$, where
$r_{\rm s}=153.2 \, {\rm Mpc}$ (that is, 107.2 $h^{-1}{\rm Mpc}$) which corresponds to the sound horizon
scale in our fiducial cosmology. 
}
\label{fig:cmass}  
\end{figure}

 When computing the pair counts, we assign a series of weights to each object in our catalogue.
First, we apply a radial weight given by 
\begin{equation}
w_{\rm r}=\frac{1}{1+P_{w}\bar{n}(z)},
\label{eq:wradial}
\end{equation}   
where $\bar{n}(z)$ is the expected number density of the catalogue at the given redshift and $P_{w}$ is
a scale-independent parameter which we set to $P_w=2\times 10^4\,h^{-3}{\rm Mpc}^3$.
We include additional weights to correct for non-random contributions to the sample
incompleteness, such as redshift failures and fibre collisions, and the systematic effect introduced
by the local stellar density, as described in detail in \citet{Ross2012}.

Panel a) of Fig.~\ref{fig:cmass} shows the clustering wedges $\xi_{\perp}(s)$ (circles) and 
$\xi_{\parallel}(s)$ (squares) of the CMASS sample obtained through the procedure described above. 
The dashed lines correspond to the best-fitting $\Lambda$CDM model obtained from the combination of 
these measurements with CMB observations as described in Section~\ref{sec:lcdm}.
Panel b) of Fig.~\ref{fig:cmass} displays the same measurements rescaled by the ratio $(s/r_{\rm s})^2$,
where $r_{\rm s}=153.2\,{\rm Mpc}$ corresponds to the sound horizon scale in our fiducial cosmology.
The BAO peak can be clearly seen in both clustering wedges.

To obtain an estimate of the covariance matrix of the CMASS clustering wedges,
we use the mock catalogues of \citet{Manera2012}\footnote{These mock catalogues are publicly
available in http://www.marcmanera.net/mocks/}.
These are a set of $N_{\rm m}=600$ independent mock catalogues corresponding to our fiducial cosmology, which
are based on a method similar to {\sc PTHalos} \citep{Scoccimarro2002} and were designed to follow the
selection function of the CMASS sample in the northern and southern Galactic survey areas.
We measured the clustering wedges of each mock catalogue using the same
binning scheme as for the real data and the radial weights of equation~(\ref{eq:wradial}).
These measurements were used to obtain an estimate of the full covariance matrix $\mathbfss{C}$ 
of the pair $(\xi_{\perp}(s),\xi_{\parallel}(s))$, that is, taking into account the covariance
between the two clustering wedges. The error bars in Fig.~\ref{fig:cmass} correspond to the square root
of the diagonal entries in $\mathbfss{C}$. 

We restrict our analysis of the full shape of the CMASS clustering wedges to 
$44\,h^{-1}{\rm Mpc}< s < 180 \,h^{-1}{\rm Mpc}$, where the model described in Section~\ref{sec:model} gives a
good description of the results from our mock catalogues.
We assume a Gaussian likelihood function of the form ${\cal L}\propto\exp(-\chi^2/2)$
 when comparing these measurements with theoretical predictions. 
The calculation of the $\chi^2$ value of a given model requires the knowledge of the inverse covariance 
matrix. As our estimation of $\mathbfss{C}$ is inferred from our mock catalogues, its inverse,
$\mathbfss{ C}^{-1}$, provides a biased estimate of the true inverse covariance matrix \citep{Hartlap2007}.
To correct for this bias we rescale the inverse covariance matrix as
\begin{equation}
\hat{\mathbfss{C}}^{-1}=\frac{N_{\rm m}-p-2}{N_{\rm m}-1}\,\mathbfss{C}^{-1},
\end{equation}
where $p=78$ corresponds to the total number of bins in the $(\xi_{\perp}(s),\xi_{\parallel}(s))$ pair,
leading to a correction factor of approximately 0.87.

\subsection{Additional data-sets}
\label{sec:moredata}

We combine the information encoded in the full shape of the clustering wedges with additional CMB, BAO and SN
observations in order to improve the obtained cosmological constraints. Here we give a brief description of
each additional dataset. 

Our CMB dataset combines the measurements of the temperature and polarization fluctuations of the CMB
from the nine-year of observations of the WMAP satellite \citep{Bennett2012,Hinshaw2012} 
in the range $2 \leq \ell \leq 1000$, the South Pole Telescope \citep[SPT,][]{Keisler2011}
for $650 \leq \ell \leq 3000$, and the Atacama Cosmology Telescope \citep[ACT,][]{Das2011}
for $500 \leq \ell \leq 10000$. 
We follow the treatment of \citet{Hinshaw2012} and account for the effect of secondary anisotropies
by including the contributions from the Sunyaev-Zel'dovich (SZ) effect, and Poisson and clustered point sources
in the form of templates whose amplitudes are treated as nuisance parameters and marginalized over. 

Our BAO dataset combines the results from independent BAO analyses based on angle-averaged clustering measurements
at lower redshifts than the CMASS sample. These constrain the parameter combination
$D_{\rm V}(z)/r_{\rm s}$, where 
\begin{equation}
D_{\rm V}(z)=\left((1+z)^2D_{\rm A}(z)^2\frac{cz}{H(z)}\right)^{1/3}. 
\label{eq:dv}
\end{equation}
We use the results of \citet{Beutler2011}, who obtained an estimate of
$D_{\rm V}(z=0.106)/r_{\rm s}= 0.336 \pm 0.015$
from the large-scale correlation function of the 6dF Galaxy Survey \citep[6dFGS,][]{Jones2009},
and the 2\% distance measurement of $D_{\rm V}(z=0.35)/r_{\rm s} = 8.88 \pm 0.17$ obtained by 
\citet{Padmanabhan2012} and \citet{Xu2012} from the application of the
reconstruction technique \citep{Eisenstein2007b} to the final SDSS-II LRG sample. 
We do not include the measurements of \citet{Blake2011} from the final WiggleZ Dark Energy Survey
\citep{Drinkwater2010} in our analysis given the significant overlap of the WiggleZ data with the
CMASS sample. 

We also include information from the type Ia supernovae (SN) compilation of \citet{Conley2011}, which
includes the high-redshift SN from the first three years of the 
Supernova Legacy Survey (SNLS). \citet{Conley2011} performed a detailed analysis of the systematic effects
affecting this sample. We follow their recipe to take into account these systematic errors
in our cosmological constraints, which requires the introduction of two additional nuisance parameters,
$\alpha$ and $\beta$, related to the stretch-luminosity and colour-luminosity relationships.
When quoting our cosmological constraints, 
the values of these parameters are marginalized over.

With the exception of Section~\ref{sec:dist}, we use the CMASS clustering wedges in combination with our CMB dataset. 
We refer to this combination as CMB+$(\xi_{\perp}(s),\xi_{\parallel}(s))$. 
Our tightest constraints are obtained including also the additional BAO and SN data in our analysis.

In order to quantify the impact of the additional information contained in the clustering wedges with respect to 
the angle-averaged correlation function, we also compare the results obtained by means of the CMB+$(\xi_{\perp}(s),\xi_{\parallel}(s))$
combination with those obtained by replacing the clustering wedges with the CMASS monopole $\xi_0(s)$ from \citet{Sanchez2012}. 

\section{Modelling clustering wedges}
\label{sec:model}

In this section we describe our model of the full shape of the clustering wedges,
taking into account the effects of non-linear evolution, redshift-space distortions and bias. 
In section \ref{sec:modelxi2d} we describe a simple recipe to compute the first multipoles of
$\xi(\mu,s)$. In section \ref{sec:wedges} we use this recipe to construct our 
model of the clustering wedges. In section~\ref{sec:da_h}
we test this model against our ensemble of mock catalogues and use these results to illustrate 
the way in which these measurements can be used to constrain $H(z)$ and $D_{\rm A}(z)$.

\begin{figure}
\includegraphics[width=0.45\textwidth]{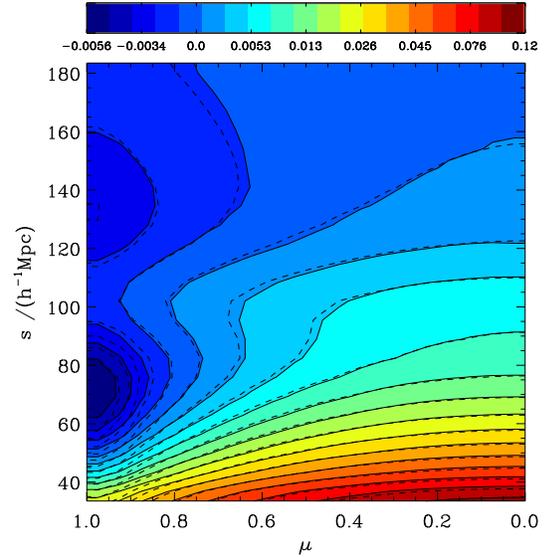}
\caption{
Mean two-dimensional correlation functions $\xi(\mu,s)$ from our ensemble of mock 
catalogues (solid contour lines following the colour scheme). The contours deviate from the
horizontal lines that would correspond to isotropic clustering due to redshift-space
distortions. The BAO feature can be noticed at $s\sim 110\mpc$. The most significant features
of $\xi(\mu,s)$ can be well described by taking into account the contributions from 
multipoles $\xi_{\ell}(s)$ with $\ell\leq2$ (dashed lines), computed as described in
Section~\ref{sec:modelxi2d}.
}
\label{fig:xi2d}  
\end{figure}

\subsection{The anisotropic correlation function}
\label{sec:modelxi2d}

The clustering wedges $\xi_{\perp}(s)$ and $\xi_{\parallel}(s)$ can be obtained by integrating
$\xi(\mu,s)$ over two wide bins of $\Delta \mu=0.5$. Thus the starting point of our
description of these measurements should be a model of the full two-dimensional correlation
function. Figure \ref{fig:xi2d} shows the mean redshift-space $\xi(\mu,s)$ from our ensemble of
mock catalogues (solid lines following the colour scheme).
The effect of redshift-space distortions can be clearly seen in these contours, which strongly 
deviate from the horizontal lines that would correspond to the true underlying isotropic clustering.
Although it is strongly affected by these distortions, the BAO feature is clearly
noticeable at $s\sim 110\mpc$.

\begin{figure}
\includegraphics[width=0.5\textwidth]{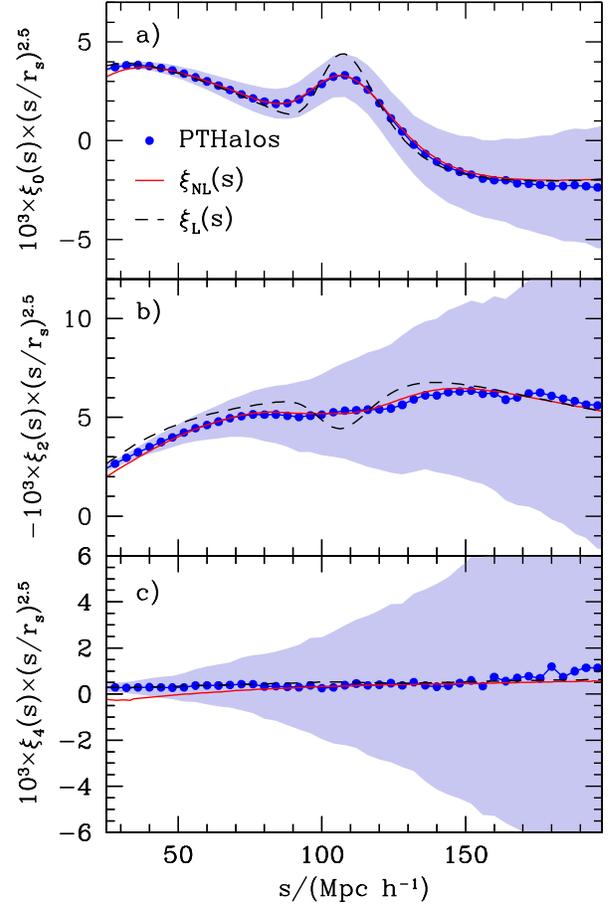}
\caption{
The points correspond to the mean monopole (panel a), quadrupole (panel b), and hexadecapole (panel c)
from our ensemble of mock catalogues. The shaded regions indicate the variance from the different realizations. 
Non-linear evolution distorts the shape of these multipoles which deviate from the linear theory predictions
(dashed lines). These distortions are well described by the parametrization presented in Section~\ref{sec:model},
shown by the solid lines.
To highlight the features on large scales, these measurements are rescaled by $(s/r_{\rm s})^{2.5}$,
where $r_{\rm s}=107.4\mpc$ corresponds to the sound horizon at the drag redshift for our fiducial cosmology.
}
\label{fig:multipoles}  
\end{figure}

The anisotropic correlation function $\xi(\mu,s)$ can be decomposed as a linear combination of
Legendre polynomials, $L_\ell(\mu)$, as 
\begin{equation}
 \xi(\mu,s)=\sum_{{\rm even}\ \ell} L_{\ell}(\mu)\xi_{\ell}(s),
\label{eq:expansion}
\end{equation}
where the multipoles $\xi_\ell(s)$ are given by
\begin{equation}
\xi_\ell(s)\equiv \frac{2\ell+1}{2}\int^1_{-1} L_\ell(\mu)\xi(\mu,s)\,{\rm d}\mu.
\label{eq:multipoles}
\end{equation}
In practice, only a small number of multipoles of $\xi(\mu,s)$ have non-negligible values on large scales. 
This can be seen in Figure \ref{fig:multipoles}, where the points correspond to the mean monopole
(panel a), quadrupole (panel b), and hexadecapole (panel c) from our ensemble of mock catalogues.
The shaded regions indicate the variance from the different realizations, corresponding to the uncertainties
associated with one CMASS DR9 volume. 
To highlight the features on large scales these measurements have been rescaled by $(s/r_{\rm s})^{2.5}$, where
$r_{\rm s}=107.4\mpc$ corresponds to the sound horizon at the drag redshift for our fiducial cosmology.
The hexadecapole $\xi_4(s)$ is consistent with zero over a wide range of scales and higher multipoles can
be safely neglected. By modelling these multipoles, equation~(\ref{eq:expansion})
can be used to describe the full $\xi(\mu,s)$.

In order to obtain a description of the multipoles $\xi_\ell(s)$, it is convenient to
work with the two-dimensional power spectrum, $P(\mu,k)$. This quantity can also be decomposed in 
terms of Legendre polynomials, with multipoles given by
\begin{equation}
P_\ell(k)\equiv \frac{2\ell+1}{2}\int^1_{-1} L_\ell(\mu)P(\mu,k)\,{\rm d}\mu,
\label{eq:multipoles_pk}
\end{equation}
from which the multipoles $\xi_{\ell}(s)$ can be obtained as
\begin{equation}
\xi_{\ell}(s)\equiv \frac{i^{\ell}}{2\pi^2}\int^{\infty}_{0} P_{\ell}(k) j_{\ell}(ks)\,k^2{\rm d}k,
\label{eq:pl2xil}
\end{equation}
where $j_{\ell}(x)$ is the spherical Bessel function of order $\ell$ \citep{Hamilton1997}. 

In the linear perturbation theory regime, and assuming the distant observer approximation, the two-dimensional
power spectrum $P(\mu,k)$ can be described by the simple formula \citep{Kaiser1987}
\begin{equation}
P(\mu,k)=b^2(1+\beta\mu^2)^2P_{\rm L}(k),
\label{eq:p2d_lin}
\end{equation}
where $P_{\rm L}(k)$ is the linear-theory real-space power spectrum, $b$ is the bias factor, and
$\beta=f/b$, with $f=\frac{{\rm d}\ln D }{{\rm d}\ln a }$, i.e., the logarithmic derivative of
the growth factor $D(a)$. In this case all multipoles with $\ell>4$ vanish exactly. 
Even though this simple picture will be approximately valid when the amplitude of the density fluctuations
is small, non-linear evolution introduces deviations from this 
behaviour \citep{Smith2008,Crocce2008,Sanchez2008}.
This can be clearly seen in Figure \ref{fig:multipoles}, where the dashed lines correspond to the
linear theory predictions for the multipoles $\xi_{\ell}(s)$. 
Although it is located at large scales, the differences in the appearance of
the BAO signal are significant, as non-linear growth damps the BAO feature.
This is particularly noticeable in the quadrupole,
where the BAO signal is almost completely erased. These effects must 
be taken into account when attempting to extract
precision cosmological information from these statistics.

Much work has been devoted over recent years to modelling the effects of non-linear evolution and
redshift-space distortions.
Pioneered by the work of \citet{Crocce2006} on Renormalized Perturbation Theory (hereafter RPT),
several new approaches to perturbation theory have been developed in recent years
\citep[e.g.][]{Matsubara2008a,Matsubara2008b,Matarrese2007,Matarrese2008,Pietroni2008,Taruya2008,Anselmi2011,
Anselmi2012,Wang2012}.
In these methods, the series expansion describing the power spectrum of standard perturbation theory is
reorganized and some of the terms are re-summed into a function $G(k)$, usually called propagator, that can be
factorized out of the series. The remaining terms contain mode-coupling contributions, $P_{\rm MC}(k)$,
to the final non-linear power spectrum, which can then be written as $P(k)=P_{\rm L}(k)G(k)^2+P_{\rm MC}(k)$.
These approaches provide a better understanding of the effects of non-linear evolution on the shape
of the two-point statistics, such as the power spectrum and the correlation function, in real-space.
However, the extension of these results to the halo clustering in redshift space is somewhat 
more complicated. Although several recent studies have provided non-linear descriptions of redshift-space distortions
for the matter and halo density fields 
\citep{Scoccimarro2004, Tinker2006, Tinker2007, Matsubara2008a, Matsubara2008b, Taruya2010, Taruya2013, Jennings2011,
Reid2011, delaTorre2012, Okumura2012},
the range of validity of these models is limited and they rely on free parameters
to fit the results from N-body simulations.

In this work we follow a simple approach and parametrize the non-linear two-dimensional power spectrum as
\begin{equation}
P(\mu,k)=\left(\frac{1}{1+(kf\sigma_{\rm v}\mu)^2}\right)^2(1+\beta\mu^2)^2P_{\rm NL}(k),
\label{eq:p2d}
\end{equation}
where 
\begin{equation}
% P_{\rm NL}(k) = b^2\left[P_{\rm L}(k)\,{\rm e}^{-(k/k_{\star})^2} + A_{\rm MC} \,P_{\rm 1loop}(k)\right],
P_{\rm NL}(k) = b^2\left[P_{\rm L}(k)\,{\rm e}^{-(k\sigma_{\rm v})^2} + A_{\rm MC} \,P_{\rm 1loop}(k)\right],  
\label{eq:pnl}
\end{equation}
and $b$, $\sigma_{\rm v}$, and $A_{\rm MC}$, are free parameters. Here $P_{\rm 1loop}(k)$
is given by 
\begin{equation}
 P_{\rm 1loop}(k) = \frac{1}{2\pi^{2}} \int \rm d^{3}q \,|F_{\rm2}(\mathbf k-\mathbf q,\mathbf q)|^{2} P(|\mathbf k-\mathbf q|)P(q),
\label{eq:pmc}
\end{equation}
where $F_{2}(\mathbf k,\mathbf q)$ is the standard second order kernel of perturbation theory. 

The description of the non-linear power spectrum of equation~(\ref{eq:pnl}) is motivated by RPT.
To a good approximation, the non-linear propagator $G(k)$ is of Gaussian form,
while, at large scales, equation~(\ref{eq:pmc}) contains the leading order contribution to the full
$P_{\rm MC}(k)$ \citep[see][for a more detailed description of these functions]{Crocce2012}.
The description of $P_{\rm NL}(k)$ given by equation~(\ref{eq:pnl}) is the basis of the parametrization
of the non-linear correlation function proposed by \citet{Crocce2008}, and has been shown to give an 
accurate description of the power spectra and correlation functions measured from N-body simulations
\citep[e.g.][]{Sanchez2008,Montesano2010} and real galaxy samples \citep{Sanchez2009,Montesano2012,Beutler2011,Blake2011}.
In particular, this parametrization was used by \citet{Sanchez2012} to describe the CMASS monopole $\xi_0(s)$.
The Lorentzian pre-factor in equation~(\ref{eq:p2d}) represents a damping function which mimics the Finger-of-God
effect corresponding to the assumption of an exponential galaxy velocity distribution function \citep{Park1994,Cole1995}.

The solid lines in Fig.~\ref{fig:multipoles} correspond to the multipoles $\xi_{\ell}(s)$ obtained using
the parametrization of equation~(\ref{eq:p2d}), by fitting the free parameters in the model.
These give an accurate description of the full shape of the mean monopole and 
quadrupole from our mock catalogues on large scales. On the other hand, while the shape of the mean
hexadecapole from the mock catalogues is well described by the linear theory prediction, the results
obtained from the parametrization of equation~(\ref{eq:p2d})
only reproduce these measurements for scales larger than $80\,h^{-1}{\rm Mpc}$.
These differences indicate the limitations of this model to describe the shape of the
full anisotropic power spectrum $P(\mu,k)$.
However, as we will see in Section~\ref{sec:da_h}, despite the simplicity of this recipe,
its use as the basis of the modelling of the clustering wedges can provide
unbiased cosmological constraints even for surveys probing volumes much larger than the
SDSS-DR9 CMASS sample.

The monopole-quadrupole pair contains most of the information in the full $\mu-s$ plane. This can be
seen in Fig.~\ref{fig:xi2d}, where the dashed lines correspond to the contours of $\xi(\mu,s)$
obtained by considering only the non-linear monopole and quadrupole terms of the multipole expansion of 
equation~(\ref{eq:expansion}).
These show a good agreement with the full measurement, describing most of its features.
This result suggests that the monopole-quadrupole pair may contain the
most relevant information for the description of the clustering wedges, a fact that we will exploit 
in the following section to construct a model for $\xi_{\perp}(s)$ and $\xi_{\parallel}(s)$.

\begin{figure}
\includegraphics[width=0.48\textwidth]{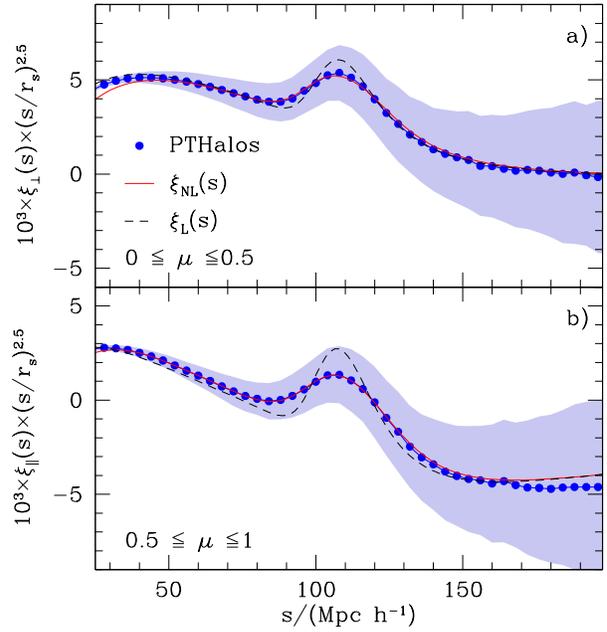}
\caption{
The points represent the mean clustering wedges $\xi_{\perp}(s)$ (panel a) and $\xi_{\parallel}(s)$ (panel b)
from our ensemble of mock catalogues, rescaled by $(s/r_{\rm s})^{2.5}$. The shaded regions correspond to
the variance from the different realizations.
The dashed lines represent the predictions of linear perturbation theory, while the solid
lines correspond to the clustering wedges obtained from the parametrization of the non-linear power
spectrum given in equation~(\ref{eq:p2d}).
}
\label{fig:wedges}  
\end{figure}

\subsection{From $\xi(\mu,s)$ to the clustering wedges}
\label{sec:wedges}

Figure \ref{fig:wedges} shows the mean clustering wedges $\xi_{\perp}(s)$ (panel a) and $\xi_{\parallel}(s)$
(panel b) from our mock catalogues, rescaled by $(s/r_{\rm s})^{2.5}$. The variance from the individual
realizations is shown by the shaded region.
The anisotropic clustering pattern generated by redshift-space distortions leads to significant differences
in the amplitude and shape of the two clustering wedges, with $\xi_{\parallel}(s)$ showing a lower amplitude
and a stronger damping of the BAO peak than $\xi_{\perp}(s)$.
Here we use the description of the multipoles $\xi_\ell(s)$ of the previous section
to construct a model for the full shape of the clustering wedges.

The multipole description of $\xi(\mu,s)$ can be used to compute the clustering wedges $\xi_{\perp}(s)$
and $\xi_{\parallel}(s)$. Discarding contributions from multipoles with $\ell>4$, equation (\ref{eq:wedges}) 
implies that \citep{Kazin2012}
\begin{eqnarray}
\xi_{\perp}(s) &=& \xi_0(s) - \frac{3}{8} \xi_2(s) + \frac{15}{128}\xi_4(s),\label{eq:multi2perp}\\
\xi_{\parallel}(s) &=& \xi_0(s) + \frac{3}{8} \xi_2(s) - \frac{15}{128}\xi_4(s).\label{eq:multi2para}
\end{eqnarray}
These equations demonstrate that the contribution from $\xi_4(s)$ to the final clustering wedges is small and can be 
safely neglected. 

The dashed lines in Fig.~\ref{fig:wedges} correspond to the linear theory predictions for 
$\xi_{\parallel}(s)$ and $\xi_{\perp}(s)$. These are obtained using the
multipoles $\xi_{\ell}(s)$ in equations~(\ref{eq:multi2perp}) and
(\ref{eq:multi2para}). Non-linear evolution causes the shape of the clustering wedges to deviate from
these predictions, with the most notable differences at the scales of the BAO peak.
The extraction of unbiased cosmological information from a measurement of the clustering
wedges requires an accurate modelling of these distortions.

The solid lines in Figure \ref{fig:wedges} show the predictions for $\xi_{\perp}(s)$ and $\xi_{\parallel}(s)$
obtained from equations (\ref{eq:multi2perp}) and (\ref{eq:multi2para}) by considering the contributions from 
the multipoles $\xi_{\ell}(s)$ with $\ell\leq2$ inferred from our model of the non-linear redshift-space power
spectrum (equation~\ref{eq:p2d}).
This simple recipe provides an accurate description of the full shape of the two clustering wedges,
implying that the monopole-quadrupole pair contains the most relevant information required to describe
these measurements.

\subsection{Measuring $H(z)$ and $D_{\rm A}(z)$ from the clustering wedges}
\label{sec:da_h}

\begin{figure}
\includegraphics[width=0.45\textwidth]{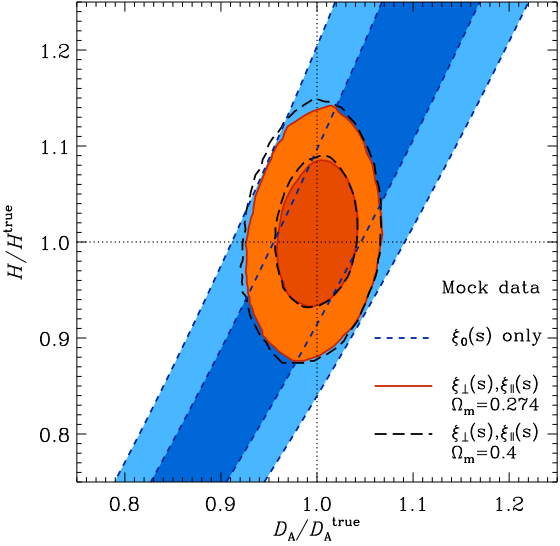}
\caption{Two-dimensional marginalized constraints on $D_{\rm A}/D_{\rm A}^{\rm true}$ and $H/H^{\rm true}$ at
the mean redshift of the sample $z_{\rm m}=0.57$, derived from the mean $\xi_0(s)$ (short-dashed lines),
and clustering wedges $\xi_{\perp}(s)$ and $\xi_{\parallel}(s)$ (solid lines) from our ensemble of mock 
catalogues.
The long-dashed lines correspond to the constraints from the clustering wedges of the same set of mock catalogues,
but measured assuming a fiducial cosmology with $\Omega_{\rm m}=0.4$. Despite the large difference
with the true value of this parameter, the effect of the fiducial cosmology is correctly 
taken into account by the treatment described in Section~\ref{sec:da_h}, leading to completely consistent
constraints.
}
\label{fig:da_h}  
\end{figure}

As shown in Fig.~\ref{fig:wedges}, the model presented in Section~\ref{sec:modelxi2d}
gives an accurate description of the full shape of the mean clustering wedges from our ensemble of mock
catalogues. Here we test the ability of this model to recover unbiased cosmological constraints from
these measurements by analysing the effect of the fiducial cosmology on $\xi_{\perp}(s)$ and $\xi_{\parallel}(s)$.

As described in Section~\ref{sec:clustering}, the measurement of the clustering wedges requires the
assumption of a fiducial cosmology to map the observed redshifts into distances. This choice has a 
significant effect on the obtained results. Different fiducial cosmologies will lead to a
rescaling of the components parallel and perpendicular to the line-of-sight, 
$s_{\parallel}$ and $s_{\perp}$, of the separation vector ${\mathbf s}$ 
\citep{Padmanabhan2008,Kazin2012,Xu2012}.
The relation between the true separations and those measured in the fiducial cosmology can be written as 
\begin{eqnarray}
 s_{\perp} &=& \alpha_{\perp}s'_{\perp},\label{eq:scaling1}\\
 s_{\parallel} &=& \alpha_{\parallel}s'_{\parallel},\label{eq:scaling2}
\end{eqnarray}
where the primes denote the quantities in the fiducial cosmology and the scaling factors are given by
\begin{eqnarray}
 \alpha_{\perp} &=& \frac{D_{\rm A}(z_{\rm m})}{D'_{\rm A}(z_{\rm m})},\\
 \alpha_{\parallel} &=& \frac{H'(z_{\rm m})}{H(z_{\rm m})},
\end{eqnarray}
i.e., the ratios of the angular diameter distance and the Hubble parameter evaluated at the mean redshift of the
survey, $z_{\rm m}=0.57$, in the true and fiducial cosmologies.
Equations (\ref{eq:scaling1}) and (\ref{eq:scaling2}) are the basis of the Alcock--Paczynski test \citep{Alcock1979}.
In terms of $s$ and $\mu$, these equations can be written as %\citep{Ballinger1996}
\begin{eqnarray}
 s &=& s'\sqrt{\alpha_{\parallel}^2(\mu')^2+\alpha_{\perp}^2(1-(\mu')^2)},\label{eq:sfid}\\
\mu &=& \frac{\alpha_{\parallel}\mu'}{\sqrt{\alpha_{\parallel}^2(\mu')^2+\alpha_{\perp}^2(1-(\mu')^2)}}.\label{eq:mufid}
\end{eqnarray}
These relations completely describe the impact of the fiducial cosmology on the clustering wedges,
as they can be used to transform the integral in equation (\ref{eq:wedges}) from the
fiducial cosmology space to the true cosmology as
\begin{equation}
\xi'_{\Delta \mu}(s')\equiv \frac{1}{\Delta \mu'}\int^{\mu'_{\rm max}}_{\mu'_{\rm min}}\xi(\mu(\mu',s'),s(\mu',s'))\,{{\rm d}\mu'}.
\label{eq:wedges_fid}
\end{equation}

Equation~(\ref{eq:wedges_fid}) can be used to perform a simple test of the accuracy of our model of the clustering wedges.
For this test we treat the parameters $\alpha_{\perp}$ and $\alpha_{\parallel}$ in equations (\ref{eq:sfid})
and (\ref{eq:mufid}) as free parameters and fit for them using the mean clustering wedges from our mocks, 
while fixing all cosmological parameters to their true underlying values.
As the fiducial cosmology used to obtain these measurements corresponds to the correct cosmology of the 
mocks, a deviation of the best fitting values of these parameters from $\alpha_{\perp,\parallel}=1$
would indicate a systematic bias in the model, which would then be unable to reproduce
the correct shape of the clustering wedges.
The result of this exercise is shown in Fig.~\ref{fig:da_h}, where the solid lines
correspond to the 68 and 95 per cent two-dimensional marginalized constraints in the 
$D_{\rm A}(z_{\rm m})/D_{\rm A}^{\rm true}(z_{\rm m})-H(z_{\rm m})/H^{\rm true}(z_{\rm m})$
plane obtained in this way. These results have been marginalized over the fiducial parameters of the 
model, $b$, $\sigma_{\rm v}$, and $A_{\rm MC}$.
The values obtained in this case are
$D_{\rm A}(z_{\rm m})/D_{\rm A}^{\rm true}(z_{\rm m})=0.998\pm0.028$ and 
$H(z_{\rm m})/H^{\rm true}(z_{\rm m})=1.001\pm0.052$, showing
that the model described in the previous sections gives an accurate description of the
full shape of the clustering wedges, providing unbiased constraints on $D_{\rm A}(z_{\rm m})$ and $H(z_{\rm m})$.

As a further test of the ability of equations~(\ref{eq:sfid}) and (\ref{eq:mufid}) to describe
the effect of the fiducial cosmology on the clustering wedges
we repeated this exercise, using the mean clustering wedges obtained from the same mock catalogues 
but assuming a different fiducial cosmology, with $\Omega_{\rm m}=0.4$. The results obtained in this 
case are indicated by the long-dashed lines in Fig.~\ref{fig:da_h}. 
Despite the large difference between this fiducial cosmology and the true one of our mock catalogues,
we recover the correct values of these parameters,
with $D_{\rm A}(z_{\rm m})/D_{\rm A}^{\rm true}(z_{\rm m})=0.997\pm0.029$ and 
$H(z_{\rm m})/H^{\rm true}(z_{\rm m})=1.006\pm0.54$. This exercise demonstrates that these equations
can be used to describe the 
effect of the fiducial cosmology when comparing a given model with 
measurements of $\xi_{\perp}(s)$ and $\xi_{\parallel}(s)$. 

For comparison, the short-dashed lines in Fig.~\ref{fig:da_h} correspond to the constraints on
$D_{\rm A}(z_{\rm m})/D_{\rm A}^{\rm true}(z_{\rm m})$ and $H(z_{\rm m})/H^{\rm true}(z_{\rm m})$
obtained from $\xi_0(s)$ alone. These follow a degeneracy of constant 
$\alpha=D_{\rm V}(z_{\rm m})/D_{\rm V}^{\rm true}(z_{\rm m})\propto \left(D_{\rm A}(z_{\rm m})^2H(z_{\rm m})\right)^{1/3}$,
where $D_{\rm V}(z)$ is given by equation~(\ref{eq:dv}).
The comparison of these constraints with the ones obtained from $\xi_{\perp}(s)$ and $\xi_{\parallel}(s)$ 
clearly illustrates the extra information contained in the clustering wedges with respect of that of the monopole.

When dealing with clustering measurements obtained from observational data the true underlying cosmology is, of course,
not known. Different cosmological models will predict different values of the acoustic scale $r_{\rm s}(z_{\rm d})$.
In this case, the angle-averaged correlation function provides constraints on the dimensionless quantity \begin{equation}
 d_{\rm s}\equiv \frac{D_{\rm V}(z_{\rm m})}{r_{\rm s}(z_{\rm d})}.
\label{eq:ds}
\end{equation}
Analogously, as shown by \citet{Kazin2012}, the clustering wedges provide constraints on the parameter combinations
\begin{equation}
 d_{\perp}\equiv \frac{D_{\rm A}(z_{\rm m})}{r_{\rm s}(z_{\rm d})},
\label{eq:dperp}
\end{equation}
and
\begin{equation}
d_{\parallel}\equiv \frac{cz_{\rm m}}{r_{\rm s}(z_{\rm d})H(z_{\rm m})}.
\label{eq:dpara}
\end{equation}
Therefore, when combined with a measurement of $r_{\rm s}(z_{\rm d})$ inferred from CMB observations, the
clustering wedges can provide separate constraints on $H(z_{\rm m})$ and $D_{\rm A}(z_{\rm m})$.

\begin{figure*}
\includegraphics[width=0.32\textwidth]{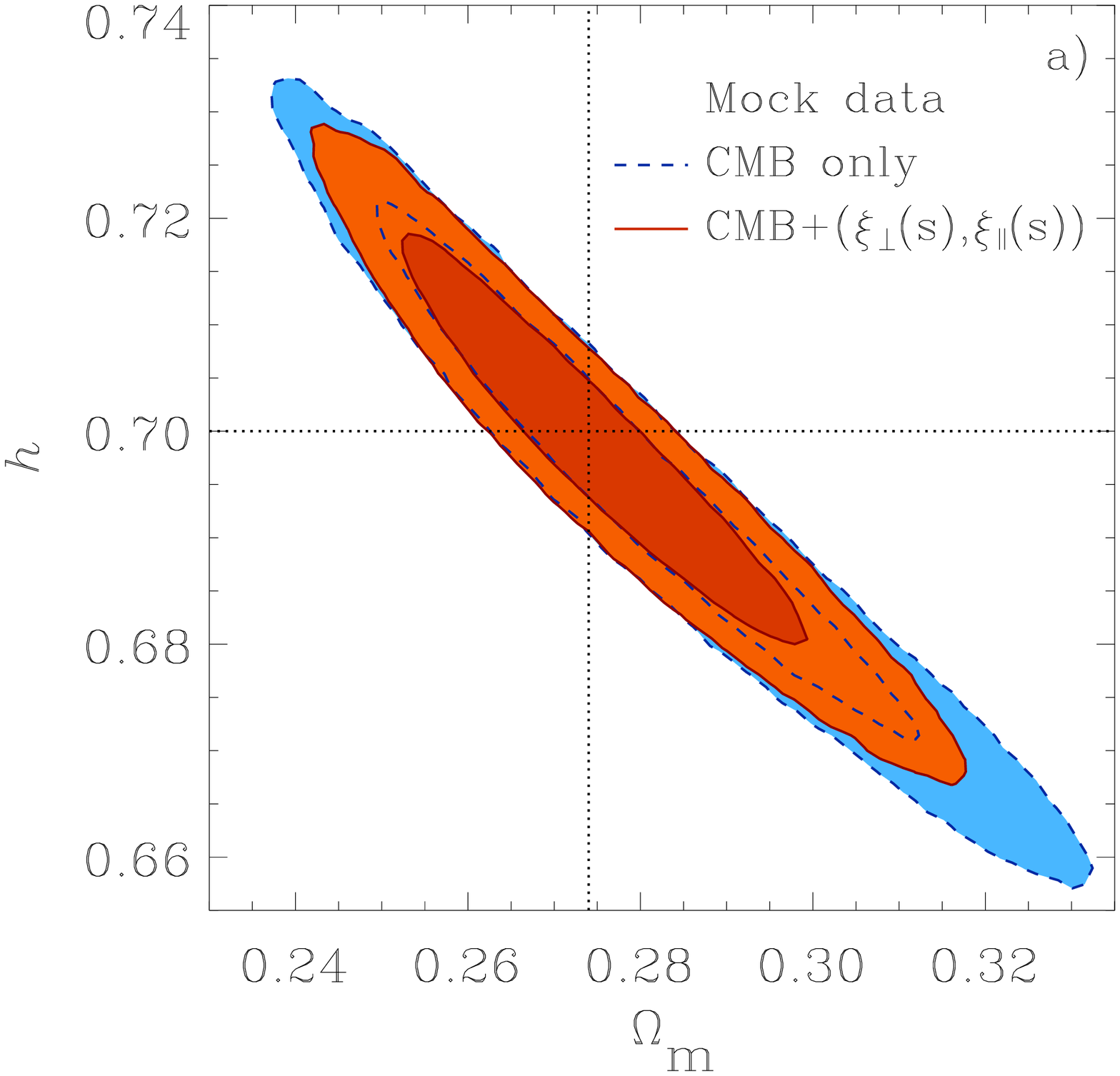}
\includegraphics[width=0.32\textwidth]{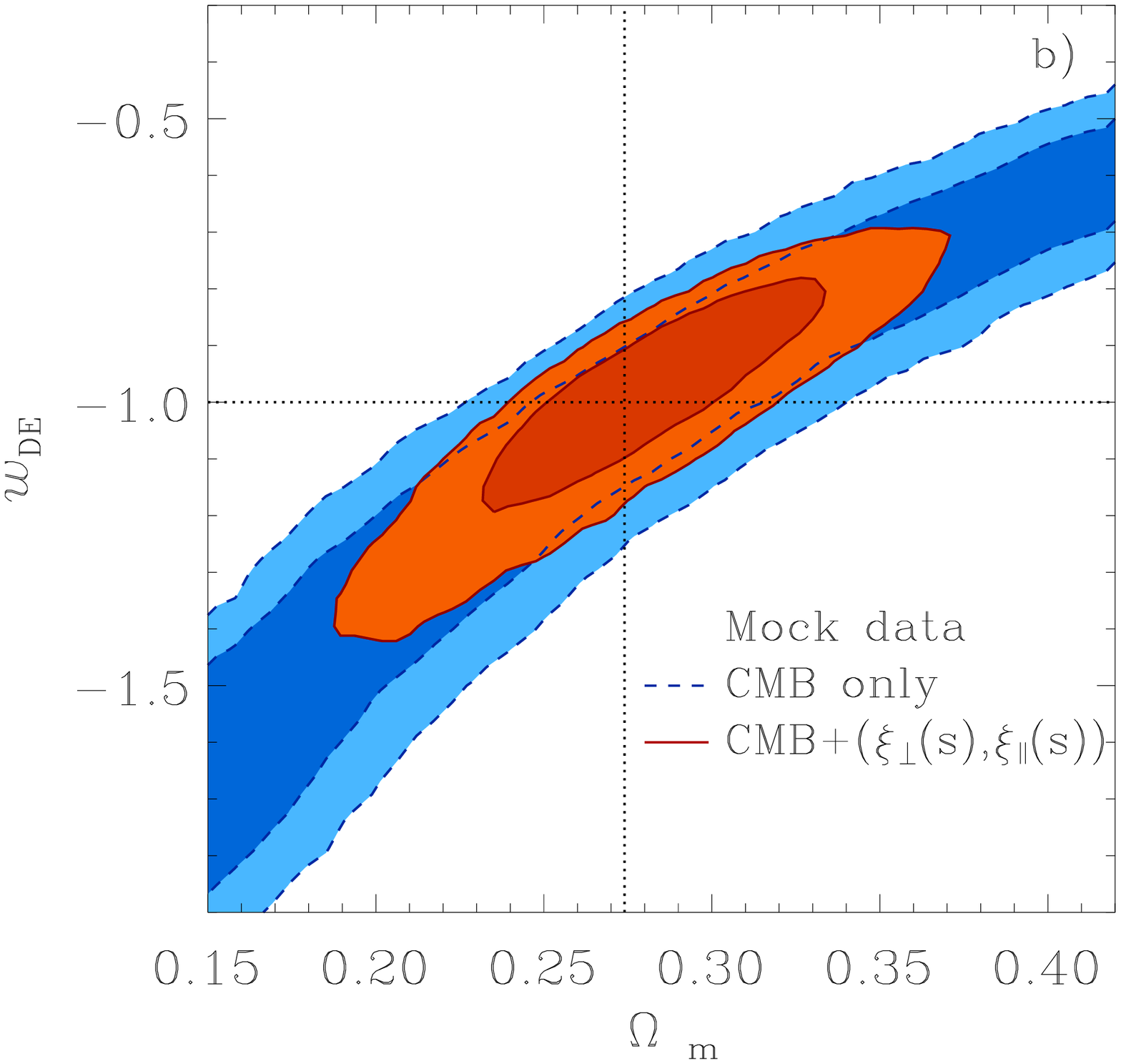}
\includegraphics[width=0.32\textwidth]{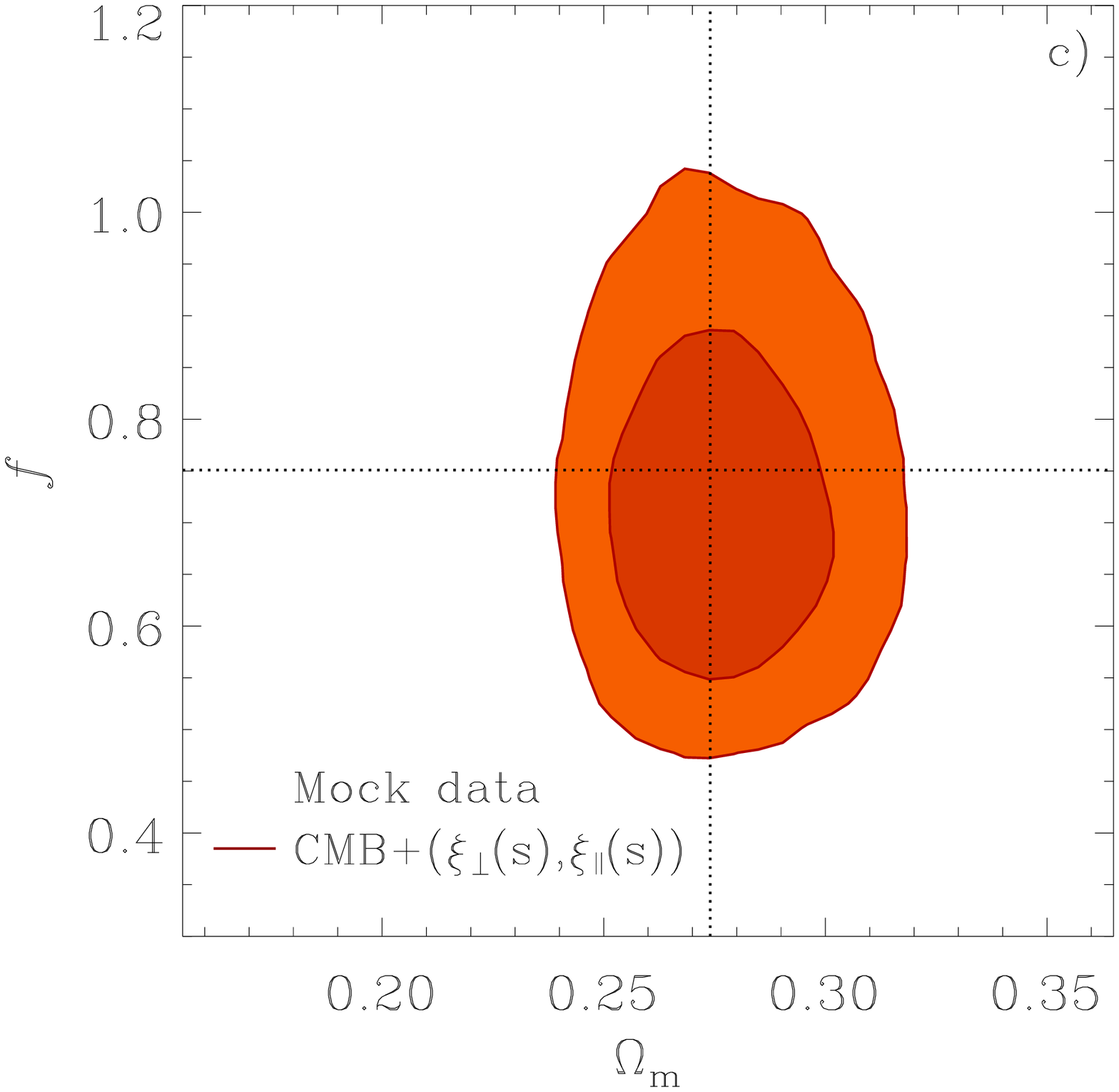}
\caption{
Two-dimensional 68 and 95 per cent CL obtained on various cosmological parameters recovered from our mock CMB
and CMASS $(\xi_{\perp}(s),\xi_{\parallel}(s))$ datasets for the basic $\Lambda$CDM parameter space (panel a), 
and its extensions obtained by allowing for variations on $w_{\rm DE}$ (panel b), and the growth factor
$f(z_{\rm m})$ (panel c). In all cases the obtained constraints are in excellent agreement with the fiducial
underlying values, indicated by the dotted lines. 
}
\label{fig:pthalos2d}
\end{figure*}

\section{Methodology}
\label{sec:method}

\subsection{Cosmological parameter spaces}
\label{sec:cospar}

We explore similar sets of cosmological parameters as in \citet{Sanchez2012}. Here we describe these
parameter spaces and the methodology used to explore them.

We assume that primordial fluctuations are adiabatic, Gaussian, and have
a power-law spectra of Fourier amplitudes, with a negligible tensor component. 
With these hypotheses a given cosmological model can be specified by the following sets of parameters,
\begin{equation}
{\bf P}_{\rm main} = (\omega_{\rm b}, \omega_{\rm dm}, \Theta, \Omega_k, f_{\nu},w_{\rm DE},A_{\rm s}, n_{\rm s}). 
\label{eq:pmain}
\end{equation}
These are the baryon and dark matter densities, $\omega_{\rm b} = \Omega_{\rm b}h^2$ and 
$\omega_{\rm dm} = \Omega_{\rm dm}h^2$, the angular size of the sound horizon at recombination, $\Theta$, 
given by the ratio between the horizon scale at recombination and the angular diameter distance to the corresponding redshift,
the curvature of the Universe, $\Omega_k$, the dark matter fraction in the form of massive neutrinos,
$f_{\nu}=\Omega_{\nu}/\Omega_{\rm dm}$, the dark energy equation of state parameter, $w_{\rm DE}$. 
and the amplitude, $A_{\rm s}$, and spectral index, $n_{\rm s}$ of the scalar primordial power spectrum, which we 
quote at the pivot wavenumber of $k_0= 0.02\,{\rm Mpc}^{-1}$.

We explore both the case of a constant dark energy
equation of state and when it is allowed to vary with time, in which case we assume the standard linear
parametrization of \citet{Chevallier2001} and \citet{Linder2003} given by
\begin{equation}
w_{\rm DE}(a) = w_0 + w_{a}(1-a), 
\label{eq:wa}
\end{equation}
where $a$ is the expansion factor and $w_0$ and $w_a$ are free parameters corresponding to the values
of $w_{\rm DE}$ today and (minus) its derivative with respect to the scale factor, respectively.

The analysis of the CMB data requires an additional parameter, the optical depth to the last scattering
surface, $\tau$. This parameter is constrained by the CMB data alone and the inclusion of additional
datasets leaves these results almost unchanged. As our CMB dataset was also used by \citet{Hinshaw2012},
who present constraints on this parameter, we do not include them here.

The parameters of equation~(\ref{eq:pmain}) allow us to derive constraints on other 
important quantities, such as
\begin{equation}
{\bf P}_{\rm der} = (\Omega_{\rm DE}, \Omega_{\rm m},\sum m_{\nu},h,\sigma_8, f(z_m)).
\label{eq:paramder} 
\end{equation}
This set contains the dark energy and total matter densities, the sum of the neutrino masses, the Hubble parameter,
the rms linear perturbation theory variance in spheres of radius $8\,h^{-1}{\rm Mpc}$, and the logarithmic
derivative of the growth factor, $f(z_m)={\rm d}\ln D/{\rm d}\ln a$.
In Section~\ref{sec:fg} we explore the constraints on potential deviations from general relativity by
treating $f(z_m)$ as a free parameter, instead of computing its value as a derived quantity.

In Section~\ref{sec:dist} we explore the constraints on the geometrical quantities
\begin{equation}
{\bf P}_{\rm geom} = (d_{\perp},d_{\parallel},d_{\rm s}),
\label{eq:paramgeo} 
\end{equation}
given by equations~(\ref{eq:ds})--(\ref{eq:dpara}). These parameters contain the combinations of the sound horizon at the drag redshift,
$r_{\rm s}(z_{\rm d})$, and the Hubble parameter, $H(z_{\rm m})$, angular diameter distance,
$D_{\rm A}(z_{\rm m})$, and average distance, $D_{\rm V}(z_{\rm m})$,
to the mean redshift of the sample.

The starting point of our analysis is the basic $\Lambda$CDM parameter space which corresponds to a
flat universe where the energy budget contains contributions from cold dark matter (CDM), baryons,
and dark energy, described by $w_{\rm DE} = -1$. 
Our constraints on the $\Lambda$CDM parameter space are described in 
Section~\ref{sec:lcdm}. In Sections~\ref{sec:omk}--\ref{sec:fg} we explore a number of
possible extensions of this parameter space by allowing for variations on the remaining parameters
of equations (\ref{eq:pmain}).

We explore these parameter spaces using the {\sc CosmoMC} code of \citet{Lewis2002}, which 
uses {\sc camb} to compute power spectra for the CMB and matter fluctuations \citep{Lewis2000}.
We use a generalized version of {\sc camb} which supports a time-dependent 
dark energy equation of state \citep{Fang2008}.
We included additional modifications from \citet{Keisler2011}, \citet{Das2011} and \citet{Conley2011}
to compute the likelihood of the SPT, ACT and SNLS datasets. 

\subsection{Testing the model of $\xi_{\perp}(s)$ and $\xi_{\parallel}(s)$}
\label{sec:test2}

In section~\ref{sec:da_h} we compared our model of $\xi_{\perp}(s)$ and $\xi_{\parallel}(s)$
against the results from our mock catalogues in a simplified case, when
the true underlying cosmology was known.
Here we test the ability of the model to recover the true cosmological parameters under
the same conditions in which we apply it to the CMASS measurements.

With the exception of Section~\ref{sec:dist}, we use the clustering wedges in combination with our CMB dataset.
The information provided by the CMB data can be well described using the following
set of parameters
\citep[see e.g.][]{Komatsu2009,Komatsu2011}
\begin{equation}
 {\mathbf P}_{\rm CMB}\equiv (z_{*},r_{\rm s}(z_{\rm d}),\ell_{\rm A},R,\omega_{\rm b},A_{\rm s},n_{\rm s}),
\label{eq:distp}
\end{equation}
This set includes the redshifts of recombination, $z_{*}$, 
the sound horizon at the drag redshift, $r_{\rm s}(z_{\rm d})$,
the CMB acoustic angular scale, $\ell_{\rm A}$, defined as
\begin{equation}
 \ell_{\rm A}=\pi(1+z_{*})D_{\rm A}(z_{*})\frac{1}{r_{\rm s}(z_{*})},
\end{equation}
the shift parameter, $R$, given by
\begin{equation}
 R=(1+z_{*})D_{\rm A}(z_{*})\frac{\sqrt{\Omega_{\rm m}H^2_0}}{c},
\end{equation}
the baryon density, and the amplitude and spectral index of the primordial scalar fluctuations.
The redshift of recombination and the drag redshift are computed
using the formulas of \citet{Eisenstein1998}.

A good approximation of the full CMB likelihood can be obtained from the best fitting values of the
parameters of equation~(\ref{eq:distp}) and their covariance matrix $\mathbfss{ C}_{\rm CMB}$.
We use this parameter set to estimate the likelihood function of a mock CMB dataset, with equivalent
characteristics to the real one, but corresponding to our fiducial cosmology, as 
${\mathcal L}({\mathbf P}_{\rm CMB})\propto \exp(-\chi^2({\mathbf P}_{\rm CMB})/2)$ with
\begin{equation}
 \chi^2({\mathbf P}_{\rm CMB})=({\mathbf P}_{\rm CMB}-{\mathbf P}^{\rm fid}_{\rm CMB})^{\rm t}\mathbfss{ C}_{\rm CMB}^{-1}({\mathbf P}_{\rm CMB}-{\mathbf P}^{\rm fid}_{\rm CMB}),
\label{eq:lcmb}
\end{equation}
where ${\mathbf P}^{\rm fid}_{\rm CMB}$ corresponds to the values of the parameters of equation~(\ref{eq:distp})
for our fiducial cosmology and $\mathbfss{C}_{\rm CMB}$ is the corresponding covariance matrix inferred from
our true CMB dataset. Using the approximated ${\mathcal L}({\mathbf P}_{\rm CMB})$ of equation~(\ref{eq:lcmb}) 
we can test our model of the clustering wedges by applying it to the mean $\xi_{\perp}(s)$ and
$\xi_{\parallel}(s)$ from our mock catalogues in combination with CMB data.

\begin{table*} 
\centering
  \caption{
    The marginalized 68\% constraints on the cosmological parameters of the $\Lambda$CDM model
    obtained using different combinations of the datasets described in Section~\ref{sec:data}.}
    \begin{tabular}{@{}lcccc@{}}
    \hline
& \multirow{2}{*}{CMB}  & \multirow{2}{*}{CMB+$\xi_0(s)$} & \multirow{2}{*}{CMB+$(\xi_{\perp}(s),\xi_{\parallel}(s))$}& CMB+$(\xi_{\perp}(s),\xi_{\parallel}(s))$  \\
&                       &                           &      &    +BAO+SN         \\  
\hline
$100\,\Theta$           & $1.0412\pm0.0014$  & $1.0410\pm 0.0014$  &  $1.0409\pm0.0014$  &  $    1.0408\pm0.0014$   \\[0.5mm] 
$100\,\omega_{\rm b}$   & $2.230\pm0.038$  & $ 2.224 \pm 0.035$  &  $2.225\pm0.035$  &  $    2.225\pm0.034$     \\[0.5mm] 
$100\,\omega_{\rm c}$   & $11.38\pm0.41$  & $11.54\pm0.27$  &  $11.53\pm0.27$  &  $ 11.50\pm0.21$  \\[0.5mm] 
$n_{\rm s}$             & $0.966\pm0.010$  & $0.9632\pm0.0087$  &  $0.9633\pm 0.0088$  &  $ 0.9636\pm0.0083$    \\[0.5mm] 
$\ln(10^{10}A_{\rm s})$ & $3.114\pm0.027$  & $3.116\pm0.025$  &  $3.115\pm0.025$  &  $ 3.114\pm 0.024$   \\ 
\hline
$\Omega_{\rm DE}$       & $0.723\pm0.022$  & $0.715\pm0.015$  &  $0.715\pm0.015$  &  $    0.717\pm0.010$  \\[0.5mm]
$\Omega_{\rm m}$        & $0.277\pm0.022$  & $0.285\pm0.015$  &  $0.285\pm0.015$  &  $    0.283\pm0.010$  \\[0.5mm]
$\sigma_{8}$            & $0.823\pm0.020$  & $0.828\pm0.016$  &  $0.827\pm0.016$  &  $0.826\pm0.014$ \\[0.5mm]
$t_{0}/{\rm Gyr}$       & $13.729\pm0.081$  & $13.753\pm0.065$  &  $13.752\pm0.066$  &  $13.753\pm0.062$ \\[0.5mm] 
$h$                     & $0.703\pm0.019 $  & $0.695\pm0.013$  &   $0.695\pm0.012$  &  $   0.6962\pm0.0088$  \\[0.5mm]
$f(z_{\rm m})$          & $0.752\pm0.019$  & $0.760\pm0.012$ &   $0.759\pm0.012$  &  $0.7585\pm0.0085$ \\
\hline
\end{tabular}
\label{tab:lcdm_full}
\end{table*}

Panel a) of Fig.~\ref{fig:pthalos2d} shows the two-dimensional marginalized constraints in
the $\Omega_{\rm m}$--$h$ plane obtained when exploring the parameters of the $\Lambda$CDM model 
using our mock CMB data (dashed lines), and its combination with the mean clustering wedges of our mock CMASS
catalogues. The information in the clustering wedges alleviates the
degeneracy in the CMB constraints, leading to a reduction of the allowed ranges of these parameters.
In this case we find $\Omega_{\rm m}=0.276\pm0.015$ and $h=0.698\pm0.012$, in excellent agreement with their true
underlying values of $\Omega_{\rm m}=0.274$ and $h=0.7$. 

More stringent tests of our methodology can be obtained extending the parameter space by including $w_{\rm DE}$
as a free parameter, where degeneracies in the CMB data allow for significant deviations from the 
true cosmology. The results obtained in this case can be seen in Panel b) of Fig.~\ref{fig:pthalos2d}, which
shows the two-dimensional marginalized constraints in the $\Omega_{\rm m}$--$w_{\rm DE}$ plane obtained from our
mock CMB+CMASS dataset (solid lines). 
For this parameter space we find $\Omega_{\rm m}=0.278\pm0.036$ and $w_{\rm DE}=-1.00\pm0.14$.
The allowed region for these parameters is centred in the correct underlying values,
indicated by the dotted lines. 

The model of the clustering wedges described in Section~\ref{sec:model} depends on the value of the growth 
index $f(z_{\rm m})$. This fact offers the opportunity to obtain constraints on this quantity by treating it as
a free parameter and checking its consistency with the value corresponding to GR.
To test the performance of our model for the clustering wedges in this case, we explored the $\Lambda$CDM
parameter space extended by including $f(z_{\rm m})$ as a free parameter.
The solid lines in panel c) of Fig.~\ref{fig:pthalos2d} correspond to the two-dimensional marginalized
constraints in the $\Omega_{\rm m}$--$f(z_{\rm m})$ plane obtained from the combination of our mock CMB and
CMASS datasets.
Also in this case the obtained constraints are in good agreement with the true underlying values for these
cosmological parameters.
We find $f(z_{\rm m})=0.73\pm 0.11$, slightly lower than the fiducial value of
$f=0.75$ but consistent within one~$\sigma$. 

The assumption that $f(z_{\rm m})$ follows the predictions of GR has a significant impact on the constraints
on $w_{\rm DE}$.
Under this assumption, the relative amplitude of the two clustering wedges, which depends on $f(z_{\rm m})$,
contains information on $\Omega_{\rm m}$, helping to constrain $w_{\rm DE}$ by reducing the CMB-only degeneracy
that can be seen in the dashed lines of panel b) of Fig.~\ref{fig:pthalos2d}.
When $f(z_{\rm m})$ is treated as a free parameter, this extra information is lost, leading to a degradation of the 
constraints.
When both $f(z_{\rm m})$ and $w_{\rm DE}$ are allowed to float, the constraints on the dark energy 
equation of state change to $w_{\rm DE}=-1.11\pm0.23$. Although a tail in the posterior distribution of this 
parameter shifts the mean value of this parameter towards lower values, 
its maximum is close to the true fiducial one, $w_{\rm DE}=-1$.

In Section~\ref{sec:dist} we focus on the constraints on the dimensionless parameter combinations 
$d_{\perp}\equiv D_{\rm A}(z_{\rm m})/r_{\rm s}(z_{\rm m})$ and
$d_{\parallel}\equiv cz/(r_{\rm s}(z_{\rm m})H(z_{\rm m}))$   
obtained from the clustering wedges in isolation, that is, without combining them with any other dataset.
We explored the constraints on these quantities using the mean clustering wedges
of our mock catalogues, marginalizing over the remaining cosmological parameters.
In this case we obtain the constraints $d_{\perp}=8.89\pm0.29$ and
$d_{\parallel}=11.93\pm0.54$. These results are in complete agreement with the
values of these parameter combinations in our fiducial cosmology, of 8.87 and 11.92, respectively.

These tests show that the model of the clustering wedges described in Section~\ref{sec:model}
can be used as a tool to extract unbiased cosmological constraints from the clustering
wedges of the CMASS sample. 

\section{Cosmological constraints}
\label{sec:results}

In this section we present the cosmological constraints obtained from the full shape of the CMASS clustering
wedges $\xi_{\perp}(s)$ and $\xi_{\parallel}(s)$.
Section~\ref{sec:lambda} presents a summary of the results obtained on the $\Lambda$CDM model and its
extensions assuming that the dark energy component can be characterized by $w_{\rm DE}=-1$, while in
Section~\ref{sec:de} we explore more general models to obtain clues on the origin of the 
accelerated expansion of the Universe. Our results show the impact of replacing the information
from the monopole correlation function by that of the clustering wedges. In the 
cases where $w_{\rm DE}$ is fixed, the information contained in $\xi_0(s)$ is sufficient to
obtain accurate constraints, with the clustering wedges providing only a slight improvement.
When this assumption is relaxed, the extra information provided by the clustering wedges is more useful,
leading to substantial improvements with respect to the results obtained by means of $\xi_0(s)$.

\subsection{Dark energy as a cosmological constant}
\label{sec:lambda}

\begin{table*} 
\centering
  \caption{
    The marginalized 68\% constraints on the most relevant cosmological parameters of the extensions of the $\Lambda$CDM model 
    analysed in Sections~\ref{sec:omk} to \ref{sec:wa}, obtained using different combinations of the datasets described in Section~\ref{sec:data}.
    A complete list of the constrains obtained in each case can be found in Appendix~\ref{sec:tables}.
}
    \begin{tabular}{@{}lcccc@{}}
    \hline
 &\multirow{2}{*}{CMB}  & \multirow{2}{*}{CMB+$\xi_0(s)$} & \multirow{2}{*}{CMB+$(\xi_{\perp}(s),\xi_{\parallel}(s))$}& CMB+$(\xi_{\perp}(s),\xi_{\parallel}(s))$  \\
 &                       &                           &      &    +BAO+SN         \\  
\hline
\multicolumn{2}{l}{Non-flat models}& & &  \\[0.6mm]
$\Omega_k$                &  $-1.118\pm0.021$   &  $-0.0033_{-0.0044}^{+0.0046}$   &  $-0.0040\pm0.0045$   &  $-0.0041\pm0.0039$ \\[0.4mm]
$\Omega_{\rm DE}$       &$0.690\pm0.072$   &  $0.715\pm0.0145$   &  $0.715\pm0.015$   &  $0.721\pm0.011$  \\[0.4mm]
$\Omega_{\rm m}$        & $0.321\pm0.093$   &  $0.288\pm0.016$   &  $0.288\pm-0.016$   &  $0.283\pm0.010$  \\[0.4mm]
\hline
\multicolumn{2}{l}{Massive neutrinos}& & &  \\[0.6mm]
$f_\nu$  &  $< 0.12 $ (95\% CL) & $< 0.054$ (95\% CL)  & $ <0.051$ (95\% CL)  & $ < 0.043 $ (95\% CL) \\[0.4mm]
$\sum m_\nu$  &  $ < 1.6\,{\rm eV} $ (95\% CL) & $ < 0.68\,{\rm eV} $ (95\% CL)  & $ < 0.62\,{\rm eV} $ (95\% CL)  & $ < 0.50\,{\rm eV}$ (95\% CL)\\ 
$\Omega_{\rm m}$        & $0.385_{-0.072}^{+0.069}$   & $0.302_{-0.020}^{+0.021}$  &  $0.302\pm0.018$  &  $0.291\pm0.012$    \\
\hline
\multicolumn{3}{l}{Constant dark energy equation of state} & &  \\[0.6mm]
$w_{\rm DE}$            &$-1.14\pm0.42$  &  $-0.99_{-0.20}^{+0.21}$  &  $-0.93\pm0.11$  &  $-1.013\pm0.064$ \\[0.4mm]
$\Omega_{\rm m}$        & $0.26\pm0.10$  &  $0.291\pm0.042$  &  $0.299\pm0.028$  &  $0.283\pm0.012$   \\[0.4mm]
\hline
\multicolumn{2}{l}{Dark energy and curvature}& &  \\[0.6mm]
$w_{\rm DE}$            & $-0.89_{-0.45}^{+0.44}$ &   $-0.96_{-0.28}^{+29}$ &  $-0.97\pm0.16$  &   $-1.042\pm0.068$ \\[0.4mm]
$\Omega_k$             & $-0.022_{-0.031}^{+0.027}$ &   $0.0012_{-0.0077}^{+0.0091}$ &  $-0.0023_{-0.0060}^{+0.0061}$  &   $-0.0047\pm0.0042$ \\[0.4mm]
$\Omega_{\rm m}$        & $0.265_{-0.094}^{+0.097}$ & $0.280_{-0.083}^{+0.093}$ & $0.297\pm0.046$  & $0.278\pm0.013$   \\
\hline%\hline
\multicolumn{3}{l}{Time-dependent dark energy equation of state} & &  \\[0.6mm]
$w_0$            &$-1.01_{-0.53}^{+0.56}$ & $-1.11_{-0.60}^{+0.63}$ & $-0.96_{-0.39}^{+0.40}$  & $-1.10_{-0.12}^{+0.12}$ \\[0.4mm]
$w_a$            & $-0.4_{-1.5}^{+1.1}$ & $0.2\pm1.0$ & $0.03_{-0.97}^{+0.96}$  & $0.31\pm0.40$ \\[0.4mm]
$\Omega_{\rm m}$        & $0.285\pm0.015$           & $0.296\pm0.037$   & $0.284\pm0.011$             &  $0.282\pm0.012$  \\
\hline
\end{tabular}
\label{tab:extra}
\end{table*}

Here we investigate the impact of the clustering wedges on the constraints on  
the $\Lambda$CDM parameter space and its extensions assuming that 
dark energy behaves as a cosmological constant, that is, keeping the dark energy
equation of state parameter fixed to $w_{\rm DE}=-1$. 

\subsubsection{The $\Lambda$CDM parameter space}
\label{sec:lcdm}

Due to its ability to describe a wide variety of cosmological observations, 
the $\Lambda$CDM model has become the standard cosmological model.
Here we study the constraints on this parameter space obtained from the datasets
described in Section~\ref{sec:data}.
Table~\ref{tab:lcdm_full} lists the constraints obtained in this case from
different dataset combinations.

With the inclusion of the new WMAP9 data, the CMB-only constraints on this parameter space 
have changed slightly from those of \citet{Sanchez2012}. The preferred values of $\Omega_{\rm m}$
and $h$ show a shift of approximately 0.5~$\sigma$ towards higher and lower values, respectively, 
with $\Omega_{\rm m}=0.277\pm0.022$ and $h=0.703\pm 0.019$. 
Although this parameter space is well constrained by the CMB data, including the CMASS monopole
in the analysis substantially improves the obtained constraints by breaking the degeneracy
between these parameters, leading to $\Omega_{\rm m}=0.285\pm0.015$ and $h=0.695\pm 0.013$.
Replacing the information in the CMASS $\xi_0(s)$ by that of the full shape of the clustering wedges
leads to essentially identical results.
The consistency between the constraints obtained in the CMB+$\xi_0(s)$ and
CMB+$(\xi_{\perp}(s),\xi_{\parallel}(s))$ cases is a confirmation of the validity of the treatment of
redshift-space distortions implemented in our modelling of the clustering wedges.

The best-fit $\Lambda$CDM model to the CMB+$(\xi_{\perp}(s),\xi_{\parallel}(s))$ combination is shown by 
the dashed lines in Fig.~\ref{fig:cmass}. This model provides an excellent description of the full shape of
both CMASS clustering wedges and, in particular, of the shape and location of their BAO features.
This model implies a real-space bias factor of $b=1.94\pm0.08$,
in good agreement with the one inferred from the CMASS monopole, $b=1.96\pm0.09$, and with the results of 
\citet{Nuza2012}, who found a value of $b\simeq2$ for this galaxy sample.
Within this parameter space, and with the assumption of GR, the combination of the CMB data with our CMASS
clustering measurements provides a constraint on the growth factor of $f(z_{\rm m})=0.759\pm0.012$.
As we will see in Section~\ref{sec:fg}, when $f(z_{\rm m})$ is treated as a free parameter, the constraints
obtained from the CMB+$(\xi_{\perp}(s),\xi_{\parallel}(s))$ combination are in agreement with this result.

Including the additional BAO and SN datasets provides an improvement on the constraints on this parameter
space beyond the results found in the CMB+$(\xi_{\perp}(s),\xi_{\parallel}(s))$ case.
Using the full dataset combination we obtain the constraints $\Omega_{\rm m}=0.283_{-0.010}^{+0.010}$ and $h=0.6962\pm0.0088$. 
These results imply that, within the $\Lambda$CDM model, the present-day dark energy density can 
be constrained to $\rho_{\rm DE}=(6.53\pm0.25)\times10^{-30}\,{\rm g}\,{\rm cm}^{-3}$,
and the current acceleration of cosmic expansion to $\ddot{a}=(4.09\pm0.16)\times10^{-11}{\rm year}^{-2}$, that is, 
at a 4 per cent accuracy.

\subsubsection{Spatial Curvature}
\label{sec:omk}

\begin{figure}
\includegraphics[width=0.45\textwidth]{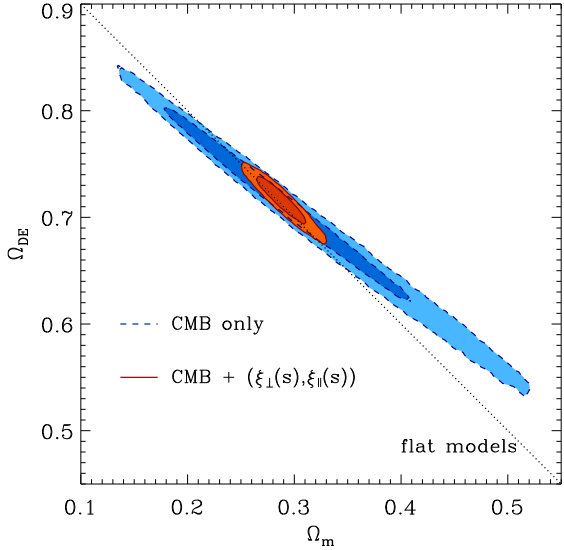}
\caption{
The marginalized 68 and 95 per cent CL in the $\Omega_{\rm m}$--$\Omega_{\rm DE}$ plane 
for the $\Lambda$CDM parameter set extended by allowing for non-flat models.
The dashed lines correspond to the results obtained using
CMB information alone. The solid lines show to the results obtained from the combination of CMB
data plus the CMASS clustering wedges.
}
\label{fig:omk}  
\end{figure}

When the $\Lambda$CDM parameter space is extended by allowing for non-flat models (i.e. $\Omega_k\neq0$)
the CMB-only constraints are significantly degraded due to the so-called geometric degeneracy \citep{Efstathiou1999},
corresponding to models with a constant CMB acoustic scale $\ell$ (given by equation~\ref{eq:lcmb}).
This degeneracy is shown by the dashed lines in Fig.~\ref{fig:omk}, which correspond to the 
two-dimensional marginalized constraints in the $\Omega_{\rm m}$--$\Omega_{\rm DE}$ plane obtained
using CMB information alone. The solid contours correspond to the 
results obtained when the CMB data is combined with the CMASS clustering wedges.
The information in the shape of the clustering wedges breaks the geometrical
degeneracy, providing much tighter constraints on these parameters. In this case we find 
the constraints $\Omega_{\rm m}=0.288\pm0.016$ and $\Omega_k=-0.0040\pm0.0045$, in good
agreement with a flat Universe.

As can be seen in Table~\ref{tab:extra}, the CMB+$\xi_0(s)$ combination leads to a similar constraint than
the one obtained in the CMB+$(\xi_{\perp}(s),\xi_{\parallel}(s))$ case, with
$\Omega_k=-0.0033_{-0.0044}^{+0.0046}$ and identical constraints on $\Omega_{\rm m}$. This result demonstrates 
that the measurement of $D_{\rm V}/r_{\rm s}(z_{\rm d})$ obtained from the
monopole is sufficient to break the CMB degeneracies, with the extra information from the clustering wedges
leading to similar constraints on $\Omega_k$.

Using the geometric constraints from the monopole-quadrupole of the same galaxy sample derived by \citet{Reid2012}
in combination with the seven-year data of the WMAP satellite \citep[WMAP7][]{Larson2011}, \citet{Samushia2013} 
obtained the constraint $\Omega_k=-0.0085_{-0.0054}^{+0.0054}$. This value is consistent with our 
results from the CMB+$(\xi_{\perp}(s),\xi_{\parallel}(s))$ combination within 1$\sigma$, but points towards a larger deviation from
a flat Universe. %As we will see in Section~\ref{sec:dist}, 
This difference can be traced back to the slightly higher value of $D_{\rm V}/r_{\rm s}$
found by \citet{Reid2012}, which cuts the geometrical degeneracy
of the CMB at a different location.

\subsubsection{Neutrino mass}
\label{sec:fnu}

\begin{figure}
\includegraphics[width=0.45\textwidth]{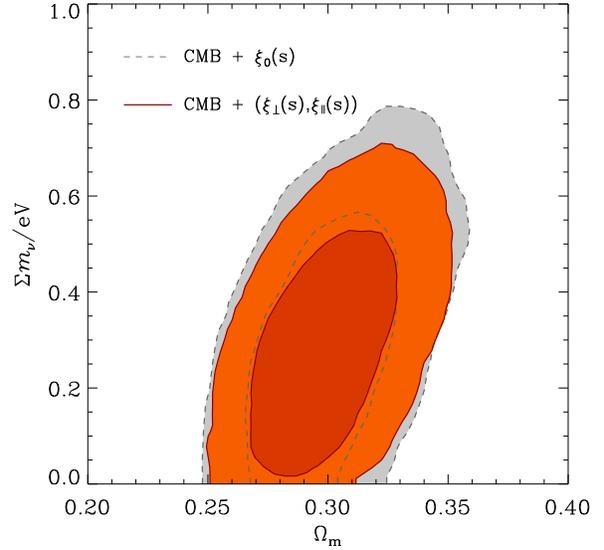}
\caption{
The marginalized constraints in the $\Omega_{\rm m}$--$\Sigma m_{\nu}$ plane
for the $\Lambda$CDM parameter set extended by allowing for massive neutrinos.
The dashed and solid lines correspond to the 68 and 95 per cent CL derived
by combining our CMB data with the full shapes of the CMASS monopole (dashed lines)
and clustering wedges (solid lines).
}
\label{fig:mnu}  
\end{figure}

In this section we study how the constraints on the
total neutrino mass change when using the information from the CMASS clustering wedges
by exploring the $\Lambda$CDM parameter space extended by including 
the neutrino fraction, $f_{\nu}$, as a free parameter.
A more detailed analysis of the constraints on neutrino masses inferred from the CMASS sample,
paying special attention to the effects of the different priors and datasets 
used in the analysis, is presented in \citet{Zhao2012}.

Fig.~\ref{fig:mnu} shows the two-dimensional marginalized constraints in the $\Omega_{\rm m}$--$\sum m_{\nu}$
plane obtained by means of the CMB+$\xi_0(s)$ (dashed lines) and CMB+$(\xi_{\perp}(s),\xi_{\parallel}(s))$
(solid lines) dataset combinations. The use of the clustering wedges leads to a slight improvement on the constraints
with respect to those obtained from the CMASS monopole. 
The CMB+$\xi_0(s)$ combination gives $\Omega_{\rm m}=0.302_{-0.020}^{+0.021}$ and
$f_{\nu} < 0.054$ (95 per cent CL). Replacing the angle-averaged correlation function by the clustering wedges
leads to a slight improvement of the constraints, with
$\Omega_{\rm m}=0.302 \pm0.018$ and $f_{\nu} < 0.051$ (95 per cent CL).
These results imply a final limit on the sum of the neutrino masses of $\sum m_{\nu} < 0.68$ eV
(95 per cent CL) for the
CMB+$\xi_0(s)$ case and  $\sum m_{\nu} < 0.62$ eV (95 per cent CL) when combining the CMB dataset with the CMASS clustering wedges.
Including the additional BAO and SN information helps to improve these limits to
$f_{\nu} < 0.043$ and $\sum m_{\nu} < 0.50$ eV (95 per cent CL).

\subsection{Understanding cosmic acceleration} 
\label{sec:de}

In the previous sections we assumed that the dark energy component was given by vacuum energy or a
cosmological constant, with $w_{\rm DE}=-1$. 
In this section we investigate alternative explanations of the observed accelerated expansion 
of the Universe by exploring constraints on the dark energy equation of state and its time evolution. 
As we will see, the constraints on these parameter spaces are substantially improved when the
information from the CMASS monopole is replaced by that of the clustering wedges.
We also analyse potential deviations from general relativity by exploring the
constraints on the growth factor $f(z_{\rm m})$, which can only be obtained from anisotropic clustering
measurements.

\subsubsection{The dark energy equation of state}
\label{sec:wde}

\begin{figure}
\includegraphics[width=0.45\textwidth]{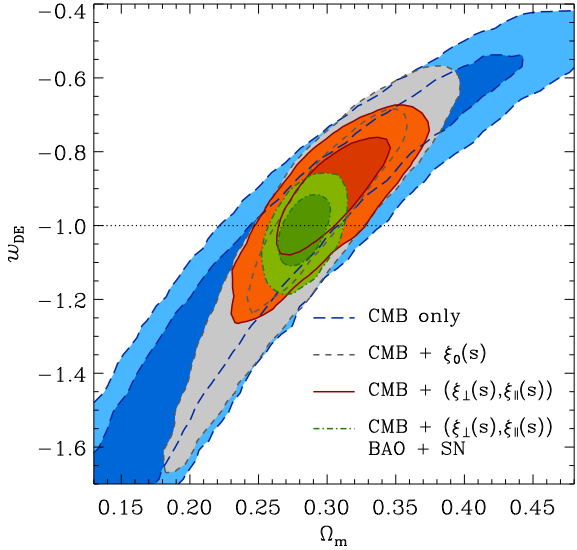}
\caption{
The marginalized 68 and 95 per cent CL in the $\Omega_{\rm m}$--$w_{\rm DE}$ plane 
for the $\Lambda$CDM parameter set extended by including the redshift-independent value of $w_{\rm DE}$
as an additional parameter. The different sets of contours correspond to the results obtained using the
CMB-only (long-dashed lines), the CMB+$\xi_0(s)$ combination (short-dashed lines), 
the CMB+$(\xi_{\perp}(s),\xi_{\parallel}(s))$ (solid lines), and when this information is combined
with our BAO and SN datasets (dot-dashed lines). The dotted line corresponds to the $\Lambda$CDM model value
of $w_{\rm DE}=-1$.
}
\label{fig:wde}  
\end{figure}

We start our exploration of more general dark energy models by extending the basic $\Lambda$CDM
parameter space by including the redshift-independent value of $w_{\rm DE}$ as a free parameter.
In this case, the constraints derived from CMB data alone exhibit a strong degeneracy
between $\Omega_{\rm m}$ and $w_{\rm DE}$. 
This is illustrated by the long-dashed lines in Fig.~\ref{fig:wde}, which correspond to the two-dimensional
marginalized constraints on these parameters obtained from our CMB dataset.
As shown in \citet{Sanchez2012}, this degeneracy is
partially broken when this information is combined with the CMASS monopole correlation function. 
Using this combination we find the marginalized constraints
$\Omega_{\rm m}=0.291\pm0.042$ and $w_{\rm DE}=-0.99_{-0.20}^{+0.21}$, in agreement with the fiducial value of
the $\Lambda$CDM model but with significant room for alternative dark energy scenarios. 
The short-dashed lines in Fig.~\ref{fig:wde} correspond to the constraints in 
the $\Omega_{\rm m}-w_{\rm DE}$ plane obtained in this case, while the solid lines
show the effect of replacing the information of the CMASS monopole
by that of the full shape of the CMASS $(\xi_{\perp}(s),\xi_{\parallel}(s))$ pair.
This information is much more efficient at breaking the CMB degeneracy than $\xi_0(s)$, leading to
a significant improvement of the obtained constraints. In this case we obtain  
$\Omega_{\rm m}=0.299\pm0.028$ and $w_{\rm DE}=-0.93\pm0.11$.
These results, which are consistent with a cosmological constant at a one~$\sigma$ level, represent 
a reduction of the allowed range of these parameters by a factor two with respect to the
ones obtained by means of the CMB+$\xi_0(s)$ combination. 

Using the consensus anisotropic BAO measurements from the CMASS clustering wedges and multipoles, 
\citet{Abalone2013} found a constraint of $w_{\rm DE}=-0.90\pm0.22$, quite similar to the results
obtained using the isotropic BAO results of \citet{Anderson2012}.
The comparison of this result with the ones from the CMB+$(\xi_{\perp}(s),\xi_{\parallel}(s))$
combination highlights the importance of using information from the full shape of the anisotropic 
clustering measurements to increase the information extracted from galaxy surveys.
As we will see in Section~\ref{sec:fg}, this extra information is degraded when $f(z_{\rm m})$
is treated as a free parameter. 

Our results are in excellent agreement with those derived from the full shape of the CMASS
monopole-quadrupole pair in our companion paper \citet{Chuang2013}, who find $w_{\rm DE}=-0.94\pm0.13$.
\citet{Samushia2013} obtained the constraints $\Omega_{\rm m}=0.313\pm0.017$ and $w_{\rm DE}=-0.87\pm0.05$
from the combination of the anisotropic clustering measurements of \citet{Reid2012} and WMAP7 data.
%The difference between these constraints and our results from the 
%CMB+$(\xi_{\perp}(s),\xi_{\parallel}(s))$ combination is related to the values of
%$d_{\rm s}(z)=D_{\rm V}(z)/r_{\rm s}$ and $F(z)=D_{\rm A}(z)H(z)/c$ obtained by \citet{Reid2012}. 
By including smaller scales than in our analysis, with a different binning scheme, and imposing a
stronger prior on the finger-of-god parameter $\sigma_{\rm v}$, \citet{Reid2012} found
slightly different, but consistent, geometrical constraints.
These values cut the CMB-only degeneracy in a
different region than our results, corresponding to slightly higher values or
$w_{\rm DE}$, with a smaller allowed range for this parameter.

\begin{figure}
\includegraphics[width=0.45\textwidth]{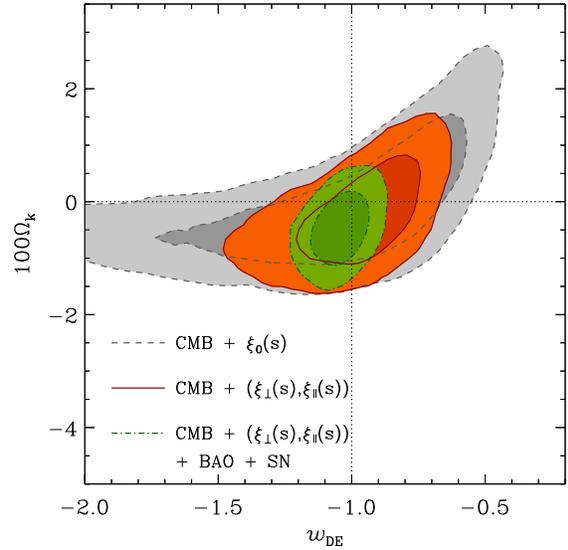}
\caption{
The marginalized constraints in the $w_{\rm DE}$--$\Omega_k$ plane for the $\Lambda$CDM 
parameter set extended by allowing for simultaneous variations on both of these parameters.
The contours correspond to the  68 and 95 per cent CL derived from the combination of CMB
data with the CMASS monopole (dashed lines), the CMB plus the clustering wedges (solid lines), and 
when the additional BAO and SN datasets are added to the later combination (dot-dashed lines).
The dotted lines correspond to the values of these parameters in the $\Lambda$CDM model.
}
\label{fig:wok}  
\end{figure}

Our final constraints, obtained by including the additional BAO and SN data in the analysis, are shown by
the dot-dashed lines in Fig.~\ref{fig:wde}, corresponding to $\Omega_{\rm m}=0.283\pm0.012$ and $w_{\rm DE}=-1.013\pm0.064$.
This result is in excellent agreement with the standard $\Lambda$CDM model value of $w_{\rm DE}=-1$,
indicated by a dotted line in Fig.\ref{fig:wde}.

\subsubsection{Dark energy and curvature}
\label{sec:wok}

When the dark energy equation of state parameter and $\Omega_k$ are varied simultaneously, 
the geometric degeneracy seen in the CMB-only results of Figs.~\ref{fig:omk} and
\ref{fig:wde} gains an extra degree of freedom, leading to poor constraints on both of these parameters.
For this reason, the flatness hypothesis has strong implications on the derived constraints on the
dark energy equation of state. In this section we explore how the constraints on $w_{\rm DE}$ are degraded 
if this assumption is relaxed. 

The dashed lines in Fig.~\ref{fig:wok} show the two-dimensional marginalized constraints
in the $\Omega_k$--$w_{\rm DE}$ plane obtained by combining our CMB dataset with the 
CMASS monopole. The information encoded in $\xi_0(s)$ reduces the two-dimensional
degeneracy obtained from the CMB data to an approximately one-dimensional degeneracy,
which allows for values of $w_{\rm DE}$ significantly different from the $\Lambda$CDM one,
with $w_{\rm DE}=-0.96^{+0.29}_{-0.28}$. 
The solid contours in Fig.~\ref{fig:wok} correspond to the results obtained when
the CMB data is combined with the CMASS clustering wedges. This dataset is much more efficient 
at breaking the degeneracy obtained from the CMB results, leading to a significant reduction
of the allowed region of this parameter space, with the marginalized constraints 
$\Omega_{k}=-0.0023\pm0.0061$ and $w_{\rm DE}=w=-0.97\pm0.16$, in excellent agreement with 
the $\Lambda$CDM model values, indicated by the dotted lines.
Including the SN and additional BAO datasets  
improves the constraints even further, leading to $\Omega_{k}=-0.0047\pm0.0042$ and
$w_{\rm DE}=w=-1.042\pm0.068$.

\begin{figure}
\includegraphics[width=0.45\textwidth]{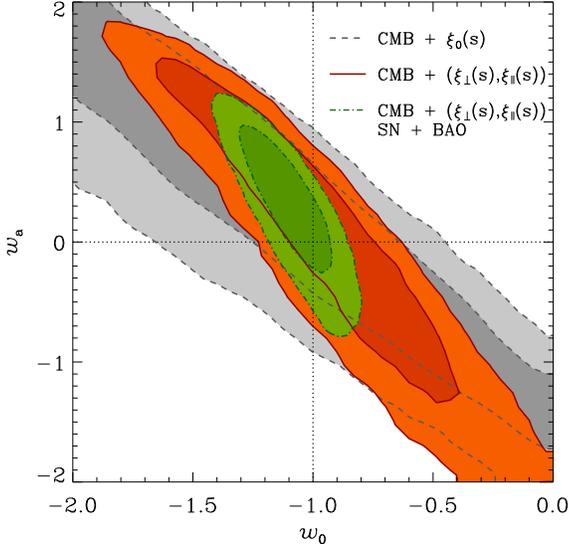}
\caption{
The marginalized constraints in the $w_0$--$w_a$ plane obtained when 
we explore the redshift dependence of the dark energy equation of state, 
parametrized as in equation~(\ref{eq:wa}).
The contours correspond to the  68 and 95 per cent CL derived from the combination of CMB
data with the CMASS monopole (dashed lines), the CMB plus the clustering wedges (solid lines), and 
when the additional BAO and SN datasets are added to the later combination (dot-dashed lines).
The dotted lines correspond to the values of these parameters in the $\Lambda$CDM model.
}
\label{fig:w0wa}  
\end{figure}

\subsubsection{The time evolution of $w_{\rm DE}$}
\label{sec:wa}

\begin{figure}
\includegraphics[width=0.48\textwidth]{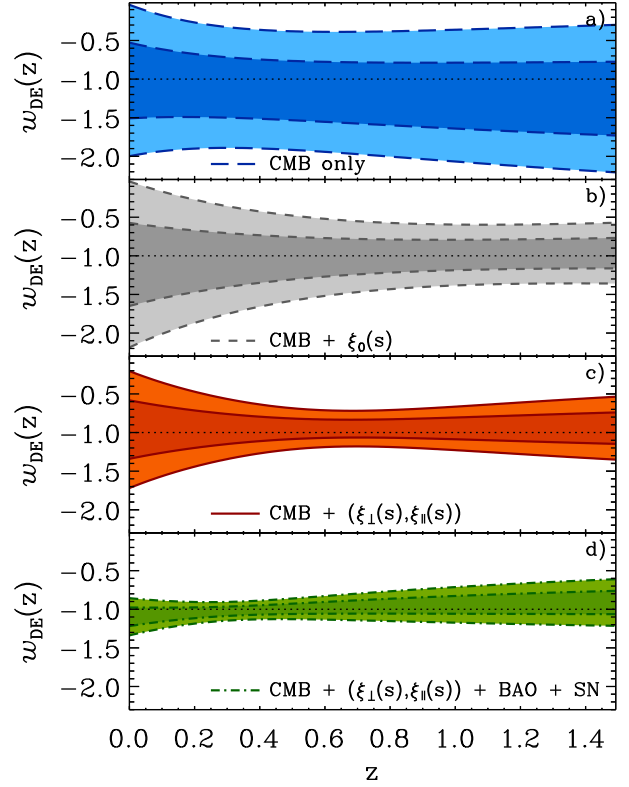}
\caption{
The marginalized 68 and 95 per cent CL on the dark energy equation of state as a function of redshift
derived from the CMB data alone (panel a), the CMB combined with the CMASS monopole 
(panel b), the CMB combined with the clustering wedges (solid lines),
and when the BAO and SN datasets are added to the later combination (panel d).
}
\label{fig:wa}  
\end{figure}

In this section we study the constraints on the time evolution of $w_{\rm DE}$, 
parametrized as in equation~(\ref{eq:wa}). Only by combining the CMB information with additional
datasets it is possible to obtain meaningful constraints on these parameters.
\begin{table*} 
\centering
  \caption{
    The marginalized 68\% constraints on the most relevant cosmological parameters of the $\Lambda$CDM model extended 
    by treating $f(z_{\rm m})$ as a free parameter, and when $f(z_{\rm m})$ and $w_{\rm DE}$ are varied simultaneously.
    The second and third columns correspond to the constraints obtained by combining the CMB data with the CMASS clustering
    wedges, while the last two columns show the result of including also the SN and additional BAO measurements.
    A complete list of the constrains obtained in each case can be found in Appendix~\ref{sec:tables}.}
    \begin{tabular}{@{}lcccc@{}}
    \hline
& \multicolumn{2}{c}{\multirow{2}{*}{CMB+$(\xi_{\perp}(s),\xi_{\parallel}(s))$}}    &\multicolumn{2}{c}{CMB+$(\xi_{\perp}(s),\xi_{\parallel}(s))$}   \\
&                       &                                    &      \multicolumn{2}{c}{+BAO+SN}         \\  
\hline
$f(z_{\rm m})$          &$0.719_{-0.096}^{+0.092}$  & $0.76\pm0.14$     & $0.715_{-0.098}^{+0.095}$   &  $0.706_{-0.099}^{+0.096}$\\[0.5mm]
$w_{\rm DE}$            &     --                    & $-0.95\pm0.17$    &         --                  &  $-1.035_{-0.069}^{+0.071}$ \\[0.5mm]
\hline
$\gamma$                & $0.59\pm0.23$             & $0.50\pm0.33$     & $0.60\pm0.23$               &  $ 0.64\pm0.26$  \\[0.5mm]
$\Omega_{\rm m}$        & $0.285\pm0.015$           & $0.296\pm0.037$   & $0.284\pm0.011$             &  $0.282\pm0.012$  \\[0.5mm]
$h$                     & $0.695\pm0.012$           & $0.684\pm0.048$   & $0.6956\pm0.0088$           &  $0.701\pm0.015$  \\
\hline
\end{tabular}
\label{tab:fg}
\end{table*}

The dashed lines in Fig.~\ref{fig:w0wa} correspond to the two-dimensional marginalized constraints 
in the $w_0$--$w_a$ plane obtained from the combination of our CMB dataset with the 
CMASS monopole. These results exhibit a strong degeneracy which spans the 
entire range of values allowed by our priors on these parameters. This degeneracy can be described 
by a linear combination of $w_0$ and $w_a$, corresponding to the value of the dark energy equation of
state at the pivot redshift $z_{\rm p}$, that is, the point where the uncertainty in $w_{\rm DE}(z)$ is minimized.
In this case we find $w_{\rm DE}(z_{\rm p}= 1.22)=-0.98\pm0.19$.
The solid lines in Fig.~\ref{fig:w0wa}
correspond to the results obtained by combining the CMB data with the CMASS clustering wedges, showing a 
significant reduction of the allowed region for these parameters. In this case we find the marginalized limits
$w_0=-0.96_{-0.39}^{+0.40}$ and $w_a=0.03_{-0.97}^{+0.96}$.
As it is a common feature resulting from the parametrization of equation~(\ref{eq:wa}),
the results obtained from the CMB+$(\xi_{\perp}(s),\xi_{\parallel}(s))$ combination 
also exhibit an approximately linear degeneracy between $w_0$ and $w_a$, which corresponds to the constraint
$w_{\rm DE}(z_{\rm p}=0.69)=-0.95\pm0.11$. 
The inclusion of the BAO and SN datasets tightens the constraints, leading to
$w_0=-1.10\pm0.12$ and $w_a=0.31\pm0.40$, consistent with the $\Lambda$CDM model.

 The constraints obtained by means of different dataset combinations can be characterized by the Figure-of-Merit,
FoM, defined as \citep{Albrecht2006,Wang2008}
\begin{equation}
 {\rm FoM}=\left({\rm det}\,{\rm Cov}(w_0,w_a)\right)^{-1/2},
\end{equation}
where ${\rm Cov}(w_0,w_a)$ corresponds to the $2\times2$ covariance matrix of the parameters $w_0$ and $w_a$.
While the results obtained using the CMB+$\xi_0(s)$combination are described by ${\rm FoM}_{{\rm CMB+}\xi_0(s)}=5.7$,
combining the CMB data with the CMASS clustering wedges instead leads to an increase of the FoM by nearly a factor two,
with ${\rm FoM}_{{\rm CMB+}(\xi_{\perp}(s),\xi_{\parallel}(s))}=9.8$. This improvement clearly illustrates the 
importance of including anisotropic clustering measurements when constraining this parameter space.
The final combination of all datasets leads to a further improvement by a factor four in the FoM,
with ${\rm FoM}_{{\rm All}}=41.2$.

Our constraints on $w_0$ and $w_a$ can be translated into constraints on the value of $w_{\rm DE}$ as a function of $z$.
Fig.~\ref{fig:wa} shows the 68 and 95 per cent CL $w_{\rm DE}(z)$ obtained using different dataset combinations.
Although the CMB-only constraints (panel a) allow for significant deviations from $w_{\rm DE}(z)=-1$, the inclusion of the 
CMASS $\xi_0(s)$ restricts these variations, especially for $z>0.6$ (panel b). 
The result of combining the CMB data with the clustering wedges is shown in panel c). This dataset combination substantially
reduces the allowed range of variations of $w_{\rm DE}(z)$.
Panel d) shows the results obtained when the additional BAO and SN datasets are included in the analysis, which 
are completely consistent with the $\Lambda$CDM model, showing no evidence of a deviation from the value $w_{\rm DE}=-1$ at any redshift.

\subsubsection{Constraining deviations from general relativity}
\label{sec:fg}

The current phase of accelerated expansion of the Universe can be interpreted as the signature of a failure
of general relativity to describe the behaviour of gravity on large scales.
This scenario cannot be distinguished from that of a dark energy component solely on the basis of
geometrical measurements. However, the combination of these tests with quantities sensitive to the growth of
density fluctuations can break this degeneracy.

The model of the clustering wedges described in Section~\ref{sec:model} depends on the value of 
$f(z_{\rm m})$, as it affects the pattern of redshift-space distortions in $\xi(\mu,s)$.
As shown in \citet{Linder2007}, in the context of general relativity the redshift evolution of 
this function can be well described by $f(z)=\Omega_{\rm m}(z)^{\gamma}$, with
\begin{equation}
 \gamma=
  \cases{ 0.55+0.05(1+w_{\rm DE}) &if $w_{\rm DE}\ge-1$, \cr
    0.55+0.02(1+w_{\rm DE}) &if $w_{\rm DE}<-1$. \cr
  }
\end{equation}
In this way, assuming $w_{\rm DE}=-1$, a detection of a deviation from $\gamma=0.55$
can be interpreted as the ``smoking gun'' of a failure of general relativity, as it
would indicate that the growth of density fluctuations is not consistent with its predictions.
In this section we explore the constraints obtained by treating $f(z_{\rm m})$ as a free parameter, instead
of using its derived value. By modelling the full shape of anisotropic clustering measurements, such as the
clustering wedges, it is possible to combine the geometrical BAO test with a measurement of structure growth
from the redshift-space distortions. 
Table~\ref{tab:fg} lists the constraints on the most relevant parameters obtained in this case, while a complete
list can be found in Appendix~\ref{sec:tables}.
This analysis is not possible when using angle-averaged measurements, where the effect of
varying $f(z_{\rm m})$ is completely degenerate with that of the bias parameter.

\begin{figure}
\centering
\centerline{\includegraphics[width=0.43\textwidth]{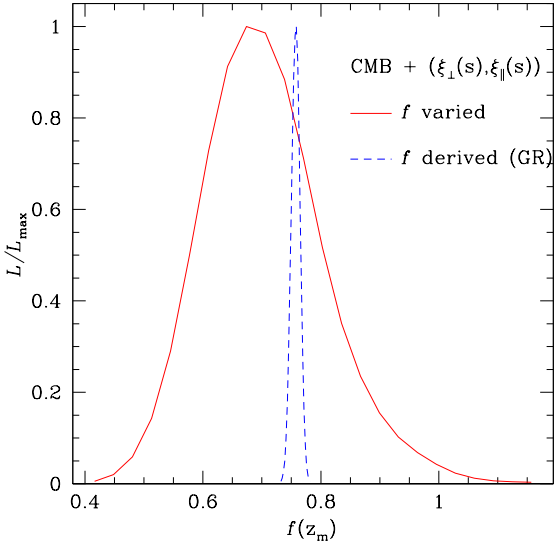}}
\caption{
Marginalized constraints on $f(z_{\rm m})$ obtained from the combination of our CMB dataset 
combined with the CMASS clustering wedges. The solid line corresponds to the result obtained 
when the $\Lambda$CDM model is extended by including $f(z_{\rm m})$ as a free parameter. This distribution
is consistent with the derived constraints on this parameter that are obtained when assuming that
it follows the predictions of general relativity, shown by the dashed line.
}
\label{fig:fg1d}  
\end{figure}

The solid line in Fig.~\ref{fig:fg1d} corresponds to the one-dimensional marginalized constraints on
$f(z_{\rm m})$ obtained from the CMB+$(\xi_{\perp}(s),\xi_{\parallel}(s))$ combination.
In this case we obtain $f(z_{\rm m})=0.719_{-0.096}^{+0.092}$. Although a wide range of values of this
parameter are allowed by the data, these results are consistent with the constraints derived when
assuming GR in the context of the $\Lambda$CDM model,
which are shown by the dashed line and correspond to $f(z_{\rm m})=0.759\pm 0.012$. The CMB+$(\xi_{\perp}(s),\xi_{\parallel}(s))$ result can be translated into a constraint of $\gamma=0.59\pm0.23$, consistent with the
GR prediction of $\gamma=0.55$.

\begin{figure}
\includegraphics[width=0.45\textwidth]{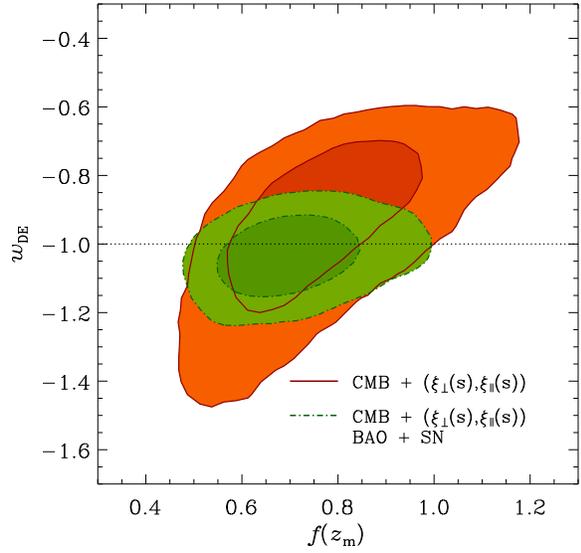}
\caption{
The marginalized 68 and 95 per cent CL in the $f(z_{\rm m})$--$w_{\rm DE}$ plane 
for the $\Lambda$CDM parameter set extended by including these as additional parameters. The solid lines 
correspond to the results obtained by combining the CMB data with the CMASS clustering wedges. The dot-dashed
lines show the result of including also the additional BAO and SN datasets in the analysis.
}
\label{fig:fgw0} 
\end{figure}

We also tested the effect of extending the $\Lambda$CDM parameter space by allowing for simultaneous variations
of $w_{\rm DE}$ (assumed time independent) and $f(z_{\rm m})$. Fig.~\ref{fig:fgw0} presents the two-dimensional
marginalized constraints in the $f(z_{\rm m})$--$w_{\rm DE}$ plane obtained in this case
by means of the CMB+$(\xi_{\perp},\xi_{\parallel})$ combination (solid lines), and when these data are combined
with the BAO and SN datasets (dot-dashed lines). Including $f(z_{\rm m})$ as a free parameter leads
to a degeneracy
between this quantity and the dark energy equation of state, degrading the constraints on these parameters.
In this case we find $w_{\rm DE}=-0.95\pm0.17$  and $f(z_{\rm m})=0.76\pm0.14$. 
As discussed in Section~\ref{sec:test2}, assuming that $f(z_{\rm m})$ follows the predictions of GR implies that 
the relative amplitude of $\xi_{\perp}(s)$ and $\xi_{\parallel}(s)$ provides information on $\Omega_{\rm m}$ which 
improves the constraints. However, treating $f(z_{\rm m})$ as a free parameter implies that this extra
constraining powers is lost, leading to weaker constraints.

The effect of treating $f(z_{\rm m})$ as a free parameter on the constraints on $w_{\rm DE}$ is
general to all anisotropic clustering measurements, as it degrades the information on $\Omega_{\rm m}$
than can be extracted from the observed redshift-space distortions.
This can be seen in the results of \citet{Samushia2013} and \citet{Chuang2013}, who find that the 
the constraints on $w_{\rm DE}$ derived from the analysis of the CMASS $\xi_0(s)$--$\xi_2(s)$ pair are
affected in a similar 
way as those obtained from the clustering wedges.

Including the SN and additional BAO measurements breaks the degeneracy present in the
CMB+$(\xi_{\perp},\xi_{\parallel})$ constraints,
leading to $w_{\rm DE}=-1.035_{-0.069}^{+0.071}$  and $f(z_{\rm m})=0.706_{-0.099}^{+0.096}$,
similar to the ones derived when these parameters are varied separately and are
in good agreement with the standard $\Lambda$CDM+GR cosmological model.

\subsection{Distance measurements}
\label{sec:dist}

In this section we focus on the constraints on geometrical quantities obtained from the 
CMASS clustering wedges and their combination with our CMB dataset. 
Angle-averaged quantities such as the monopole $\xi_0(r)$ provide constraints on the dimensionless quantity 
$d_{\rm s}(z_{\rm m}) = D_{\rm V}(z_{\rm m})/r_{\rm s}(z_{\rm d})$.
However, as $D_{\rm V}(z_{\rm m})\propto \left(D_{\rm A}(z_{\rm m})^2/H(z_{\rm m})\right)$, this measurement
represents a degeneracy between the angular diameter distance and the Hubble parameter which limits its power as
a tool to derive cosmological constraints. As discussed in Section~\ref{sec:da_h}, the clustering wedges provide
constraints on the parameter combinations $d_{\perp}(z_{\rm m})$ and $d_{\parallel}(z_{\rm m})$ given by equations~(\ref{eq:dperp})
and (\ref{eq:dpara}), breaking the degeneracy obtained from angle-averaged measurements.

\begin{figure}
\includegraphics[width=0.45\textwidth]{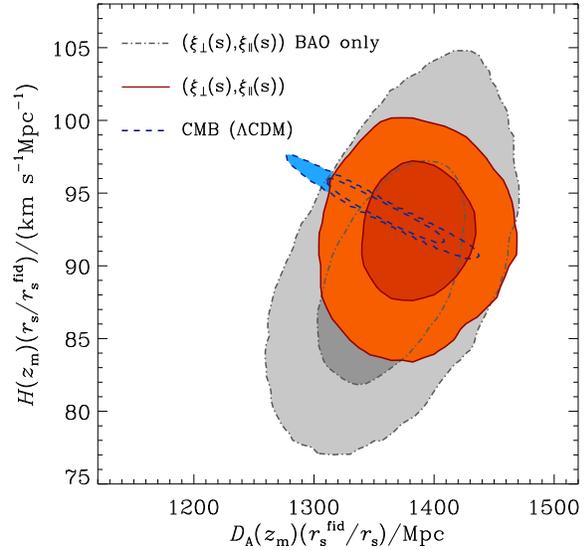}
\caption{
Two-dimensional marginalized constraints in the
$D_{\rm A}(z_{\rm m})\left(r_{\rm s}(z_{\rm d})^{\rm fid}/r_{\rm s}(z_{\rm d})\right)$--$H(z_{\rm m})\left(r_{\rm s}(z_{\rm d})/r_{\rm s}(z_{\rm d})^{\rm fid}\right)$
plane derived from the CMASS clustering wedges $\xi_{\perp}(s)$ and $\xi_{\parallel}(s)$ 
using the information from their full shape (solid lines) and from the BAO signal alone \citep[dashed lines, from][]{Kazin2013}.
The dashed contours correspond to the prediction for these parameters from the $\Lambda$CDM model fits to our CMB dataset.
}
\label{fig:dist} 
\end{figure}

Using the full shape of the CMASS clustering wedges alone, and assuming that
$f(z_{\rm m})$ follows the predictions of GR, we obtain the constraints
$d_{\perp}(z_{\rm m})=9.03\pm0.21$ and $d_{\parallel}(z_{\rm m})=12.14\pm0.43$, with a weak correlation
coefficient of $-7.8\times10^{-2}$. 
The two-dimensional marginalized constraints on these quantities are shown by the solid lines
in Fig.~\ref{fig:dist}, where we have rescaled our results by the sound horizon at the drag redshift
for our fiducial cosmology, $r_{\rm s}^{\rm fid}=153.2\,{\rm Mpc}$, to express them in
units of Mpc and ${\rm km}\,{\rm s}^{-1}{\rm Mpc}^{-1}$. 
These results are in good agreement with the constraints obtained from a BAO-only analysis of the same
CMASS clustering wedges in our companion paper \citet{Kazin2013}, which are shown by
the dot-dashed lines in the same figure.
This comparison illustrates the additional constraining power of the full shape of $\xi_{\perp}(s)$ and
$\xi_{\parallel}(s)$ beyond that in the scale of the BAO peak alone. While a pre-reconstruction BAO-only analysis gives
$D_{\rm A}(z_{\rm m})\left(r_{\rm s}(z_{\rm d})^{\rm fid}/r_{\rm s}(z_{\rm d})\right)= 1366\pm41\,{\rm Mpc}$
and $H(z_{\rm m})\left(r_{\rm s}(z_{\rm d})/r_{\rm s}(z_{\rm d})^{\rm fid}\right)=89.9\pm5.4\,{\rm km}\,{\rm s}^{-1}{\rm Mpc}^{-1}$,
we obtain $D_{\rm A}(z_{\rm m})\left(r_{\rm s}(z_{\rm d})^{\rm fid}/r_{\rm s}(z_{\rm d})\right)= 1384\pm32\,{\rm Mpc}$ and $H(z_{\rm m})\left(r_{\rm s}(z_{\rm d})/r_{\rm s}(z_{\rm d})^{\rm fid}\right)=92.0\pm3.3\,{\rm km}\,{\rm s}^{-1}{\rm Mpc}^{-1}$.

The constraints on $d_{\parallel}(z_{\rm m})$ and $d_{\perp}(z_{\rm m})$ obtained from the 
clustering wedges are degraded when $f(z_{\rm m})$ is treated as a free parameter.
In this case we find $d_{\perp}(z_{\rm m})=9.04\pm0.25$ and $d_{\parallel}(z_{\rm m})=12.23\pm0.56$.
These results are in agreement with those of \citet{Chuang2013}, who find $d_{\perp}=8.95\pm0.27$ and $d_{\parallel}=12.55\pm0.85$ when fitting the full-shape of $\xi_0(s)$ and $\xi_2(s)$ while simultaneously 
varying $D_{\rm A}(z_{\rm m})$, $H(z_{\rm m})$ and $f(z_{\rm m})\sigma_8$.

 Using the full shape of the CMASS angle-averaged correlation function, \citet{Sanchez2012} derived
the constraint $d_{\rm s}(z_{\rm m})=13.42\pm0.25$, in good agreement with the pre-reconstruction results of \citet{Anderson2012}.
From the analysis of the clustering wedges of this sample we find $d_{\rm s}(z_{\rm m})=13.46\pm0.25$,
showing that the same information is contained in the clustering wedges. The agreement between these
results also provides a 
consistency test of the explicit treatment of redshift-space distortions implemented in our modelling
of $\xi_{\perp}(s)$ and $\xi_{\parallel}(s)$.	

The geometric constraints obtained from the clustering wedges are in agreement with those derived by  
\citet{Reid2012} from the full shape of the CMASS monopole-quadrupole pair. 
Using a WMAP7-based prior on the primordial power spectrum to calibrate the BAO ruler,
\citet{Reid2012} found 
$D_{\rm A}(z_{\rm m})= 1403\pm28\,{\rm Mpc}$ and $H(z_{\rm m})=92.9_{-3.3}^{+3.6}\,{\rm km}\,{\rm s}^{-1}{\rm Mpc}^{-1}$.
Applying a similar prior we find $D_{\rm A}(z_{\rm m})= 1387\pm31\,{\rm Mpc}$ and $H(z_{\rm m})=92.3\pm3.3\,{\rm km}\,{\rm s}^{-1}{\rm Mpc}^{-1}$, showing the consistency between our analysis techniques.
By including smaller scales than in our analysis ($25 < s < 160\, h^{-1}{\rm Mpc}$) and imposing a stronger prior on the finger-of-god parameter $\sigma_{\rm v}$, \citet{Reid2012} found slightly different results than
in our general analysis (i.e., without imposing a WMAP7 prior), but in agreement.
This is the cause of many of the small differences in the cosmological constraints derived here and
those of \citet{Samushia2013}. When combined with CMB measurements, these geometric constraints cut
the CMB degeneracies in slightly different regions, leading to distinct, but consistent, constraints.

The dashed lines in Fig.~\ref{fig:dist} correspond to the constraints obtained from our CMB dataset under the assumption of a $\Lambda$CDM model.
The results obtained from the clustering wedges are in good agreement with the predictions of the $\Lambda$CDM model that best describes the CMB data.
These results are a strong indication of the consistency of these datasets and their good agreement with the 
concordance $\Lambda$CDM cosmological model.

When the clustering wedges are combined with CMB observations, the extra information leads to tighter constraints,
with $d_{\perp}= 9.06\pm0.19$ and $d_{\parallel}=12.05\pm0.28$. The information on $r_{\rm s}(z_{\rm d})$
provided by the CMB data makes it possible to derive direct constraints on the 
angular diameter distance and the Hubble parameter of $D_{\rm A}(z_{\rm m})= 1388\pm30\,{\rm Mpc}$ and
$H(z_{\rm m})=92.6\pm2.1\,{\rm km}\,{\rm s}^{-1}{\rm Mpc}^{-1}$.

\section{Conclusions}
\label{sec:conclusions}

We have presented an analysis of the cosmological implications of the
clustering wedges $\xi_{\perp}(s)$ and $\xi_{\parallel}(s)$, measured from the BOSS-DR9 CMASS sample.
These statistics are defined by averaging the full two-dimensional correlation function, $\xi(\mu,s)$,
over the ranges $0<\mu<0.5$ and $0.5<\mu<1$, respectively. We apply a simple model of the full
shape of these measurements in the mildly non-linear regime to derive constraints on the
parameter combinations $d_{\perp}(z)=D_{\rm A}(z)/r_{\rm s}(z_{\rm d})$ and
$d_{\parallel}(z)=cz/(H(z)r_{\rm s}(z_{\rm d}))$, breaking the degeneracy between
$D_{\rm A}(z)$ and $H(z)$ obtained when using angle-averaged clustering measurements.
We combine this information with additional cosmological probes,
including CMB, SN, and BAO measurements from other datasets, to derive constraints on
cosmological parameters.
Our analysis is an extension of that of \citet{Sanchez2012} based on the monopole of
the same galaxy sample analysed here. By comparing the constraints obtained from the clustering wedges with 
those derived by means of $\xi_0(s)$, we can quantify the impact of the extra
information provided by the anisotropic clustering measurements.

Although using different measurements and modelling details,
our results are in good agreement
with those inferred from the full shape of the 
CMASS monopole-quadrupole pair by \citet{Reid2012}, \citet{Samushia2013} and \citet{Chuang2013},
and from the ``consensus''  anisotropic BAO measurements by \citet{Abalone2013},
indicating the robustness of our results.

We find that the $\Lambda$CDM model provides an excellent description of the full shape of the 
CMASS clustering wedges. When restricting our analysis to this parameter space, the CMB+$\xi_0(s)$ and
CMB+$(\xi_{\perp}(s),\xi_{\parallel}(s))$ combinations give almost identical results. 
The consistency of the constraints obtained in these cases is a reassuring indication 
of the validity of the description of redshift-space distortions implemented in our
modelling of the clustering wedges.

We explored a number of possible extensions of the $\Lambda$CDM model.
In some of these cases, such as when constraining non-flat models or the massive neutrino fraction
while assuming that $w_{\rm DE}=-1$, %dark energy behaves as a cosmological constant,
the extra information in the clustering wedges only slightly improves the constraints with respect to those
obtained by means of $\xi_0(s)$.
However, this information proved to be most useful when allowing for deviations of the dark
energy equation of state from the canonical $\Lambda$CDM value.
For example, when assuming a time-independent dark energy equation of state, 
using the CMB+$\xi_0(s)$ combination, we find $w_{\rm DE}=-0.99_{-0.20}^{+0.21}$.
Instead, if the monopole is replaced by the clustering wedges, the allowed range for this parameter is 
reduced by a factor of two, leading to $w_{\rm DE}=-0.93\pm0.11$, consistent with a cosmological constant
at a one~$\sigma$ level. Including also the BAO and SN datasets leads to 
the marginalized constraint $w_{\rm DE}=-1.013\pm0.064$, in excellent agreement with the $\Lambda$CDM model.

When we explore the redshift dependence of the dark energy equation of state, 
parametrized as in equation~(\ref{eq:wa}), the CMB+$(\xi_{\perp}(s),\xi_{\parallel}(s))$ combination 
implies $w_0=-0.96_{-0.39}^{+0.40}$ and $w_a=0.03_{-0.97}^{+0.96}$, which can be translated into
the constraint $w_{\rm DE}(z_{\rm p})=-0.95\pm0.11$ at the pivot redshift $z_{\rm p}=0.69$.
These results represent an increase of the FoM by a factor of two with respect to the one found in 
the CMB+$\xi_0(s)$ case.
Our final limit on these parameters is obtained from our full dataset combination, 
which shows no evidence for a redshift evolution of $w_{\rm DE}$.
In this case we find $w_0=-1.10\pm0.12$ and $w_a=0.31\pm0.40$ and a constraint of 
$w_{\rm DE}(z_{\rm p})=-1.018\pm0.060$ at the pivot redshift $z_{\rm p}=0.35$.

As they are sensitive to the value of $f(z_{\rm m})$,
the clustering wedges offer the opportunity to constrain potential deviations from 
the predictions of general relativity. This is not possible for angle-averaged clustering measurements,
where $f(z_{\rm m})$ is completely degenerate with the bias factor.
If we extend the $\Lambda$CDM model to include $f(z_{\rm m})$ as a free parameter, we find
 $f(z_{\rm m})=0.719_{-0.096}^{+0.092}$. 
Assuming that this quantity behaves as $f(z)=\Omega_{\rm m}(z)^{\gamma}$ our results imply
a constraint of $\gamma=0.59\pm0.23$, consistent with no deviation from the GR prediction of $\gamma=0.55$.

The assumption that $f(z_{\rm m})$ follows the predictions of GR implies that 
the relative amplitude of the clustering wedges contains information on $\Omega_{\rm m}$.
When $f(z_{\rm m})$ is allowed to vary freely this additional constraining power is lost,
affecting the constraints on other parameters.
For example, if $f(z_{\rm m})$ and $w_{\rm DE}$ are varied simultaneously, 
the CMB+$(\xi_{\perp}(s),\xi_{\parallel}(s))$ combination implies that
$f(z_{\rm m})=0.76\pm0.14$ and  $w_{\rm DE}=-0.95\pm0.17$. 
However, when the additional BAO and SN measurements are included in the analysis 
we obtain $w_{\rm DE}=-1.035_{-0.069}^{+0.071}$ and $f(z_{\rm m})=0.706_{-0.099}^{+0.096}$,
with similar accuracies to the constraints derived when these parameters are varied independently and 
in good agreement with the standard $\Lambda$CDM+GR cosmological model.

The geometric constraints on $H(z)$ and $D_{\rm A}(z)$ from the clustering wedges are in perfect agreement
with the best-fitting $\Lambda$CDM model to the CMB data. This result shows the consistency between these datasets
and with the standard cosmological model. Assuming the predictions of GR, using the information of the full
shape of the clustering wedges we obtain
$D_{\rm A}(z_{\rm m})\left(r_{\rm s}(z_{\rm d})^{\rm fid}/r_{\rm s}(z_{\rm d})\right)= 1384\pm32\,{\rm Mpc}$ and
$H(z_{\rm m})\left(r_{\rm s}(z_{\rm d})/r_{\rm s}(z_{\rm d})^{\rm fid}\right)=92.0\pm3.3\,{\rm km}\,{\rm s}^{-1}{\rm Mpc}^{-1}$.
These values are in good agreement with the pre-reconstruction results of \citet{Kazin2013}, who used only the BAO signal in the same
measurements, of $D_{\rm A}(z_{\rm m})\left(r_{\rm s}(z_{\rm d})^{\rm fid}/r_{\rm s}(z_{\rm d})\right)= 1366\pm41\,{\rm Mpc}$
and $H(z_{\rm m})\left(r_{\rm s}(z_{\rm d})/r_{\rm s}(z_{\rm d})^{\rm fid}\right)=89.9\pm5.4\,{\rm km}\,{\rm s}^{-1}{\rm Mpc}^{-1}$.
The comparison of these results shows the effect of the extra information in the full shape of $\xi_{\perp}(s)$
and $\xi_{\parallel}(s)$ beyond that encoded in the scale of the BAO peak.

Our results illustrate the extra constraining power of anisotropic clustering measurements with 
respect to that of angle-averaged quantities. The large volume and high number density of the CMASS DR9 sample
make it possible to explore these measurements with a sufficiently high signal-to-noise ratio to derive meaningful
cosmological constraints. By probing larger volumes, the galaxy samples from subsequent SDSS data releases will
provide more accurate anisotropic clustering measurements. 
The availability of these new samples will be accompanied by the release of the CMB measurements from the Planck 
satellite. The combination of these datasets will undoubtedly push the achievable precision on our cosmological
constraints to a new level, allowing us to put the $\Lambda$CDM paradigm under even stricter scrutiny.

\section*{Acknowledgements}

AGS acknowledges support by the Trans-regional Collaborative Research 
Centre TR33 `The Dark Universe' of the German Research Foundation (DFG).
EK is supported by the Australian Research Council Centre of Excellence for
All-sky Astrophysics (CAASTRO), through project number CE110001020.

Numerical computations for the PTHalos mocks were done on the Sciama
High Performance Compute (HPC) cluster which is supported by the ICG,
SEPNet and the University of Portsmouth

Funding for SDSS-III has been provided by the Alfred P. Sloan Foundation, the Participating Institutions, 
the National Science Foundation, and the U.S. Department of Energy. 

SDSS-III is managed by the Astrophysical Research Consortium for the
Participating Institutions of the SDSS-III Collaboration including the
University of Arizona,
the Brazilian Participation Group,
Brookhaven National Laboratory,
University of Cambridge,
Carnegie Mellon University,
University of Florida,
the French Participation Group,
the German Participation Group,
Harvard University,
the Instituto de Astrofisica de Canarias,
the Michigan State/Notre Dame/JINA Participation Group,
Johns Hopkins University,
Lawrence Berkeley National Laboratory,
Max Planck Institute for Astrophysics,
Max Planck Institute for Extraterrestrial Physics,
New Mexico State University,
New York University,
Ohio State University,
Pennsylvania State University,
University of Portsmouth,
Princeton University,
the Spanish Participation Group,
University of Tokyo,
University of Utah,
Vanderbilt University,
University of Virginia,
University of Washington,
and Yale University.

We acknowledge the use of the Legacy Archive for Microwave Background 
Data Analysis (LAMBDA). Support for LAMBDA is provided by the NASA
Office of Space Science.

%\begin{thebibliography}{99}

%\bibitem[\protect\citeauthoryear{Abazajian et al.}{Abazajian et al.}{2009}]{Abazajian2009}
%Abazajian K., et al. 2009, ApJS, 182, 543

%\end{thebibliography}

%\bibliographystyle{mn2e}
%\bibliography{biblio}

\begin{thebibliography}{119}
\expandafter\ifx\csname natexlab\endcsname\relax\def\natexlab#1{#1}\fi

\bibitem[{{Ahn} {et~al}\mbox{.}(2012){Ahn}, {Alexandroff}, {Allende Prieto},
  {Anderson}, {Anderton}, {Andrews}, {Aubourg}, {Bailey}, {Balbinot}, {Barnes},
  {Bautista}, {Beers}, {Beifiori}, {Berlind}, {Bhardwaj}, {Bizyaev}, {Blake},
  {Blanton}, {Blomqvist}, {Bochanski}, {Bolton}, {Borde}, {Bovy}, {Brandt},
  {Brinkmann}, {Brown}, {Brownstein}, {Bundy}, {Busca}, {Carithers}, {Carnero},
  {Carr}, {Casetti-Dinescu}, {Chen}, {Chiappini}, {Comparat}, {Connolly},
  {Crepp}, {Cristiani}, {Croft}, {Cuesta}, {da Costa}, {Davenport}, {Dawson},
  {de Putter}, {De Lee}, {Delubac}, {Dhital}, {Ealet}, {Ebelke}, {Edmondson},
  {Eisenstein}, {Escoffier}, {Esposito}, {Evans}, {Fan}, {Femen{\'i}a
  Castell{\'a}}, {Fern{\'a}ndez Alvar}, {Ferreira}, {Filiz Ak}, {Finley},
  {Fleming}, {Font-Ribera}, {Frinchaboy}, {Garc{\'i}a-Hern{\'a}ndez},
  {Garc{\'i}a P{\'e}rez}, {Ge}, {G{\'e}nova-Santos}, {Gillespie}, {Girardi},
  {Gonz{\'a}lez Hern{\'a}ndez}, {Grebel}, {Gunn}, {Guo}, {Haggard}, {Hamilton},
  {Harris}, {Hawley}, {Hearty}, {Ho}, {Hogg}, {Holtzman}, {Honscheid},
  {Huehnerhoff}, {Ivans}, \v{Z}. {Ivezi{\'c}}, {Jacobson}, {Jiang},
  {Johansson}, {Johnson}, {Kauffmann}, {Kirkby}, {Kirkpatrick}, {Klaene},
  {Knapp}, {Kneib}, {Le Goff}, {Leauthaud}, {Lee}, {Lee}, {Long}, {Loomis},
  {Lucatello}, {Lundgren}, {Lupton}, {Ma}, {Ma}, {MacDonald}, {Mack},
  {Mahadevan}, {Maia}, {Majewski}, {Makler}, {Malanushenko}, {Malanushenko},
  {Manchado}, {Mandelbaum}, {Manera}, {Maraston}, {Margala}, {Martell},
  {McBride}, {McGreer}, {McMahon}, {M{\'e}nard}, {Meszaros},
  {Miralda-Escud{\'e}}, {Montero-Dorta}, {Montesano}, {Morrison}, {Muna},
  {Munn}, {Murayama}, {Myers}, {Neto}, {Nguyen}, {Nichol}, {Nidever},
  {Noterdaeme}, {Nuza}, {Ogando}, {Olmstead}, {Oravetz}, {Owen}, {Padmanabhan},
  {Palanque-Delabrouille}, {Pan}, {Parejko}, {Parihar}, {P{\^a}ris},
  {Pattarakijwanich}, {Pepper}, {Percival}, {P{\'e}rez-Fournon},
  {P{\'e}rez-R{\`a}fols}, {Petitjean}, {Pforr}, {Pieri}, {Pinsonneault}, {Porto
  de Mello}, {Prada}, {Price-Whelan}, {Raddick}, {Rebolo}, {Rich}, {Richards},
  {Robin}, {Rocha-Pinto}, {Rockosi}, {Roe}, {Ross}, {Ross}, {Rossi},
  {Rubi{\~n}o-Martin}, {Samushia}, {Sanchez Almeida}, {S{\'a}nchez},
  {Santiago}, {Sayres}, {Schlegel}, {Schlesinger}, {Schmidt}, {Schneider},
  {Schultheis}, {Schwope}, {Sc{\'o}ccola}, {Seljak}, {Sheldon}, {Shen}, {Shu},
  {Simmerer}, {Simmons}, {Skibba}, {Skrutskie}, {Slosar}, {Sobreira}, {Sobeck},
  {Stassun}, {Steele}, {Steinmetz}, {Strauss}, {Streblyanska}, {Suzuki},
  {Swanson}, {Tal}, {Thakar}, {Thomas}, {Thompson}, {Tinker}, {Tojeiro},
  {Tremonti}, {Vargas Maga{\~n}a}, {Verde}, {Viel}, {Vikas}, {Vogt}, {Wake},
  {Wang}, {Weaver}, {Weinberg}, {Weiner}, {West}, {White}, {Wilson},
  {Wisniewski}, {Wood-Vasey}, {Yanny}, {Y{\`e}che}, {York}, {Zamora},
  {Zasowski}, {Zehavi}, {Zhao}, {Zheng}, {Zhu}, \& {Zinn}}]{Ahn2012}
{Ahn} C.~P. {et~al.}, 2012, \apjs, 203, 21

\bibitem[{{Aihara} {et~al}\mbox{.}(2011){Aihara}, {Allende Prieto}, {An},
  {Anderson}, {Aubourg}, {Balbinot}, {Beers}, {Berlind}, {Bickerton},
  {Bizyaev}, {Blanton}, {Bochanski}, {Bolton}, {Bovy}, {Brandt}, {Brinkmann},
  {Brown}, {Brownstein}, {Busca}, {Campbell}, {Carr}, {Chen}, {Chiappini},
  {Comparat}, {Connolly}, {Cortes}, {Croft}, {Cuesta}, {da Costa}, {Davenport},
  {Dawson}, {Dhital}, {Ealet}, {Ebelke}, {Edmondson}, {Eisenstein},
  {Escoffier}, {Esposito}, {Evans}, {Fan}, {Femen{\'i}a Castell{\'a}},
  {Font-Ribera}, {Frinchaboy}, {Ge}, {Gillespie}, {Gilmore}, {Gonz{\'a}lez
  Hern{\'a}ndez}, {Gott}, {Gould}, {Grebel}, {Gunn}, {Hamilton}, {Harding},
  {Harris}, {Hawley}, {Hearty}, {Ho}, {Hogg}, {Holtzman}, {Honscheid}, {Inada},
  {Ivans}, {Jiang}, {Johnson}, {Jordan}, {Jordan}, {Kazin}, {Kirkby}, {Klaene},
  {Knapp}, {Kneib}, {Kochanek}, {Koesterke}, {Kollmeier}, {Kron}, {Lampeitl},
  {Lang}, {Le Goff}, {Lee}, {Lin}, {Long}, {Loomis}, {Lucatello}, {Lundgren},
  {Lupton}, {Ma}, {MacDonald}, {Mahadevan}, {Maia}, {Makler}, {Malanushenko},
  {Malanushenko}, {Mandelbaum}, {Maraston}, {Margala}, {Masters}, {McBride},
  {McGehee}, {McGreer}, {M{\'e}nard}, {Miralda-Escud{\'e}}, {Morrison},
  {Mullally}, {Muna}, {Munn}, {Murayama}, {Myers}, {Naugle}, {Neto}, {Nguyen},
  {Nichol}, {O'Connell}, {Ogando}, {Olmstead}, {Oravetz}, {Padmanabhan},
  {Palanque-Delabrouille}, {Pan}, {Pandey}, {P{\^a}ris}, {Percival},
  {Petitjean}, {Pfaffenberger}, {Pforr}, {Phleps}, {Pichon}, {Pieri}, {Prada},
  {Price-Whelan}, {Raddick}, {Ramos}, {Reyl{\'e}}, {Rich}, {Richards}, {Rix},
  {Robin}, {Rocha-Pinto}, {Rockosi}, {Roe}, {Rollinde}, {Ross}, {Ross},
  {Rossetto}, {S{\'a}nchez}, {Sayres}, {Schlegel}, {Schlesinger}, {Schmidt},
  {Schneider}, {Sheldon}, {Shu}, {Simmerer}, {Simmons}, {Sivarani}, {Snedden},
  {Sobeck}, {Steinmetz}, {Strauss}, {Szalay}, {Tanaka}, {Thakar}, {Thomas},
  {Tinker}, {Tofflemire}, {Tojeiro}, {Tremonti}, {Vandenberg}, {Vargas
  Maga{\~n}a}, {Verde}, {Vogt}, {Wake}, {Wang}, {Weaver}, {Weinberg}, {White},
  {White}, {Yanny}, {Yasuda}, {Yeche}, \& {Zehavi}}]{Aihara2011}
{Aihara} H. {et~al.}, 2011, \apjs, 193, 29

\bibitem[{{Albrecht} {et~al}\mbox{.}(2006){Albrecht}, {Bernstein}, {Cahn},
  {Freedman}, {Hewitt}, {Hu}, {Huth}, {Kamionkowski}, {Kolb}, {Knox}, {Mather},
  {Staggs}, \& {Suntzeff}}]{Albrecht2006}
{Albrecht} A. {et~al.}, 2006, ArXiv Astrophysics e-prints

\bibitem[{{Alcock} \& {Paczynski}(1979)}]{Alcock1979}
{Alcock} C., {Paczynski} B., 1979, \nat, 281, 358

\bibitem[{{Anderson} {et~al}\mbox{.}(2012){Anderson}, {Aubourg}, {Bailey},
  {Bizyaev}, {Blanton}, {Bolton}, {Brinkmann}, {Brownstein}, {Burden},
  {Cuesta}, {da Costa}, {Dawson}, {de Putter}, {Eisenstein}, {Gunn}, {Guo},
  {Hamilton}, {Harding}, {Ho}, {Honscheid}, {Kazin}, {Kirkby}, {Kneib},
  {Labatie}, {Loomis}, {Lupton}, {Malanushenko}, {Malanushenko}, {Mandelbaum},
  {Manera}, {Maraston}, {McBride}, {Mehta}, {Mena}, {Montesano}, {Muna},
  {Nichol}, {Nuza}, {Olmstead}, {Oravetz}, {Padmanabhan},
  {Palanque-Delabrouille}, {Pan}, {Parejko}, {P{\^a}ris}, {Percival},
  {Petitjean}, {Prada}, {Reid}, {Roe}, {Ross}, {Ross}, {Samushia},
  {S{\'a}nchez}, {Schlegel}, {Schneider}, {Sc{\'o}ccola}, {Seo}, {Sheldon},
  {Simmons}, {Skibba}, {Strauss}, {Swanson}, {Thomas}, {Tinker}, {Tojeiro},
  {Maga{\~n}a}, {Verde}, {Wagner}, {Wake}, {Weaver}, {Weinberg}, {White}, {Xu},
  {Y{\`e}che}, {Zehavi}, \& {Zhao}}]{Anderson2012}
{Anderson} L. {et~al.}, 2012, \mnras, 427, 3435


\bibitem[{{Anderson et al.}(2013)}]{Abalone2013}
{Anderson} L. {et~al.}, 2013, submitted



\bibitem[{{Anselmi} {et~al}\mbox{.}(2011){Anselmi}, {Matarrese}, \&
  {Pietroni}}]{Anselmi2011}
{Anselmi} S., {Matarrese} S., {Pietroni} M., 2011, \jcap, 6, 15

\bibitem[{{Anselmi} \& {Pietroni}(2012)}]{Anselmi2012}
{Anselmi} S., {Pietroni} M., 2012, \jcap, 12, 13

\bibitem[{{Bennett} {et~al}\mbox{.}(2012){Bennett}, {Larson}, {Weiland},
  {Jarosik}, {Hinshaw}, {Odegard}, {Smith}, {Hill}, {Gold}, {Halpern},
  {Komatsu}, {Nolta}, {Page}, {Spergel}, {Wollack}, {Dunkley}, {Kogut},
  {Limon}, {Meyer}, {Tucker}, \& {Wright}}]{Bennett2012}
{Bennett} C.~L. {et~al.}, 2012, ArXiv e-prints

\bibitem[{{Beutler} {et~al}\mbox{.}(2011){Beutler}, {Blake}, {Colless},
  {Jones}, {Staveley-Smith}, {Campbell}, {Parker}, {Saunders}, \&
  {Watson}}]{Beutler2011}
{Beutler} F. {et~al.}, 2011, \mnras, 416, 3017

\bibitem[{{Blake} \& {Glazebrook}(2003)}]{Blake2003}
{Blake} C., {Glazebrook} K., 2003, \apj, 594, 665

\bibitem[{{Blake} {et~al}\mbox{.}(2011){Blake}, {Kazin}, {Beutler}, {Davis},
  {Parkinson}, {Brough}, {Colless}, {Contreras}, {Couch}, {Croom}, {Croton},
  {Drinkwater}, {Forster}, {Gilbank}, {Gladders}, {Glazebrook}, {Jelliffe},
  {Jurek}, {Li}, {Madore}, {Martin}, {Pimbblet}, {Poole}, {Pracy}, {Sharp},
  {Wisnioski}, {Woods}, {Wyder}, \& {Yee}}]{Blake2011}
{Blake} C. {et~al.}, 2011, \mnras, 418, 1707

\bibitem[{{Bolton} {et~al}\mbox{.}(2012){Bolton}, {Schlegel}, {Aubourg},
  {Bailey}, {Bhardwaj}, {Brownstein}, {Burles}, {Chen}, {Dawson}, {Eisenstein},
  {Gunn}, {Knapp}, {Loomis}, {Lupton}, {Maraston}, {Muna}, {Myers}, {Olmstead},
  {Padmanabhan}, {P{\^a}ris}, {Percival}, {Petitjean}, {Rockosi}, {Ross},
  {Schneider}, {Shu}, {Strauss}, {Thomas}, {Tremonti}, {Wake}, {Weaver}, \&
  {Wood-Vasey}}]{Bolton2012}
{Bolton} A.~S. {et~al.}, 2012, \aj, 144, 144

\bibitem[{{Busca} {et~al}\mbox{.}(2012){Busca}, {Delubac}, {Rich}, {Bailey},
  {Font-Ribera}, {Kirkby}, {Le Goff}, {Pieri}, {Slosar}, {Aubourg}, {Bautista},
  {Bizyaev}, {Blomqvist}, {Bolton}, {Bovy}, {Brewington}, {Borde}, {Brinkmann},
  {Carithers}, {Croft}, {Dawson}, {Ebelke}, {Eisenstein}, {Hamilton}, {Ho},
  {Hogg}, {Honscheid}, {Lee}, {Lundgren}, {Malanushenko}, {Malanushenko},
  {Margala}, {Maraston}, {Mehta}, {Miralda-Escud{\'e}}, {Myers}, {Nichol},
  {Noterdaeme}, {Olmstead}, {Oravetz}, {Palanque-Delabrouille}, {Pan},
  {P{\^a}ris}, {Percival}, {Petitjean}, {Roe}, {Rollinde}, {Ross}, {Rossi},
  {Schlegel}, {Schneider}, {Shelden}, {Sheldon}, {Simmons}, {Snedden},
  {Tinker}, {Viel}, {Weaver}, {Weinberg}, {White}, {Y{\`e}che}, {York}, \&
  {Zhao}}]{Busca2012}
{Busca} N.~G. {et~al.}, 2012, ArXiv e-prints

\bibitem[{{Cabr{\'e}} \& {Gazta{\~n}aga}(2009)}]{Cabre2009}
{Cabr{\'e}} A., {Gazta{\~n}aga} E., 2009, \mnras, 393, 1183

\bibitem[{{Chevallier} \& {Polarski}(2001)}]{Chevallier2001}
{Chevallier} M., {Polarski} D., 2001, International Journal of Modern Physics
  D, 10, 213

\bibitem[{{Chuang} \& {Wang}(2012)}]{Chuang2012}
{Chuang} C.-H., {Wang} Y., 2012, \mnras, 426, 226

\bibitem[{{Chuang et al.}(2013)}]{Chuang2013}
{Chuang et al.}, 2013, submitted

\bibitem[{{Cole} {et~al}\mbox{.}(1995){Cole}, {Fisher}, \&
  {Weinberg}}]{Cole1995}
{Cole} S., {Fisher} K.~B., {Weinberg} D.~H., 1995, \mnras, 275, 515

\bibitem[{{Cole} {et~al}\mbox{.}(2005){Cole}, {Percival}, {Peacock}, {Norberg},
  {Baugh}, {Frenk}, {Baldry}, {Bland-Hawthorn}, {Bridges}, {Cannon}, {Colless},
  {Collins}, {Couch}, {Cross}, {Dalton}, {Eke}, {De Propris}, {Driver},
  {Efstathiou}, {Ellis}, {Glazebrook}, {Jackson}, {Jenkins}, {Lahav}, {Lewis},
  {Lumsden}, {Maddox}, {Madgwick}, {Peterson}, {Sutherland}, \&
  {Taylor}}]{Cole2005}
{Cole} S. {et~al.}, 2005, \mnras, 362, 505

\bibitem[{{Colless} {et~al}\mbox{.}(2001){Colless}, {Dalton}, {Maddox},
  {Sutherland}, {Norberg}, {Cole}, {Bland-Hawthorn}, {Bridges}, {Cannon},
  {Collins}, {Couch}, {Cross}, {Deeley}, {De Propris}, {Driver}, {Efstathiou},
  {Ellis}, {Frenk}, {Glazebrook}, {Jackson}, {Lahav}, {Lewis}, {Lumsden},
  {Madgwick}, {Peacock}, {Peterson}, {Price}, {Seaborne}, \&
  {Taylor}}]{Colless2001}
{Colless} M. {et~al.}, 2001, \mnras, 328, 1039

\bibitem[{{Colless} {et~al}\mbox{.}(2003){Colless}, {Peterson}, {Jackson},
  {Peacock}, {Cole}, {Norberg}, {Baldry}, {Baugh}, {Bland-Hawthorn}, {Bridges},
  {Cannon}, {Collins}, {Couch}, {Cross}, {Dalton}, {De Propris}, {Driver},
  {Efstathiou}, {Ellis}, {Frenk}, {Glazebrook}, {Lahav}, {Lewis}, {Lumsden},
  {Maddox}, {Madgwick}, {Sutherland}, \& {Taylor}}]{Colless2003}
{Colless} M. {et~al.}, 2003, ArXiv Astrophysics e-prints

\bibitem[{{Conley} {et~al}\mbox{.}(2011){Conley}, {Guy}, {Sullivan},
  {Regnault}, {Astier}, {Balland}, {Basa}, {Carlberg}, {Fouchez}, {Hardin},
  {Hook}, {Howell}, {Pain}, {Palanque-Delabrouille}, {Perrett}, {Pritchet},
  {Rich}, {Ruhlmann-Kleider}, {Balam}, {Baumont}, {Ellis}, {Fabbro},
  {Fakhouri}, {Fourmanoit}, {Gonz{\'a}lez-Gait{\'a}n}, {Graham}, {Hudson},
  {Hsiao}, {Kronborg}, {Lidman}, {Mourao}, {Neill}, {Perlmutter}, {Ripoche},
  {Suzuki}, \& {Walker}}]{Conley2011}
{Conley} A. {et~al.}, 2011, \apjs, 192, 1

\bibitem[{{Crocce} \& {Scoccimarro}(2006)}]{Crocce2006}
{Crocce} M., {Scoccimarro} R., 2006, \prd, 73, 063519

\bibitem[{{Crocce} \& {Scoccimarro}(2008)}]{Crocce2008}
{Crocce} M., {Scoccimarro} R., 2008, \prd, 77, 023533

\bibitem[{{Crocce} {et~al}\mbox{.}(2012){Crocce}, {Scoccimarro}, \&
  {Bernardeau}}]{Crocce2012}
{Crocce} M., {Scoccimarro} R., {Bernardeau} F., 2012, \mnras, 427, 2537

\bibitem[{{Das} {et~al}\mbox{.}(2011){Das}, {Sherwin}, {Aguirre}, {Appel},
  {Bond}, {Carvalho}, {Devlin}, {Dunkley}, {D{\"u}nner}, {Essinger-Hileman},
  {Fowler}, {Hajian}, {Halpern}, {Hasselfield}, {Hincks}, {Hlozek},
  {Huffenberger}, {Hughes}, {Irwin}, {Klein}, {Kosowsky}, {Lupton}, {Marriage},
  {Marsden}, {Menanteau}, {Moodley}, {Niemack}, {Nolta}, {Page}, {Parker},
  {Reese}, {Schmitt}, {Sehgal}, {Sievers}, {Spergel}, {Staggs}, {Swetz},
  {Switzer}, {Thornton}, {Visnjic}, \& {Wollack}}]{Das2011}
{Das} S. {et~al.}, 2011, Physical Review Letters, 107, 021301

\bibitem[{{Davis} \& {Peebles}(1983)}]{Davis1983}
{Davis} M., {Peebles} P.~J.~E., 1983, \apj, 267, 465

\bibitem[{{Dawson} {et~al}\mbox{.}(2013){Dawson}, {Schlegel}, {Ahn},
  {Anderson}, {Aubourg}, {Bailey}, {Barkhouser}, {Bautista}, {Beifiori},
  {Berlind}, {Bhardwaj}, {Bizyaev}, {Blake}, {Blanton}, {Blomqvist}, {Bolton},
  {Borde}, {Bovy}, {Brandt}, {Brewington}, {Brinkmann}, {Brown}, {Brownstein},
  {Bundy}, {Busca}, {Carithers}, {Carnero}, {Carr}, {Chen}, {Comparat},
  {Connolly}, {Cope}, {Croft}, {Cuesta}, {da Costa}, {Davenport}, {Delubac},
  {de Putter}, {Dhital}, {Ealet}, {Ebelke}, {Eisenstein}, {Escoffier}, {Fan},
  {Filiz Ak}, {Finley}, {Font-Ribera}, {G{\'e}nova-Santos}, {Gunn}, {Guo},
  {Haggard}, {Hall}, {Hamilton}, {Harris}, {Harris}, {Ho}, {Hogg}, {Holder},
  {Honscheid}, {Huehnerhoff}, {Jordan}, {Jordan}, {Kauffmann}, {Kazin},
  {Kirkby}, {Klaene}, {Kneib}, {Le Goff}, {Lee}, {Long}, {Loomis}, {Lundgren},
  {Lupton}, {Maia}, {Makler}, {Malanushenko}, {Malanushenko}, {Mandelbaum},
  {Manera}, {Maraston}, {Margala}, {Masters}, {McBride}, {McDonald}, {McGreer},
  {McMahon}, {Mena}, {Miralda-Escud{\'e}}, {Montero-Dorta}, {Montesano},
  {Muna}, {Myers}, {Naugle}, {Nichol}, {Noterdaeme}, {Nuza}, {Olmstead},
  {Oravetz}, {Oravetz}, {Owen}, {Padmanabhan}, {Palanque-Delabrouille}, {Pan},
  {Parejko}, {P{\^a}ris}, {Percival}, {P{\'e}rez-Fournon},
  {P{\'e}rez-R{\`a}fols}, {Petitjean}, {Pfaffenberger}, {Pforr}, {Pieri},
  {Prada}, {Price-Whelan}, {Raddick}, {Rebolo}, {Rich}, {Richards}, {Rockosi},
  {Roe}, {Ross}, {Ross}, {Rossi}, {Rubi{\~n}o-Martin}, {Samushia},
  {S{\'a}nchez}, {Sayres}, {Schmidt}, {Schneider}, {Sc{\'o}ccola}, {Seo},
  {Shelden}, {Sheldon}, {Shen}, {Shu}, {Slosar}, {Smee}, {Snedden}, {Stauffer},
  {Steele}, {Strauss}, {Streblyanska}, {Suzuki}, {Swanson}, {Tal}, {Tanaka},
  {Thomas}, {Tinker}, {Tojeiro}, {Tremonti}, {Vargas Maga{\~n}a}, {Verde},
  {Viel}, {Wake}, {Watson}, {Weaver}, {Weinberg}, {Weiner}, {West}, {White},
  {Wood-Vasey}, {Yeche}, {Zehavi}, {Zhao}, \& {Zheng}}]{Dawson2013}
{Dawson} K.~S. {et~al.}, 2013, \aj, 145, 10

\bibitem[{{de la Torre} \& {Guzzo}(2012)}]{delaTorre2012}
{de la Torre} S., {Guzzo} L., 2012, \mnras, 427, 327

\bibitem[{{Drinkwater} {et~al}\mbox{.}(2010){Drinkwater}, {Jurek}, {Blake},
  {Woods}, {Pimbblet}, {Glazebrook}, {Sharp}, {Pracy}, {Brough}, {Colless},
  {Couch}, {Croom}, {Davis}, {Forbes}, {Forster}, {Gilbank}, {Gladders},
  {Jelliffe}, {Jones}, {Li}, {Madore}, {Martin}, {Poole}, {Small}, {Wisnioski},
  {Wyder}, \& {Yee}}]{Drinkwater2010}
{Drinkwater} M.~J. {et~al.}, 2010, \mnras, 401, 1429

\bibitem[{{Efstathiou} \& {Bond}(1999)}]{Efstathiou1999}
{Efstathiou} G., {Bond} J.~R., 1999, \mnras, 304, 75

\bibitem[{{Eisenstein} \& {Hu}(1998)}]{Eisenstein1998}
{Eisenstein} D.~J., {Hu} W., 1998, \apj, 496, 605

\bibitem[{{Eisenstein} {et~al}\mbox{.}(2005){Eisenstein}, {Zehavi}, {Hogg},
  {Scoccimarro}, {Blanton}, {Nichol}, {Scranton}, {Seo}, {Tegmark}, {Zheng},
  {Anderson}, {Annis}, {Bahcall}, {Brinkmann}, {Burles}, {Castander},
  {Connolly}, {Csabai}, {Doi}, {Fukugita}, {Frieman}, {Glazebrook}, {Gunn},
  {Hendry}, {Hennessy}, {Ivezi{\'c}}, {Kent}, {Knapp}, {Lin}, {Loh}, {Lupton},
  {Margon}, {McKay}, {Meiksin}, {Munn}, {Pope}, {Richmond}, {Schlegel},
  {Schneider}, {Shimasaku}, {Stoughton}, {Strauss}, {SubbaRao}, {Szalay},
  {Szapudi}, {Tucker}, {Yanny}, \& {York}}]{Eisenstein2005}
{Eisenstein} D.~J. {et~al.}, 2005, \apj, 633, 560


\bibitem[{{Eisenstein} {et~al}\mbox{.}(2007){Eisenstein}, {Seo}, {Sirko}, \&
  {Spergel}}]{Eisenstein2007b}
{Eisenstein} D.~J., {Seo} H.-J., {Sirko} E., {Spergel} D.~N., 2007, \apj, 664,
  675

\bibitem[{{Eisenstein} {et~al}\mbox{.}(2011){Eisenstein}, {Weinberg}, {Agol},
  {Aihara}, {Allende Prieto}, {Anderson}, {Arns}, {Aubourg}, {Bailey},
  {Balbinot}, {Barkhouser}, {Beers}, {Berlind}, {Bickerton}, {Bizyaev},
  {Blanton}, {Bochanski}, {Bolton}, {Bosman}, {Bovy}, {Brandt}, {Breslauer},
  {Brewington}, {Brinkmann}, {Brown}, {Brownstein}, {Burger}, {Busca},
  {Campbell}, {Cargile}, {Carithers}, {Carlberg}, {Carr}, {Chang}, {Chen},
  {Chiappini}, {Comparat}, {Connolly}, {Cortes}, {Croft}, {Cunha}, {da Costa},
  {Davenport}, {Dawson}, {De Lee}, {Porto de Mello}, {de Simoni}, {Dean},
  {Dhital}, {Ealet}, {Ebelke}, {Edmondson}, {Eiting}, {Escoffier}, {Esposito},
  {Evans}, {Fan}, {Femen{\'i}a Castell{\'a}}, {Dutra Ferreira}, {Fitzgerald},
  {Fleming}, {Font-Ribera}, {Ford}, {Frinchaboy}, {Elia Garc{\'i}a P{\'e}rez},
  {Gaudi}, {Ge}, {Ghezzi}, {Gillespie}, {Gilmore}, {Girardi}, {Gott}, {Gould},
  {Grebel}, {Gunn}, {Hamilton}, {Harding}, {Harris}, {Hawley}, {Hearty},
  {Hennawi}, {Gonz{\'a}lez Hern{\'a}ndez}, {Ho}, {Hogg}, {Holtzman},
  {Honscheid}, {Inada}, {Ivans}, {Jiang}, {Jiang}, {Johnson}, {Jordan},
  {Jordan}, {Kauffmann}, {Kazin}, {Kirkby}, {Klaene}, {Knapp}, {Kneib},
  {Kochanek}, {Koesterke}, {Kollmeier}, {Kron}, {Lampeitl}, {Lang}, {Lawler},
  {Le Goff}, {Lee}, {Lee}, {Leisenring}, {Lin}, {Liu}, {Long}, {Loomis},
  {Lucatello}, {Lundgren}, {Lupton}, {Ma}, {Ma}, {MacDonald}, {Mack},
  {Mahadevan}, {Maia}, {Majewski}, {Makler}, {Malanushenko}, {Malanushenko},
  {Mandelbaum}, {Maraston}, {Margala}, {Maseman}, {Masters}, {McBride},
  {McDonald}, {McGreer}, {McMahon}, {Mena Requejo}, {M{\'e}nard},
  {Miralda-Escud{\'e}}, {Morrison}, {Mullally}, {Muna}, {Murayama}, {Myers},
  {Naugle}, {Neto}, {Nguyen}, {Nichol}, {Nidever}, {O'Connell}, {Ogando},
  {Olmstead}, {Oravetz}, {Padmanabhan}, {Paegert}, {Palanque-Delabrouille},
  {Pan}, {Pandey}, {Parejko}, {P{\^a}ris}, {Pellegrini}, {Pepper}, {Percival},
  {Petitjean}, {Pfaffenberger}, {Pforr}, {Phleps}, {Pichon}, {Pieri}, {Prada},
  {Price-Whelan}, {Raddick}, {Ramos}, {Reid}, {Reyle}, {Rich}, {Richards},
  {Rieke}, {Rieke}, {Rix}, {Robin}, {Rocha-Pinto}, {Rockosi}, {Roe},
  {Rollinde}, {Ross}, {Ross}, {Rossetto}, {S{\'a}nchez}, {Santiago}, {Sayres},
  {Schiavon}, {Schlegel}, {Schlesinger}, {Schmidt}, {Schneider}, {Sellgren},
  {Shelden}, {Sheldon}, {Shetrone}, {Shu}, {Silverman}, {Simmerer}, {Simmons},
  {Sivarani}, {Skrutskie}, {Slosar}, {Smee}, {Smith}, {Snedden}, {Stassun},
  {Steele}, {Steinmetz}, {Stockett}, {Stollberg}, {Strauss}, {Szalay},
  {Tanaka}, {Thakar}, {Thomas}, {Tinker}, {Tofflemire}, {Tojeiro}, {Tremonti},
  {Vargas Maga{\~n}a}, {Verde}, {Vogt}, {Wake}, {Wan}, {Wang}, {Weaver},
  {White}, {White}, {Wilson}, {Wisniewski}, {Wood-Vasey}, {Yanny}, {Yasuda},
  {Y{\`e}che}, {York}, {Young}, {Zasowski}, {Zehavi}, \&
  {Zhao}}]{Eisenstein2011}
{Eisenstein} D.~J. {et~al.}, 2011, \aj, 142, 72


\bibitem[{{Fang} {et~al}\mbox{.}(2008){Fang}, {Hu}, \& {Lewis}}]{Fang2008}
{Fang} W., {Hu} W., {Lewis} A., 2008, \prd, 78, 087303

\bibitem[{{Frieman} {et~al}\mbox{.}(2008){Frieman}, {Turner}, \&
  {Huterer}}]{Frieman2008}
{Frieman} J.~A., {Turner} M.~S., {Huterer} D., 2008, \araa, 46, 385


\bibitem[{{Gazta{\~n}aga} {et~al}\mbox{.}(2009{\natexlab{a}}){Gazta{\~n}aga},
  {Cabr{\'e}}, {Castander}, {Crocce}, \& {Fosalba}}]{Gaztanaga2009b}
{Gazta{\~n}aga} E., {Cabr{\'e}} A., {Castander} F., {Crocce} M., {Fosalba} P.,
  2009{\natexlab{a}}, \mnras, 399, 801

\bibitem[{{Gazta{\~n}aga} {et~al}\mbox{.}(2009{\natexlab{b}}){Gazta{\~n}aga},
  {Miquel}, \& {S{\'a}nchez}}]{Gaztanaga2009a}
{Gazta{\~n}aga} E., {Miquel} R., {S{\'a}nchez} E., 2009{\natexlab{b}}, Physical
  Review Letters, 103, 091302


\bibitem[{{Gunn} {et~al}\mbox{.}(1998){Gunn}, {Carr}, {Rockosi}, {Sekiguchi},
  {Berry}, {Elms}, {de Haas}, \v{Z}. {Ivezi{\'c}}, {Knapp}, {Lupton}, {Pauls},
  {Simcoe}, {Hirsch}, {Sanford}, {Wang}, {York}, {Harris}, {Annis}, {Bartozek},
  {Boroski}, {Bakken}, {Haldeman}, {Kent}, {Holm}, {Holmgren}, {Petravick},
  {Prosapio}, {Rechenmacher}, {Doi}, {Fukugita}, {Shimasaku}, {Okada}, {Hull},
  {Siegmund}, {Mannery}, {Blouke}, {Heidtman}, {Schneider}, {Lucinio}, \&
  {Brinkman}}]{Gunn1998}
{Gunn} J.~E. {et~al.}, 1998, \aj, 116, 3040

\bibitem[{{Gunn} {et~al}\mbox{.}(2006){Gunn}, {Siegmund}, {Mannery}, {Owen},
  {Hull}, {Leger}, {Carey}, {Knapp}, {York}, {Boroski}, {Kent}, {Lupton},
  {Rockosi}, {Evans}, {Waddell}, {Anderson}, {Annis}, {Barentine}, {Bartoszek},
  {Bastian}, {Bracker}, {Brewington}, {Briegel}, {Brinkmann}, {Brown}, {Carr},
  {Czarapata}, {Drennan}, {Dombeck}, {Federwitz}, {Gillespie}, {Gonzales},
  {Hansen}, {Harvanek}, {Hayes}, {Jordan}, {Kinney}, {Klaene}, {Kleinman},
  {Kron}, {Kresinski}, {Lee}, {Limmongkol}, {Lindenmeyer}, {Long}, {Loomis},
  {McGehee}, {Mantsch}, {Neilsen}, {Neswold}, {Newman}, {Nitta}, {Peoples},
  {Pier}, {Prieto}, {Prosapio}, {Rivetta}, {Schneider}, {Snedden}, \&
  i.~{Wang}}]{Gunn2006}
{Gunn} J.~E. {et~al.}, 2006, \aj, 131, 2332

\bibitem[{{Hamilton}(1997)}]{Hamilton1997}
{Hamilton} A.~J.~S., 1997, \mnras, 289, 295

\bibitem[{{Hartlap} {et~al}\mbox{.}(2007){Hartlap}, {Simon}, \&
  {Schneider}}]{Hartlap2007}
{Hartlap} J., {Simon} P., {Schneider} P., 2007, \aap, 464, 399

\bibitem[{{Hinshaw} {et~al}\mbox{.}(2012){Hinshaw}, {Larson}, {Komatsu},
  {Spergel}, {Bennett}, {Dunkley}, {Nolta}, {Halpern}, {Hill}, {Odegard},
  {Page}, {Smith}, {Weiland}, {Gold}, {Jarosik}, {Kogut}, {Limon}, {Meyer},
  {Tucker}, {Wollack}, \& {Wright}}]{Hinshaw2012}
{Hinshaw} G. {et~al.}, 2012, ArXiv e-prints

\bibitem[{{Hu} \& {Haiman}(2003)}]{Hu2003}
{Hu} W., {Haiman} Z., 2003, \prd, 68, 063004

\bibitem[{{H{\"u}tsi}(2010)}]{Hutsi2010}
{H{\"u}tsi} G., 2010, \mnras, 401, 2477

\bibitem[{{Jennings} {et~al}\mbox{.}(2011){Jennings}, {Baugh}, \&
  {Pascoli}}]{Jennings2011}
{Jennings} E., {Baugh} C.~M., {Pascoli} S., 2011, \mnras, 410, 2081

\bibitem[{{Jones} {et~al}\mbox{.}(2009){Jones}, {Read}, {Saunders}, {Colless},
  {Jarrett}, {Parker}, {Fairall}, {Mauch}, {Sadler}, {Watson}, {Burton},
  {Campbell}, {Cass}, {Croom}, {Dawe}, {Fiegert}, {Frankcombe}, {Hartley},
  {Huchra}, {James}, {Kirby}, {Lahav}, {Lucey}, {Mamon}, {Moore}, {Peterson},
  {Prior}, {Proust}, {Russell}, {Safouris}, {Wakamatsu}, {Westra}, \&
  {Williams}}]{Jones2009}
{Jones} D.~H. {et~al.}, 2009, \mnras, 399, 683

\bibitem[{{Kaiser}(1987)}]{Kaiser1987}
{Kaiser} N., 1987, \mnras, 227, 1

\bibitem[{{Kazin} {et~al}\mbox{.}(2010){Kazin}, {Blanton}, {Scoccimarro},
  {McBride}, {Berlind}, {Bahcall}, {Brinkmann}, {Czarapata}, {Frieman}, {Kent},
  {Schneider}, \& {Szalay}}]{Kazin2010}
{Kazin} E.~A. {et~al.}, 2010, \apj, 710, 1444

\bibitem[{{Kazin} {et~al}\mbox{.}(2012){Kazin}, {S{\'a}nchez}, \&
  {Blanton}}]{Kazin2012}
{Kazin} E.~A., {S{\'a}nchez} A.~G., {Blanton} M.~R., 2012, \mnras, 419, 3223

\bibitem[{{Kazin et al.}(2013)}]{Kazin2013}
{Kazin et al.}, 2013, submitted

\bibitem[{{Keisler} {et~al}\mbox{.}(2011){Keisler}, {Reichardt}, {Aird},
  {Benson}, {Bleem}, {Carlstrom}, {Chang}, {Cho}, {Crawford}, {Crites}, {de
  Haan}, {Dobbs}, {Dudley}, {George}, {Halverson}, {Holder}, {Holzapfel},
  {Hoover}, {Hou}, {Hrubes}, {Joy}, {Knox}, {Lee}, {Leitch}, {Lueker},
  {Luong-Van}, {McMahon}, {Mehl}, {Meyer}, {Millea}, {Mohr}, {Montroy},
  {Natoli}, {Padin}, {Plagge}, {Pryke}, {Ruhl}, {Schaffer}, {Shaw},
  {Shirokoff}, {Spieler}, {Staniszewski}, {Stark}, {Story}, {van Engelen},
  {Vanderlinde}, {Vieira}, {Williamson}, \& {Zahn}}]{Keisler2011}
{Keisler} R. {et~al.}, 2011, \apj, 743, 28

\bibitem[{{Komatsu} {et~al}\mbox{.}(2009){Komatsu}, {Dunkley}, {Nolta},
  {Bennett}, {Gold}, {Hinshaw}, {Jarosik}, {Larson}, {Limon}, {Page},
  {Spergel}, {Halpern}, {Hill}, {Kogut}, {Meyer}, {Tucker}, {Weiland},
  {Wollack}, \& {Wright}}]{Komatsu2009}
{Komatsu} E. {et~al.}, 2009, \apjs, 180, 330

\bibitem[{{Komatsu} {et~al}\mbox{.}(2011){Komatsu}, {Smith}, {Dunkley},
  {Bennett}, {Gold}, {Hinshaw}, {Jarosik}, {Larson}, {Nolta}, {Page},
  {Spergel}, {Halpern}, {Hill}, {Kogut}, {Limon}, {Meyer}, {Odegard}, {Tucker},
  {Weiland}, {Wollack}, \& {Wright}}]{Komatsu2011}
{Komatsu} E. {et~al.}, 2011, \apjs, 192, 18

\bibitem[{{Landy} \& {Szalay}(1993)}]{Landy1993}
{Landy} S.~D., {Szalay} A.~S., 1993, \apj, 412, 64

\bibitem[{{Larson} {et~al}\mbox{.}(2011){Larson}, {Dunkley}, {Hinshaw},
  {Komatsu}, {Nolta}, {Bennett}, {Gold}, {Halpern}, {Hill}, {Jarosik}, {Kogut},
  {Limon}, {Meyer}, {Odegard}, {Page}, {Smith}, {Spergel}, {Tucker}, {Weiland},
  {Wollack}, \& {Wright}}]{Larson2011}
{Larson} D. {et~al.}, 2011, \apjs, 192, 16

\bibitem[{{Lewis} \& {Bridle}(2002)}]{Lewis2002}
{Lewis} A., {Bridle} S., 2002, \prd, 66, 103511

\bibitem[{{Lewis} {et~al}\mbox{.}(2000){Lewis}, {Challinor}, \&
  {Lasenby}}]{Lewis2000}
{Lewis} A., {Challinor} A., {Lasenby} A., 2000, \apj, 538, 473

\bibitem[{{Linder}(2003)}]{Linder2003}
{Linder} E.~V., 2003, Physical Review Letters, 90, 091301

\bibitem[{{Linder} \& {Cahn}(2007)}]{Linder2007}
{Linder} E.~V., {Cahn} R.~N., 2007, Astroparticle Physics, 28, 481

\bibitem[{{Maddox} {et~al}\mbox{.}(1990){Maddox}, {Efstathiou}, {Sutherland},
  \& {Loveday}}]{Maddox1990}
{Maddox} S.~J., {Efstathiou} G., {Sutherland} W.~J., {Loveday} J., 1990,
  \mnras, 242, 43P

\bibitem[{{Manera} {et~al}\mbox{.}(2012){Manera}, {Scoccimarro}, {Percival},
  {Samushia}, {McBride}, {Ross}, {Sheth}, {White}, {Reid}, {S{\'a}nchez}, {de
  Putter}, {Xu}, {Berlind}, {Brinkmann}, {Nichol}, {Montesano}, {Padmanabhan},
  {Skibba}, {Tojeiro}, \& {Weaver}}]{Manera2012}
{Manera} M. {et~al.}, 2012, ArXiv e-prints

\bibitem[{{Maraston} {et~al}\mbox{.}(2012){Maraston}, {Pforr}, {Henriques},
  {Thomas}, {Wake}, {Brownstein}, {Capozzi}, {Bundy}, {Skibba}, {Beifiori},
  {Nichol}, {Edmondson}, {Schneider}, {Chen}, {Masters}, {Steele}, {Bolton},
  {York}, {Bizyaev}, {Brewington}, {Malanushenko}, {Malanushenko}, {Snedden},
  {Oravetz}, {Pan}, {Shelden}, \& {Simmons}}]{Maraston2012}
{Maraston} C. {et~al.}, 2012, ArXiv e-prints

\bibitem[{{Masters} {et~al}\mbox{.}(2011){Masters}, {Maraston}, {Nichol},
  {Thomas}, {Beifiori}, {Bundy}, {Edmondson}, {Higgs}, {Leauthaud},
  {Mandelbaum}, {Pforr}, {Ross}, {Ross}, {Schneider}, {Skibba}, {Tinker},
  {Tojeiro}, {Wake}, {Brinkmann}, \& {Weaver}}]{Masters2011}
{Masters} K.~L. {et~al.}, 2011, \mnras, 418, 1055

\bibitem[{{Matarrese} \& {Pietroni}(2007)}]{Matarrese2007}
{Matarrese} S., {Pietroni} M., 2007, \jcap, 6, 26

\bibitem[{{Matarrese} \& {Pietroni}(2008)}]{Matarrese2008}
{Matarrese} S., {Pietroni} M., 2008, Modern Physics Letters A, 23, 25

\bibitem[{{Matsubara}(2004)}]{Matsubara2004}
{Matsubara} T., 2004, \apj, 615, 573

\bibitem[{{Matsubara}(2008{\natexlab{a}})}]{Matsubara2008b}
{Matsubara} T., 2008{\natexlab{a}}, \prd, 78, 083519

\bibitem[{{Matsubara}(2008{\natexlab{b}})}]{Matsubara2008a}
{Matsubara} T., 2008{\natexlab{b}}, \prd, 77, 063530

\bibitem[{{Meiksin} {et~al}\mbox{.}(1999){Meiksin}, {White}, \&
  {Peacock}}]{Meiksin1999}
{Meiksin} A., {White} M., {Peacock} J.~A., 1999, \mnras, 304, 851

\bibitem[{{Montesano} {et~al}\mbox{.}(2010){Montesano}, {S{\'a}nchez}, \&
  {Phleps}}]{Montesano2010}
{Montesano} F., {S{\'a}nchez} A.~G., {Phleps} S., 2010, \mnras, 408, 2397

\bibitem[{{Montesano} {et~al}\mbox{.}(2012){Montesano}, {S{\'a}nchez}, \&
  {Phleps}}]{Montesano2012}
{Montesano} F., {S{\'a}nchez} A.~G., {Phleps} S., 2012, \mnras, 421, 2656

\bibitem[{{Nuza} {et~al}\mbox{.}(2012){Nuza}, {Sanchez}, {Prada}, {Klypin},
  {Schlegel}, {Gottloeber}, {Montero-Dorta}, {Manera}, {McBride}, {Ross},
  {Angulo}, {Blanton}, {Bolton}, {Favole}, {Samushia}, {Montesano}, {Percival},
  {Padmanabhan}, {Steinmetz}, {Tinker}, {Skibba}, {Schneider}, {Guo}, {Zehavi},
  {Zheng}, {Bizyaev}, {Malanushenko}, {Malanushenko}, {Oravetz}, {Oravetz}, \&
  {Shelden}}]{Nuza2012}
{Nuza} S.~E. {et~al.}, 2012, ArXiv e-prints

\bibitem[{{Okumura} {et~al}\mbox{.}(2008){Okumura}, {Matsubara}, {Eisenstein},
  {Kayo}, {Hikage}, {Szalay}, \& {Schneider}}]{Okumura2008}
{Okumura} T., {Matsubara} T., {Eisenstein} D.~J., {Kayo} I., {Hikage} C.,
  {Szalay} A.~S., {Schneider} D.~P., 2008, \apj, 676, 889

\bibitem[{{Okumura} {et~al}\mbox{.}(2012){Okumura}, {Seljak}, \&
  {Desjacques}}]{Okumura2012}
{Okumura} T., {Seljak} U., {Desjacques} V., 2012, \jcap, 11, 14

\bibitem[{{Padmanabhan} {et~al}\mbox{.}(2007){Padmanabhan}, {Schlegel},
  {Seljak}, {Makarov}, {Bahcall}, {Blanton}, {Brinkmann}, {Eisenstein},
  {Finkbeiner}, {Gunn}, {Hogg}, \v{Z}. {Ivezi{\'c}}, {Knapp}, {Loveday},
  {Lupton}, {Nichol}, {Schneider}, {Strauss}, {Tegmark}, \&
  {York}}]{Padmanabhan2007}
{Padmanabhan} N. {et~al.}, 2007, \mnras, 378, 852

\bibitem[{{Padmanabhan} \& {White}(2008)}]{Padmanabhan2008}
{Padmanabhan} N., {White} M., 2008, \prd, 77, 123540

\bibitem[{{Padmanabhan} {et~al}\mbox{.}(2012){Padmanabhan}, {Xu}, {Eisenstein},
  {Scalzo}, {Cuesta}, {Mehta}, \& {Kazin}}]{Padmanabhan2012}
{Padmanabhan} N., {Xu} X., {Eisenstein} D.~J., {Scalzo} R., {Cuesta} A.~J.,
  {Mehta} K.~T., {Kazin} E., 2012, \mnras, 427, 2132

\bibitem[{{Park} {et~al}\mbox{.}(1994){Park}, {Vogeley}, {Geller}, \&
  {Huchra}}]{Park1994}
{Park} C., {Vogeley} M.~S., {Geller} M.~J., {Huchra} J.~P., 1994, \apj, 431,
  569

\bibitem[{{Parkinson} {et~al}\mbox{.}(2012){Parkinson}, {Riemer-S{\o}rensen},
  {Blake}, {Poole}, {Davis}, {Brough}, {Colless}, {Contreras}, {Couch},
  {Croom}, {Croton}, {Drinkwater}, {Forster}, {Gilbank}, {Gladders},
  {Glazebrook}, {Jelliffe}, {Jurek}, h.~{Li}, {Madore}, {Martin}, {Pimbblet},
  {Pracy}, {Sharp}, {Wisnioski}, {Woods}, {Wyder}, \& {Yee}}]{Parkinson2012}
{Parkinson} D. {et~al.}, 2012, \prd, 86, 103518

\bibitem[{{Peebles} \& {Ratra}(2003)}]{Peebles2003}
{Peebles} P.~J., {Ratra} B., 2003, Reviews of Modern Physics, 75, 559

\bibitem[{{Percival} {et~al}\mbox{.}(2001){Percival}, {Baugh},
  {Bland-Hawthorn}, {Bridges}, {Cannon}, {Cole}, {Colless}, {Collins}, {Couch},
  {Dalton}, {De Propris}, {Driver}, {Efstathiou}, {Ellis}, {Frenk},
  {Glazebrook}, {Jackson}, {Lahav}, {Lewis}, {Lumsden}, {Maddox}, {Moody},
  {Norberg}, {Peacock}, {Peterson}, {Sutherland}, \& {Taylor}}]{Percival2001}
{Percival} W.~J. {et~al.}, 2001, \mnras, 327, 1297

\bibitem[{{Percival} {et~al}\mbox{.}(2007){Percival}, {Cole}, {Eisenstein},
  {Nichol}, {Peacock}, {Pope}, \& {Szalay}}]{Percival2007}
{Percival} W.~J., {Cole} S., {Eisenstein} D.~J., {Nichol} R.~C., {Peacock}
  J.~A., {Pope} A.~C., {Szalay} A.~S., 2007, \mnras, 381, 1053

\bibitem[{{Percival} {et~al}\mbox{.}(2010){Percival}, {Reid}, {Eisenstein},
  {Bahcall}, {Budavari}, {Frieman}, {Fukugita}, {Gunn}, \v{Z}. {Ivezi{\'c}},
  {Knapp}, {Kron}, {Loveday}, {Lupton}, {McKay}, {Meiksin}, {Nichol}, {Pope},
  {Schlegel}, {Schneider}, {Spergel}, {Stoughton}, {Strauss}, {Szalay},
  {Tegmark}, {Vogeley}, {Weinberg}, {York}, \& {Zehavi}}]{Percival2010}
{Percival} W.~J. {et~al.}, 2010, \mnras, 401, 2148

\bibitem[{{Percival} {et~al}\mbox{.}(2002){Percival}, {Sutherland}, {Peacock},
  {Baugh}, {Bland-Hawthorn}, {Bridges}, {Cannon}, {Cole}, {Colless}, {Collins},
  {Couch}, {Dalton}, {De Propris}, {Driver}, {Efstathiou}, {Ellis}, {Frenk},
  {Glazebrook}, {Jackson}, {Lahav}, {Lewis}, {Lumsden}, {Maddox}, {Moody},
  {Norberg}, {Peterson}, \& {Taylor}}]{Percival2002}
{Percival} W.~J. {et~al.}, 2002, \mnras, 337, 1068

\bibitem[{{Pietroni}(2008)}]{Pietroni2008}
{Pietroni} M., 2008, \jcap, 10, 36

\bibitem[{{Reid} {et~al}\mbox{.}(2010){Reid}, {Percival}, {Eisenstein},
  {Verde}, {Spergel}, {Skibba}, {Bahcall}, {Budavari}, {Frieman}, {Fukugita},
  {Gott}, {Gunn}, \v{Z}. {Ivezi{\'c}}, {Knapp}, {Kron}, {Lupton}, {McKay},
  {Meiksin}, {Nichol}, {Pope}, {Schlegel}, {Schneider}, {Stoughton}, {Strauss},
  {Szalay}, {Tegmark}, {Vogeley}, {Weinberg}, {York}, \& {Zehavi}}]{Reid2010}
{Reid} B.~A. {et~al.}, 2010, \mnras, 404, 60

\bibitem[{{Reid} {et~al}\mbox{.}(2012){Reid}, {Samushia}, {White}, {Percival},
  {Manera}, {Padmanabhan}, {Ross}, {S{\'a}nchez}, {Bailey}, {Bizyaev},
  {Bolton}, {Brewington}, {Brinkmann}, {Brownstein}, {Cuesta}, {Eisenstein},
  {Gunn}, {Honscheid}, {Malanushenko}, {Malanushenko}, {Maraston}, {McBride},
  {Muna}, {Nichol}, {Oravetz}, {Pan}, {de Putter}, {Roe}, {Ross}, {Schlegel},
  {Schneider}, {Seo}, {Shelden}, {Sheldon}, {Simmons}, {Skibba}, {Snedden},
  {Swanson}, {Thomas}, {Tinker}, {Tojeiro}, {Verde}, {Wake}, {Weaver},
  {Weinberg}, {Zehavi}, \& {Zhao}}]{Reid2012}
{Reid} B.~A. {et~al.}, 2012, \mnras, 426, 2719

\bibitem[{{Reid} \& {White}(2011)}]{Reid2011}
{Reid} B.~A., {White} M., 2011, \mnras, 417, 1913

\bibitem[{{Ross} {et~al}\mbox{.}(2013){Ross}, {Percival}, {Carnero}, b.~{Zhao},
  {Manera}, {Raccanelli}, {Aubourg}, {Bizyaev}, {Brewington}, {Brinkmann},
  {Brownstein}, {Cuesta}, {da Costa}, {Eisenstein}, {Ebelke}, {Guo},
  {Hamilton}, {Maga{\~n}a}, {Malanushenko}, {Malanushenko}, {Maraston},
  {Montesano}, {Nichol}, {Oravetz}, {Pan}, {Prada}, {S{\'a}nchez}, {Samushia},
  {Schlegel}, {Schneider}, {Seo}, {Sheldon}, {Simmons}, {Snedden}, {Swanson},
  {Thomas}, {Tinker}, {Tojeiro}, \& {Zehavi}}]{Ross2013}
{Ross} A.~J. {et~al.}, 2013, \mnras, 428, 1116

\bibitem[{{Ross} {et~al}\mbox{.}(2012){Ross}, {Percival}, {S{\'a}nchez},
  {Samushia}, {Ho}, {Kazin}, {Manera}, {Reid}, {White}, {Tojeiro}, {McBride},
  {Xu}, {Wake}, {Strauss}, {Montesano}, {Swanson}, {Bailey}, {Bolton}, {Dorta},
  {Eisenstein}, {Guo}, {Hamilton}, {Nichol}, {Padmanabhan}, {Prada},
  {Schlegel}, {Maga{\~n}a}, {Zehavi}, {Blanton}, {Bizyaev}, {Brewington},
  {Cuesta}, {Malanushenko}, {Malanushenko}, {Oravetz}, {Parejko}, {Pan},
  {Schneider}, {Shelden}, {Simmons}, {Snedden}, \& b.~{Zhao}}]{Ross2012}
{Ross} A.~J. {et~al.}, 2012, \mnras, 424, 564

\bibitem[{{Samushia} {et~al}\mbox{.}(2013){Samushia}, {Reid}, {White},
  {Percival}, {Cuesta}, {Lombriser}, {Manera}, {Nichol}, {Schneider},
  {Bizyaev}, {Brewington}, {Malanushenko}, {Malanushenko}, {Oravetz}, {Pan},
  {Simmons}, {Shelden}, {Snedden}, {Tinker}, {Weaver}, {York}, \&
  {Zhao}}]{Samushia2013}
{Samushia} L. {et~al.}, 2013, \mnras, 429, 1514

\bibitem[{{S{\'a}nchez} {et~al}\mbox{.}(2008){S{\'a}nchez}, {Baugh}, \&
  {Angulo}}]{Sanchez2008}
{S{\'a}nchez} A.~G., {Baugh} C.~M., {Angulo} R.~E., 2008, \mnras, 390, 1470

\bibitem[{{S{\'a}nchez} {et~al}\mbox{.}(2006){S{\'a}nchez}, {Baugh},
  {Percival}, {Peacock}, {Padilla}, {Cole}, {Frenk}, \&
  {Norberg}}]{Sanchez2006}
{S{\'a}nchez} A.~G., {Baugh} C.~M., {Percival} W.~J., {Peacock} J.~A.,
  {Padilla} N.~D., {Cole} S., {Frenk} C.~S., {Norberg} P., 2006, \mnras, 366,
  189

\bibitem[{{S{\'a}nchez} {et~al}\mbox{.}(2009){S{\'a}nchez}, {Crocce},
  {Cabr{\'e}}, {Baugh}, \& {Gazta{\~n}aga}}]{Sanchez2009}
{S{\'a}nchez} A.~G., {Crocce} M., {Cabr{\'e}} A., {Baugh} C.~M.,
  {Gazta{\~n}aga} E., 2009, \mnras, 400, 1643

\bibitem[{{S{\'a}nchez} {et~al}\mbox{.}(2012){S{\'a}nchez}, {Sc{\'o}ccola},
  {Ross}, {Percival}, {Manera}, {Montesano}, {Mazzalay}, {Cuesta},
  {Eisenstein}, {Kazin}, {McBride}, {Mehta}, {Montero-Dorta}, {Padmanabhan},
  {Prada}, {Rubi{\~n}o-Mart{\'i}n}, {Tojeiro}, {Xu}, {Maga{\~n}a}, {Aubourg},
  {Bahcall}, {Bailey}, {Bizyaev}, {Bolton}, {Brewington}, {Brinkmann},
  {Brownstein}, {Gott}, {Hamilton}, {Ho}, {Honscheid}, {Labatie},
  {Malanushenko}, {Malanushenko}, {Maraston}, {Muna}, {Nichol}, {Oravetz},
  {Pan}, {Ross}, {Roe}, {Reid}, {Schlegel}, {Shelden}, {Schneider}, {Simmons},
  {Skibba}, {Snedden}, {Thomas}, {Tinker}, {Wake}, {Weaver}, {Weinberg},
  {White}, {Zehavi}, \& {Zhao}}]{Sanchez2012}
{S{\'a}nchez} A.~G. {et~al.}, 2012, \mnras, 425, 415

\bibitem[{{Scoccimarro}(2004)}]{Scoccimarro2004}
{Scoccimarro} R., 2004, \prd, 70, 083007

\bibitem[{{Scoccimarro} \& {Sheth}(2002)}]{Scoccimarro2002}
{Scoccimarro} R., {Sheth} R.~K., 2002, \mnras, 329, 629

\bibitem[{{Seo} {et~al}\mbox{.}(2012){Seo}, {Ho}, {White}, {Cuesta}, {Ross},
  {Saito}, {Reid}, {Padmanabhan}, {Percival}, {de Putter}, {Schlegel},
  {Eisenstein}, {Xu}, {Schneider}, {Skibba}, {Verde}, {Nichol}, {Bizyaev},
  {Brewington}, {Brinkmann}, {Nicolaci da Costa}, {Gott}, {Malanushenko},
  {Malanushenko}, {Oravetz}, {Palanque-Delabrouille}, {Pan}, {Prada}, {Ross},
  {Simmons}, {de Simoni}, {Shelden}, {Snedden}, \& {Zehavi}}]{Seo2012}
{Seo} H.-J. {et~al.}, 2012, \apj, 761, 13

\bibitem[{{Shoji} {et~al}\mbox{.}(2009){Shoji}, {Jeong}, \&
  {Komatsu}}]{Shoji2009}
{Shoji} M., {Jeong} D., {Komatsu} E., 2009, \apj, 693, 1404

\bibitem[{{Smee} {et~al}\mbox{.}(2012){Smee}, {Gunn}, {Uomoto}, {Roe},
  {Schlegel}, {Rockosi}, {Carr}, {Leger}, {Dawson}, {Olmstead}, {Brinkmann},
  {Owen}, {Barkhouser}, {Honscheid}, {Harding}, {Long}, {Lupton}, {Loomis},
  {Anderson}, {Annis}, {Bernardi}, {Bhardwaj}, {Bizyaev}, {Bolton},
  {Brewington}, {Briggs}, {Burles}, {Burns}, {Castander}, {Connolly},
  {Davenport}, {Ebelke}, {Epps}, {Feldman}, {Friedman}, {Frieman}, {Heckman},
  {Hull}, {Knapp}, {Lawrence}, {Loveday}, {Mannery}, {Malanushenko},
  {Malanushenko}, {Merrelli}, {Muna}, {Newman}, {Nichol}, {Oravetz}, {Pan},
  {Pope}, {Ricketts}, {Shelden}, {Sandford}, {Siegmund}, {Simmons}, {Smith},
  {Snedden}, {Schneider}, {Strauss}, {SubbaRao}, {Tremonti}, {Waddell}, \&
  {York}}]{Smee2012}
{Smee} S. {et~al.}, 2012, ArXiv e-prints

\bibitem[{{Smith} {et~al}\mbox{.}(2008){Smith}, {Scoccimarro}, \&
  {Sheth}}]{Smith2008}
{Smith} R.~E., {Scoccimarro} R., {Sheth} R.~K., 2008, \prd, 77, 043525

\bibitem[{{Spergel} {et~al}\mbox{.}(2007){Spergel}, {Bean}, {Dor{\'e}},
  {Nolta}, {Bennett}, {Dunkley}, {Hinshaw}, {Jarosik}, {Komatsu}, {Page},
  {Peiris}, {Verde}, {Halpern}, {Hill}, {Kogut}, {Limon}, {Meyer}, {Odegard},
  {Tucker}, {Weiland}, {Wollack}, \& {Wright}}]{Spergel2007}
{Spergel} D.~N. {et~al.}, 2007, \apjs, 170, 377

\bibitem[{{Taruya} \& {Hiramatsu}(2008)}]{Taruya2008}
{Taruya} A., {Hiramatsu} T., 2008, \apj, 674, 617

\bibitem[{{Taruya} {et~al}\mbox{.}(2013){Taruya}, {Nishimichi}, \&
  {Bernardeau}}]{Taruya2013}
{Taruya} A., {Nishimichi} T., {Bernardeau} F., 2013, ArXiv e-prints

\bibitem[{{Taruya} {et~al}\mbox{.}(2010){Taruya}, {Nishimichi}, \&
  {Saito}}]{Taruya2010}
{Taruya} A., {Nishimichi} T., {Saito} S., 2010, \prd, 82, 063522

\bibitem[{{Tegmark} {et~al}\mbox{.}(2004){Tegmark}, {Blanton}, {Strauss},
  {Hoyle}, {Schlegel}, {Scoccimarro}, {Vogeley}, {Weinberg}, {Zehavi},
  {Berlind}, {Budavari}, {Connolly}, {Eisenstein}, {Finkbeiner}, {Frieman},
  {Gunn}, {Hamilton}, {Hui}, {Jain}, {Johnston}, {Kent}, {Lin}, {Nakajima},
  {Nichol}, {Ostriker}, {Pope}, {Scranton}, {Seljak}, {Sheth}, {Stebbins},
  {Szalay}, {Szapudi}, {Verde}, {Xu}, {Annis}, {Bahcall}, {Brinkmann},
  {Burles}, {Castander}, {Csabai}, {Loveday}, {Doi}, {Fukugita}, {Gott},
  {Hennessy}, {Hogg}, \v{Z}. {Ivezi{\'c}}, {Knapp}, {Lamb}, {Lee}, {Lupton},
  {McKay}, {Kunszt}, {Munn}, {O'Connell}, {Peoples}, {Pier}, {Richmond},
  {Rockosi}, {Schneider}, {Stoughton}, {Tucker}, {Vanden Berk}, {Yanny},
  {York}, \& {SDSS Collaboration}}]{Tegmark2004}
{Tegmark} M. {et~al.}, 2004, \apj, 606, 702

\bibitem[{{Tinker}(2007)}]{Tinker2007}
{Tinker} J.~L., 2007, \mnras, 374, 477

\bibitem[{{Tinker} {et~al}\mbox{.}(2006){Tinker}, {Weinberg}, \&
  {Zheng}}]{Tinker2006}
{Tinker} J.~L., {Weinberg} D.~H., {Zheng} Z., 2006, \mnras, 368, 85

\bibitem[{{Tojeiro} {et~al}\mbox{.}(2012){Tojeiro}, {Percival}, {Brinkmann},
  {Brownstein}, {Eisenstein}, {Manera}, {Maraston}, {McBride}, {Muna}, {Reid},
  {Ross}, {Ross}, {Samushia}, {Padmanabhan}, {Schneider}, {Skibba},
  {S{\'a}nchez}, {Swanson}, {Thomas}, {Tinker}, {Verde}, {Wake}, {Weaver}, \&
  {Zhao}}]{Tojeiro2012}
{Tojeiro} R. {et~al.}, 2012, \mnras, 424, 2339

\bibitem[{{Wagner} {et~al}\mbox{.}(2008){Wagner}, {M{\"u}ller}, \&
  {Steinmetz}}]{Wagner2008}
{Wagner} C., {M{\"u}ller} V., {Steinmetz} M., 2008, \aap, 487, 63

\bibitem[{{Wang} \& {Szalay}(2012)}]{Wang2012}
{Wang} X., {Szalay} A., 2012, \prd, 86, 043508

\bibitem[{{Wang}(2008)}]{Wang2008}
{Wang} Y., 2008, \prd, 77, 123525

\bibitem[{{White} {et~al}\mbox{.}(2011){White}, {Blanton}, {Bolton},
  {Schlegel}, {Tinker}, {Berlind}, {da Costa}, {Kazin}, {Lin}, {Maia},
  {McBride}, {Padmanabhan}, {Parejko}, {Percival}, {Prada}, {Ramos}, {Sheldon},
  {de Simoni}, {Skibba}, {Thomas}, {Wake}, {Zehavi}, {Zheng}, {Nichol},
  {Schneider}, {Strauss}, {Weaver}, \& {Weinberg}}]{White2011}
{White} M. {et~al.}, 2011, \apj, 728, 126

\bibitem[{{Xu} {et~al}\mbox{.}(2012){Xu}, {Padmanabhan}, {Eisenstein}, {Mehta},
  \& {Cuesta}}]{Xu2012}
{Xu} X., {Padmanabhan} N., {Eisenstein} D.~J., {Mehta} K.~T., {Cuesta} A.~J.,
  2012, \mnras, 427, 2146

\bibitem[{{York} {et~al}\mbox{.}(2000){York}, {Adelman}, {Anderson},
  {Anderson}, {Annis}, {Bahcall}, {Bakken}, {Barkhouser}, {Bastian}, {Berman},
  {Boroski}, {Bracker}, {Briegel}, {Briggs}, {Brinkmann}, {Brunner}, {Burles},
  {Carey}, {Carr}, {Castander}, {Chen}, {Colestock}, {Connolly}, {Crocker},
  {Csabai}, {Czarapata}, {Davis}, {Doi}, {Dombeck}, {Eisenstein}, {Ellman},
  {Elms}, {Evans}, {Fan}, {Federwitz}, {Fiscelli}, {Friedman}, {Frieman},
  {Fukugita}, {Gillespie}, {Gunn}, {Gurbani}, {de Haas}, {Haldeman}, {Harris},
  {Hayes}, {Heckman}, {Hennessy}, {Hindsley}, {Holm}, {Holmgren}, h.~{Huang},
  {Hull}, {Husby}, {Ichikawa}, {Ichikawa}, \v{Z}. {Ivezi{\'c}}, {Kent}, {Kim},
  {Kinney}, {Klaene}, {Kleinman}, {Kleinman}, {Knapp}, {Korienek}, {Kron},
  {Kunszt}, {Lamb}, {Lee}, {Leger}, {Limmongkol}, {Lindenmeyer}, {Long},
  {Loomis}, {Loveday}, {Lucinio}, {Lupton}, {MacKinnon}, {Mannery}, {Mantsch},
  {Margon}, {McGehee}, {McKay}, {Meiksin}, {Merelli}, {Monet}, {Munn},
  {Narayanan}, {Nash}, {Neilsen}, {Neswold}, {Newberg}, {Nichol}, {Nicinski},
  {Nonino}, {Okada}, {Okamura}, {Ostriker}, {Owen}, {Pauls}, {Peoples},
  {Peterson}, {Petravick}, {Pier}, {Pope}, {Pordes}, {Prosapio},
  {Rechenmacher}, {Quinn}, {Richards}, {Richmond}, {Rivetta}, {Rockosi},
  {Ruthmansdorfer}, {Sandford}, {Schlegel}, {Schneider}, {Sekiguchi}, {Sergey},
  {Shimasaku}, {Siegmund}, {Smee}, {Smith}, {Snedden}, {Stone}, {Stoughton},
  {Strauss}, {Stubbs}, {SubbaRao}, {Szalay}, {Szapudi}, {Szokoly}, {Thakar},
  {Tremonti}, {Tucker}, {Uomoto}, {Vanden Berk}, {Vogeley}, {Waddell},
  i.~{Wang}, {Watanabe}, {Weinberg}, {Yanny}, {Yasuda}, \& {SDSS
  Collaboration}}]{York2000}
{York} D.~G. {et~al.}, 2000, \aj, 120, 1579

\bibitem[{{Zhao} {et~al}\mbox{.}(2012){Zhao}, {Saito}, {Percival}, {Ross},
  {Montesano}, {Viel}, {Schneider}, {Ernst}, {Manera}, {Miralda-Escude},
  {Ross}, {Samushia}, {Sanchez}, {Swanson}, {Thomas}, {Tojeiro}, {Yeche}, \&
  {York}}]{Zhao2012}
{Zhao} G.-B. {et~al.}, 2012, ArXiv e-prints

\end{thebibliography}

\appendix

\section{Summary of the obtained constraints on extensions of the $\Lambda$CDM model} 
\label{sec:tables}

In this appendix we summarize the constraints on cosmological parameters obtained using different combinations of the datasets described in Section~\ref{sec:data}.
Tables~\ref{tab:omk_full}-\ref{tab:fg_full} list the 68\% confidence limits obtained in the extensions of the $\Lambda$CDM parameter space analysed in
Sections~\ref{sec:omk} to \ref{sec:fg}. The upper section of these tables lists the constraints on the main parameters (equation \ref{eq:pmain}) included in
the fits, while the lower section contains the results on the parameters derived from this set (equation \ref{eq:paramder}).

\begin{table*} 
\centering
  \caption{
    The marginalized 68\% constraints on the cosmological parameters of the $\Lambda$CDM model extended by including non-flat models,
    obtained using different combinations of the datasets described in Section~\ref{sec:data}.}
    \begin{tabular}{@{}lcccc@{}}
    \hline
& \multirow{2}{*}{CMB}  & \multirow{2}{*}{CMB+$\xi_0(s)$} & \multirow{2}{*}{CMB+$(\xi_{\perp}(s),\xi_{\parallel}(s))$}& CMB+$(\xi_{\perp}(s),\xi_{\parallel}(s))$  \\
&                       &                           &      &    +BAO+SN         \\  
\hline
$\Omega_k$                &  $-1.118\pm0.021$   &  $-0.0033_{-0.0044}^{+0.0046}$   &  $-0.0040\pm0.0045$   &  $-0.0041\pm0.0039$ \\[0.4mm]
$100\,\Theta$           & $1.0412\pm0.0015$   &  $1.0413\pm0.0016$   &  $1.0413\pm0.0015$   &  $1.0415\pm0.0014$ \\[0.4mm] 
$100\,\omega_{\rm b}$   & $2.230\pm0.038$   &  $2.229_{-0.038}^{+0.036}$   &  $2.235\pm0.037$   &  $2.241\pm0.037$ \\[0.4mm] 
$100\,\omega_{\rm c}$   & $11.40\pm0.41$    &  $11.36\pm0.39$   &  $11.31\pm0.36$   &  $11.23\pm0.35$\\[0.4mm] 
$n_{\rm s}$             & $0.964584\pm0.010$   &  $ 0.9653_{-0.0095}^{+0.0099}$   &  $  0.9661_{-0.0097}^{+0.0099}$   &  $0.9683_{-0.0096}^{+ 0.0097}$  \\[0.4mm] 
$\ln(10^{10}A_{\rm s})$ & $3.113\pm0.027$   &  $3.111_{-0.028}^{+0.027}$   &  $3.108\pm0.026$   &  $3.106\pm0.026$ \\ 
\hline
$\Omega_{\rm DE}$       &$0.690\pm0.072$   &  $0.715\pm0.0145$   &  $0.715\pm0.015$   &  $0.721\pm0.011$  \\[0.4mm]
$\Omega_{\rm m}$        & $0.321\pm0.093$   &  $0.288_{-0.015}^{+0.016}$   &  $0.288\pm-0.016$   &  $0.283\pm0.010$  \\[0.4mm]
$\sigma_{8}$            & $0.815\pm0.024$   &  $0.819_{-0.021}^{+0.020}$   &  $0.816\pm0.020$   &  $0.814\pm0.020$   \\[0.4mm]
$t_{0}/{\rm Gyr}$       & $14.08_{-0.98}^{+0.99}$   &  $13.90\pm0.20$   &  $13.92\pm0.20$   &  $13.91\pm0.17$  \\[0.4mm] 
$h$                     & $0.674_{-0.098}^{+0.097}$   &  $   0.687\pm0.016$   &  $0.686\pm0.017$   &  $0.689\pm0.011$  \\[0.4mm]
$f(z_{\rm m})$          & $0.778\pm0.075$   &  $0.764\pm0.013$   &  $0.764\pm0.013$   &  $0.7600_{-0.0085}^{+0.0084}$  \\
\hline
\end{tabular}
\label{tab:omk_full}
\end{table*}

\begin{table*} 
\centering
  \caption{
    The marginalized 68\% constraints on the cosmological parameters of the $\Lambda$CDM model extended by allowing for massive neutrinos,
    obtained using different combinations of the datasets described in Section~\ref{sec:data}.}
    \begin{tabular}{@{}lcccc@{}}
    \hline
& \multirow{2}{*}{CMB}  & \multirow{2}{*}{CMB+$\xi_0(s)$} & \multirow{2}{*}{CMB+$(\xi_{\perp}(s),\xi_{\parallel}(s))$}& CMB+$(\xi_{\perp}(s),\xi_{\parallel}(s))$  \\
&                       &                           &      &    +BAO+SN         \\  
\hline
$f_\nu$  &  $< 0.11 $ (95\% CL) & $< 0.055$ (95\% CL)  & $ <0.049$ (95\% CL)  & $ < 0.050 $ (95\% CL) \\[0.4mm] 
$100\,\Theta$           & $1.0406\pm0.0015$   & $1.0409\pm0.0013$  &  $1.0411\pm0.0015$  &  $1.0412\pm0.0014$ \\[0.4mm] 
$100\,\omega_{\rm b}$   & $2.191\pm0.042$   & $2.222_{-0.036}^{+0.037}$  &  $2.226\pm0.035$  &  $2.232\pm0.035$     \\[0.4mm] 
$100\,\omega_{\rm c}$   & $12.48_{-0.73}^{+0.71}$   & $11.64_{-0.32}^{+0.0035}$  &  $ 11.64\pm0.027$  &  $11.48\pm0.020$   \\[0.4mm] 
$n_{\rm s}$             & $0.953\pm0.013$   & $0.9641_{-0.0098}^{+0.0095}$  &  $0.965676_{-0.0088}^{+0.0089}$  &  $0.9675_{-0.0087}^{+0.0085}$     \\[0.4mm] 
$\ln(10^{10}A_{\rm s})$ & $3.122\pm0.027$   & $ 3.112_{-0.024}^{+0.023}$  &  $3.111\pm0.025$  &  $3.109\pm0.026$ \\ 
\hline
$\sum m_\nu$  &  $ < 1.4\,{\rm eV} $ (95\% CL) & $ < 0.61\,{\rm eV} $ (95\% CL)  & $ < 0.52\,{\rm eV} $ (95\% CL)  & $ < 0.51\,{\rm eV}$ (95\% CL) \\[0.4mm] 
$\Omega_{\rm DE}$       & $0.615_{-0.069}^{+0.072}$   & $0.698_{-0.021}^{+0.020}$  &  $0.698\pm0.018$ &  $0.709\pm0.012$   \\[0.4mm]
$\Omega_{\rm m}$        & $0.385_{-0.072}^{+0.069}$   & $0.302_{-0.020}^{+0.021}$  &  $0.302\pm0.018$  &  $0.291\pm0.012$    \\[0.4mm]
$\sigma_{8}$            & $0.673_{-0.073}^{+0.082}$   & $0.761\pm0.047$  &  $0.758_{-0.042}^{+0.043}$  &  $0.766\pm0.039$     \\[0.4mm]
$t_{0}/{\rm Gyr}$       & $14.17_{-0.25}^{+0.23}$   & $13.88\pm0.11$  &  $13.88\pm0.10$  &  $13.840_{-0.082}^{+0.081}$    \\[0.4mm] 
$h$                     & $0.623_{-0.042}^{+0.046}$   & $0.678\pm0.017$  &  $0.0678\pm0.015$  &  $0.687\pm0.010$   \\[0.4mm]
$f(z_{\rm m})$          & $0.823_{-0.042}^{+0.039}$   & $0.772_{-0.015}^{+0.016}$  &  $0.773_{-0.014}^{+0.014}$  &  $0.7644_{-0.0093}^{+0.0094}$    \\
\hline
\end{tabular}
\label{tab:fnu_full}
\end{table*}

\begin{table*} 
\centering
  \caption{
    The marginalized 68\% constraints on the cosmological parameters of the $\Lambda$CDM model extended by including $w_{\rm DE}$
    (assumed constant) as an additional parameter, obtained using different combinations of the datasets described in Section~\ref{sec:data}.}
    \begin{tabular}{@{}lcccc@{}}
    \hline
& \multirow{2}{*}{CMB}  & \multirow{2}{*}{CMB+$\xi_0(s)$} & \multirow{2}{*}{CMB+$(\xi_{\perp}(s),\xi_{\parallel}(s))$}& CMB+$(\xi_{\perp}(s),\xi_{\parallel}(s))$  \\
&                       &                           &      &    +BAO+SN         \\  
\hline
$w_{\rm DE}$            &$-1.14\pm0.42$  &  $-0.99_{-0.20}^{+0.21}$  &  $-0.93\pm0.11$  &  $-1.013\pm0.064$ \\[0.4mm]
$100\,\Theta$           & $1.0412\pm0.0014$  &  $1.0410\pm0.0015$  &  $1.0411\pm0.0014$  &  $1.0407\pm0.0014$ \\[0.4mm] 
$100\,\omega_{\rm b}$   & $2.229\pm0.038$  &  $2.227\pm0.038$  &  $2.230\pm0.036$  &  $2.223\pm0.035$       \\[0.4mm] 
$100\,\omega_{\rm c}$   & $11.43\pm0.41$  &  $11.49\pm0.042$ &  $11.42\pm0.32$  &  $11.55\pm0.28$  \\[0.4mm] 
$n_{\rm s}$             & $0.965\pm0.010$  &  $0.964\pm0.011$  &  $0.9656\pm0.0095$  &  $0.9626\pm0.0092$     \\[0.4mm] 
$\ln(10^{10}A_{\rm s})$ & $3.116\pm0.027$  &  $3.116\pm0.026$  &  $3.112\pm0.025$  &  $3.114\pm0.024$   \\ 
\hline
$\Omega_{\rm DE}$       & $0.74\pm0.10$  &  $0.709\pm0.042$  &  $0.701\pm0.028$  &  $0.717\pm0.012$    \\[0.4mm]
$\Omega_{\rm m}$        & $0.26\pm0.10$  &  $0.291\pm0.042$  &  $0.299\pm0.028$  &  $0.283\pm0.012$   \\[0.4mm]
$\sigma_{8}$            & $0.86_{-0.14}^{+0.13}$  &  $0.822_{-0.080}^{+0.078}$  &  $0.801_{-0.045}^{+0.043}$  &  $0.831\pm0.030$      \\[0.4mm]
$t_{0}/{\rm Gyr}$       & $13.70_{-0.25}^{+0.27}$  &  $13.78_{-0.13}^{+0.12}$  &  $13.794\pm0.091$  &  $13.755_{-0.064}^{+0.063}$    \\[0.4mm] 
$h$                     & $0.75\pm0.15$  &  $0.69_{-0.057}^{+0.055}$& $0.678_{-0.033}^{+0.031}$& $0.698\pm0.015$ \\ [0.4mm]
$f(z_{\rm m})$          & $0.764\pm0.023$  &  $0.761\pm0.021$  &  $0.754\pm0.015$  &  $0.761\pm0.013$    \\
\hline
\end{tabular}
\label{tab:wde_full}
\end{table*}

\begin{table*} 
\centering
  \caption{
    The marginalized 68\% constraints on the cosmological parameters of the $\Lambda$CDM model extended by allowing for simultaneous
    variations on $w_{\rm DE}$ and $\Omega_k$, obtained using different combinations of the datasets described in Section~\ref{sec:data}.}
    \begin{tabular}{@{}lcccc@{}}
    \hline
& \multirow{2}{*}{CMB}  & \multirow{2}{*}{CMB+$\xi_0(s)$} & \multirow{2}{*}{CMB+$(\xi_{\perp}(s),\xi_{\parallel}(s))$}& CMB+$(\xi_{\perp}(s),\xi_{\parallel}(s))$  \\
&                       &                           &      &    +BAO+SN         \\  
\hline
$w_{\rm DE}$            & $-0.89_{-0.45}^{+0.44}$ &   $-0.96_{-0.28}^{+29}$ &  $-0.97\pm0.16$  &   $-1.042\pm0.068$ \\[0.4mm]
$\Omega_k$             & $-0.022_{-0.031}^{+0.027}$ &   $0.0012_{-0.0077}^{+0.0091}$ &  $-0.0023_{-0.0060}^{+0.0061}$  &   $-0.0047\pm0.0042$ \\[0.4mm]
$100\,\Theta$           & $1.0412\pm0.0015$ &   $1.0413\pm0.0014$ &  $1.0413\pm0.0014$  &   $1.0413\pm0.0014$   \\[0.4mm] 
$100\,\omega_{\rm b}$   & $2.227\pm0.040$ &   $2.235_{-0.038}^{+0.040}$ &  $2.235\pm0.037$  &   $0.0224\pm0.037$       \\[0.4mm] 
$100\,\omega_{\rm c}$   & $11.41\pm0.41$  &   $11.39\pm0.38$            &  $11.34\pm0.37$   &   $11.27\pm0.37$  \\[0.4mm] 
$n_{\rm s}$             & $0.964\pm0.011$ &   $0.967\pm0.010$ &  $0.9667\pm0.0097$  &   $0.9667\pm0.0097$     \\[0.4mm] 
$\ln(10^{10}A_{\rm s})$ & $3.112\pm0.027$ &   $3.114\pm0.026$ &  $3.110\pm0.026$  &   $3.108\pm0.026$   \\ 
\hline
$\Omega_{\rm DE}$       & $0.61_{-0.18}^{+0.17}$ &   $0.699\pm0.058$ &  $0.707\pm0.035$  &   $0.725\pm0.013$  \\[0.4mm]
$\Omega_{\rm m}$        & $0.41_{-0.19}^{+0.21}$ &   $0.299\pm0.052$ &  $0.295\pm0.032$  &   $0.280\pm0.012$   \\[0.4mm]
$\sigma_{8}$            & $0.78_{-0.13}^{+0.14}$ &   $0.804_{-0.096}^{+0.091}$ &  $0.809\pm0.050$  &   $0.8279\pm0.029$   \\[0.4mm]
$t_{0}/{\rm Gyr}$       & $14.5\pm1.1$ &   $13.82_{-0.23}^{+0.22}$ &  $13.89_{-0.23}^{+0.22}$  &   $13.95\pm0.18$  \\[0.4mm] 
$h$                     & $0.63_{-0.16}^{+0.17}$ &   $0.684_{-0.064}^{+0.061}$ &  $0.681\pm0.036$  &   $0.695\pm0.014$  \\[0.4mm]
$f(z_{\rm m})$          & $0.808_{-0.072}^{+0.073}$ &   $0.759\pm0.033$ &  $0.762\pm0.021$  &   $0.768\pm0.014$   \\
\hline
\end{tabular}
\label{tab:wok_full}
\end{table*}

\begin{table*} 
\centering
  \caption{
    The marginalized 68\% constraints on the cosmological parameters of the $\Lambda$CDM model extended by allowing for
    variations on $w_{\rm DE}(a)$ (parametrized according to equation~\ref{eq:wa}),
    obtained using different combinations of the datasets described in Section~\ref{sec:data}.}
    \begin{tabular}{@{}lcccc@{}}
    \hline
& \multirow{2}{*}{CMB}  & \multirow{2}{*}{CMB+$\xi_0(s)$} & \multirow{2}{*}{CMB+$(\xi_{\perp}(s),\xi_{\parallel}(s))$}& CMB+$(\xi_{\perp}(s),\xi_{\parallel}(s))$  \\
&                       &                           &      &    +BAO+SN         \\  
\hline
$w_0$            &$-1.08_{-0.53}^{+0.55}$ & $-1.12_{-0.59}^{+0.62}$ & $-0.96_{-0.39}^{+0.40}$  & $-1.10_{-0.12}^{+0.12}$ \\[0.4mm]
$w_a$            & $-0.4_{-1.1}^{+1.1}$ & $0.2\pm1.0$ & $0.03_{-0.97}^{+0.96}$  & $0.31\pm0.40$ \\[0.4mm]
$100\,\Theta$           & $1.0411\pm0.0015$ & $1.0411\pm0.0015$ & $1.0412\pm0.0014$  & $1.0410\pm0.0014$   \\[0.4mm] 
$100\,\omega_{\rm b}$   & $2.229\pm0.037$ & $2.227\pm0.037$ &  $2.229\pm0.035$ & $2.229\pm0.035$       \\[0.4mm] 
$100\,\omega_{\rm c}$   & $11.4_{-0.40}^{+0.42}$ & $11.50\pm0.40$ &  $11.43\pm0.032$ & $11.42_{-0.34}^{+0.33}$   \\[0.4mm] 
$n_{\rm s}$             & $0.964\pm0.010$ & $0.964\pm0.010$ & $0.966_{-0.0093}^{+0.0095}$  & $0.9652_{-0.0097}^{+0.0096}$     \\[0.4mm] 
$\ln(10^{10}A_{\rm s})$ & $3.116\pm0.027$ & $3.117\pm0.027$ & $3.114\pm0.025$  & $3.111\pm0.025$   \\ 
\hline
$\Omega_{\rm DE}$       & $0.735_{-0.097}^{+0.094}$ & $0.719_{-0.093}^{+0.083}$ & $0.703\pm0.046$  & $0.722\pm0.013$    \\[0.4mm]
$\Omega_{\rm m}$        & $0.265_{-0.094}^{+0.097}$ & $0.280_{-0.083}^{+0.093}$ & $0.297\pm0.046$  & $0.278\pm0.013$   \\[0.4mm]
$\sigma_{8}$            & $0.86\pm0.12$ & $0.84\pm0.11$ & $0.804_{-0.046}^{+0.048}$  & $0.819_{-0.034}^{+0.033}$     \\[0.4mm]
$t_{0}/{\rm Gyr}$       & $13.68_{-0.23}^{+0.22}$ & $13.77_{-0.14}^{+0.15}$ & $13.797\pm0.092$  & $13.784\pm0.072$    \\[0.4mm] 
$h$                     & $0.75_{-0.13}^{+0.14}$ & $0.72\pm0.11$ & $0.684\pm0.052$  & $0.701\pm0.016$  \\[0.4mm]
$f(z_{\rm m})$          & $0.768_{-0.023}^{+0.024}$ & $0.762\pm0.020$ & $0.757\pm0.016$  & $0.754\pm0.016$    \\
\hline
\end{tabular}
\label{tab:wa_full}
\end{table*}

\begin{table*} 
\centering
  \caption{
    The marginalized 68\% constraints on the cosmological parameters of the $\Lambda$CDM model extended 
    by treating $f(z_{\rm m})$ as a free parameter, and when $f(z_{\rm m})$ and $w_{\rm DE}$ are varied simultaneously.
    The second and third columns correspond to the constraints obtained by combining the CMB data with the CMASS clustering
    wedges, while the last two columns show the result of including also the additional BAO and SN measurements.
}
    \begin{tabular}{@{}lcccc@{}}
    \hline
& \multicolumn{2}{c}{\multirow{2}{*}{CMB+$(\xi_{\perp}(s),\xi_{\parallel}(s))$}}    &\multicolumn{2}{c}{CMB+$(\xi_{\perp}(s),\xi_{\parallel}(s))$}   \\
&                       &                                    &      \multicolumn{2}{c}{+BAO+SN}         \\  
\hline
$f(z_{\rm m})$          &$0.719_{-0.096}^{+0.092}$  & $0.76\pm0.14$     & $0.715_{-0.098}^{+0.095}$   &  $0.706_{-0.099}^{+0.096}$\\[0.4mm]
$w_{\rm DE}$            &     --                    & $-0.95\pm0.17$    &         --                  &  $-1.035_{-0.069}^{+0.071}$ \\[0.4mm]
$100\,\Theta$           & $1.0409\pm0.0014$          & $1.0411\pm0.0014$ & $1.0409\pm0.0014$           &  $1.0407\pm0.0014$   \\[0.4mm] 
$100\,\omega_{\rm b}$   & $2.224\pm0.035$            & $2.231\pm0.038$   & $2.225\pm0.033$             &  $2.219\pm0.035$     \\[0.4mm] 
$100\,\omega_{\rm c}$   & $11.53_{-0.28}^{+0.27}$    & $11.42\pm0.39$    & $11.52\pm0.21$              &  $11.64\pm0.28$  \\[0.4mm] 
$n_{\rm s}$             & $0.9633\pm0.0087$          & $0.9667\pm0.011$  & $0.9633\pm0.0084$           &  $0.9617_{-0.0091}^{+0.0090}$ \\[0.4mm] 
$\ln(10^{10}A_{\rm s})$ & $3.116\pm0.025$            & $3.113\pm0.027$   & $3.116\pm0.024$             &  $3.119\pm0.025$  \\ 
\hline
$\Omega_{\rm DE}$       & $0.715\pm0.015$            & $0.704\pm0.037$   & $0.716\pm0.011$             &  $0.718\pm0.012$   \\[0.4mm]
$\Omega_{\rm m}$        & $0.285\pm0.015$           & $0.296\pm0.037$   & $0.284\pm0.011$             &  $0.282\pm0.012$  \\[0.4mm]
$\sigma_{8}$            & $0.828\pm0.016$            & $0.807_{-0.068}^{+0.067}$   & $0.828\pm0.015$   &  $0.842\pm0.032$    \\[0.4mm]
$t_{0}/{\rm Gyr}$       & $13.754\pm0.068$           & $13.79\pm0.11$    & $13.753\pm0.062$            &  $13.750\pm0.064$   \\[0.4mm] 
$h$                     & $0.695\pm0.012$            & $0.684\pm0.048$     & $0.6956\pm0.0088$         &  $0.701\pm0.015$  \\
\hline
\end{tabular}
\label{tab:fg_full}
\end{table*}

\end{document}